\documentclass{article}
\usepackage{amsfonts}
\usepackage{makeidx}
\usepackage{latexsym,amsmath,amssymb,amscd}
\usepackage{makecell}
\usepackage{xcolor}
\usepackage[all]{xy}
\usepackage{float}
\usepackage{longtable}

\allowdisplaybreaks

\newtheorem{theorem}{Theorem}

\begin{document}

\title{Integrable and superintegrable 3d Newtonian potentials using quadratic first integrals: A review}
\author{Antonios Mitsopoulos$^{1,a)}$ and Michael Tsamparlis$^{2,3,b)}$ \\
{\ \ }\\
$^{1}${\textit{Faculty of Physics, Department of
Astronomy-Astrophysics-Mechanics,}}\\
{\ \textit{University of Athens, Panepistemiopolis, Athens 157 83, Greece}}
\\
$^{2}${\textit{NITheCS, National Institute for Theoretical and Computational Sciences,}}\\
{\textit{Pietermaritzburg 3201, KwaZulu-Natal, South Africa}}\\
$^{3}${\textit{TCCMMP, Theoretical and Computational Condensed Matter and Materials Physics Group, }}\\
{\textit{School of Chemistry and Physics, University of KwaZulu-Natal,}}\\
{\textit{Pietermaritzburg 3201, KwaZulu-Natal, South Africa}}\\
\vspace{12pt} 
\\
$^{a)}$Author to whom correspondence should be addressed:
antmits@phys.uoa.gr 
\\
$^{b)}$Email: mtsampa@phys.uoa.gr \\
}
\date{}
\maketitle

\begin{abstract}
The determination of the first integrals (FIs) of a dynamical system and the subsequent assessment of their integrability or superintegrability in a systematic way is still an open subject. One method which has been developed along these lines for holonomic autonomous dynamical systems with dynamical equations $\ddot{q}^{a}= -\Gamma_{bc}^{a}(q) \dot{q}^{b}\dot{q}^{c} -Q^{a}(q)$, where $\Gamma_{bc}^{a}(q)$ are the coefficients of the Riemannian connection defined by the kinetic metric of the system and $-Q^{a}(q)$ are the generalized forces, is the so-called direct method. According to this method, one assumes a general functional form for the FI $I$ and requires the condition $\frac{dI}{dt} =0$ along the dynamical equations. This results to a system of partial differential equations (PDEs) to which one adds the necessary integrability conditions of the involved scalar quantities. It is found that the final system of PDEs breaks into two sets: a. One set containing geometric elements only and b. A second set with geometric and dynamical quantities. Then, provided the geometric quantities are known or can be found, one uses the second set to compute the FIs and, accordingly, assess on the integrability of the dynamical system. The `solution' of the system of PDEs for quadratic FIs (QFIs) has been given in a recent paper (M. Tsamparlis and A. Mitsopoulos, J. Math. Phys. \textbf{61}, 122701 (2020) ). In the present work, we consider the application of this `solution' to Newtonian autonomous conservative dynamical systems with three degrees of freedom, and compute integrable and superintegrable potentials $V(x,y,z)$ whose integrability is determined via autonomous and/or time-dependent QFIs. The geometric elements of these systems are the ones of the Euclidean space $E^{3}$ which are known. Setting various values for the parameters determining the geometric elements, we determine in a systematic way all known integrable and superintegrable potentials in $E^{3}$ together with new ones obtained in this work. For easy reference, the results are collected in tables so that the present work may act as an updated review of the QFIs of Newtonian autonomous conservative dynamical systems with three degrees of freedom. It is emphasized that by assuming different values for the parameters, other authors may find more integrable potentials of this type of systems.
\end{abstract}


\section{Introduction}

\label{sec.3dint}

According to Liouville integrability theorem \cite{Arnold 1989}, a three-dimensional (3d) Newtonian autonomous conservative system is (Liouville) integrable if it admits three (functionally) independent first integrals (FIs) in involution. Integrable systems that admit five independent FIs are called maximally superintegrable, while if they admit four independent FIs they are called minimally superintegrable. A superintegrable potential is always integrable; however, some authors \cite{Evans 1990, Miller 2005, Kalnins 2006Pr, Kalnins 2007} define superintegrability without the requirement of integrability, that is, they look only for sets of independent FIs whose number exceeds the degrees of freedom of the system.

For 3d Newtonian autonomous conservative systems one quadratic FI (QFI) is the Hamiltonian $H$; therefore, one needs two additional independent autonomous\footnote{These additional FIs must be autonomous because the Poisson bracket (PB) of the Hamiltonian with an arbitrary time-dependent FI $J(t,q,\dot{q})$ does not vanish. Indeed, we have $\{H,J\}= \frac{\partial J}{\partial t}\neq 0$.} FIs in involution in order to establish integrability. If in addition to these FIs there exist one/two more independent autonomous or time-dependent FIs, then the system is minimally/maximally superintegrable. Besides establishing superintegrability, time-dependent FIs can be used also to establish the integrability of a dynamical system provided they are in involution (see e.g. \cite{Kozlov 1983, Vozmishcheva 2005}).

The maximum number of independent autonomous FIs of a Hamiltonian dynamical system of $n$ degrees of freedom is $2n-1$. However, if time-dependent FIs are considered, this maximum limit can be exceeded. For example, the 3d potential $V =-kr^{2}$, where $r= \sqrt{x^{2} +y^{2} +z^{2}}$ and $k$ is an arbitrary constant, admits the six (five are enough) time-dependent linear FIs (LFIs) $I_{3a\pm}$, $a=1,2,3$ (see Table V in \cite{Tsamparlis 2020}):
\begin{eqnarray*}
k>0&:& I_{3a\pm}= e^{\pm\sqrt{2k}t} \left( \dot{q}_{a} \mp \sqrt{2k} q_{a} \right) \\
k<0&:& I_{3a\pm}= e^{\pm i\sqrt{-2k}t} \left( \dot{q}_{a} \mp i\sqrt{-2k} q_{a} \right)
\end{eqnarray*}
which are functionally independent. Since the three LFIs $I_{3a+}$ (or $I_{3a-}$) are also in involution, the considered 3d potential is superinetgarble.

Concerning the number of the free parameters that define a 3d superintegrable potential, the following terminology is used (see e.g. \cite{Kalnins 2007}): \newline
a. The degenerate (or three-parameter) potentials, and \newline
b. The non-degenerate (or four-parameter) potentials.

In many works \cite{Miller 2005, Kalnins 2006Pr, Kalnins 2007, Kalnins JMP2007}, the term second order superintegrable potentials is used for potentials that are superintegrable due to QFIs only. Such potentials have the following special properties \cite{Miller 2005, Kalnins 2006Pr}: \newline
1) Multi-integrability. They are integrable in multiple ways  and the comparison of ways of integration leads to new facts about the system. \newline
2) They are multi-separable. \newline
3) The second order symmetries expressed by second order Killing tensors (KTs) generate a closed quadratic algebra. In the quantum case, the representation of this algebra yields results concerning the spectral resolution of the Schr\"{o}dinger operator and the other symmetry operators.

There are two types of integrable potentials in $E^{3}$. The decomposable potentials (or 2+1 separable integrable potentials) generated from integrable potentials in $E^{2}$ and the non-decomposable ones.

Let $V(x,y)$ be a 2d integrable potential in $E^{2}$ which admits an additional autonomous FI $I_{1}$. Then, the 3d Newtonian $z$-separable potential $\bar{V}(x,y,z) =V(x,y) +F(z)$, where $F$ is an arbitrary smooth function of $z$, is a $2+1$ separable integrable potential in $E^{3}$. The integrability of these potentials is due to the three independent FIs $H, I_{1}$ and $I_{2}= \frac{1}{2}\dot{z}^{2} +F(z)$ which are in involution. If $V(x,y)$ is superintegrable with respect to (wrt) two additional FIs, say $J_{1}$ and $J_{2}$, then $\bar{V}(x,y,z)$ is minimally superintegrable because of the four independent FIs $H, J_{1}, J_{2}$, and $I_{2}$. If in addition to $J_{1}$ and $J_{2}$ the 2d superintegrable potential $V(x,y)$ admits also a time-dependent FI $J_{3}$, then $\bar{V}(x,y,z)$ is maximally superintegrable. For example, the second potential of Table II in \cite{Evans 1990} is not minimally superintegrable but maximally superintegrable because it admits in addition the time-dependent FIs $I_{73a}$ and $I_{73b}$ from the last Table of \cite{MitsTsam sym}.

The non-decomposable (i.e. non-separable) 3d Newtonian integrable potentials $V(x,y,z)$ cannot be written in the form $\bar{V}(x,y,z) =V(x,y) +F(z)$ where $V(x,y)$ is a 2d Newtonian integrable potential. In general, their determination is more difficult and various methods of escalating complexity have been proposed. Furthermore, the existing results concern autonomous FIs only and are limited in number. The purpose of the present work is to provide a systematic (i.e. algorithmic) method which enables one to determine integrable and superintegrable potentials in $E^3$ using autonomous and time-dependent QFIs. The method relies on Theorem \ref{The first integrals of an
autonomous holonomic dynamical system} \cite{Tsamparlis 2020B} (see section \ref{sec.e3.1}) which relates the QFIs of the dynamical system with the dynamical elements (i.e. the potential) and the geometry defined by the kinetic energy of the system. The structure of the paper is as follows.

In section \ref{sec.2d3d}, we determine the 3d integrable/superintegrable 2+1 decomposable potentials directly from the well-known 2d integrable/superintegrable potentials listed in the reference works \cite{MitsTsam sym} and \cite{Hietarinta 1987}. The results are presented in tables where the known potentials with the corresponding reference are listed together with the new ones determined in this work. In section \ref{sec.e3.1}, we state Theorem \ref{The first integrals of an autonomous holonomic dynamical system} from which follows that there are three types of QFIs to consider, denoted as $I_{(1,\ell)}, I_{(2,\ell)}, I_{(3)}$, which are expressed in terms of the geometric elements of the kinetic metric and the potential function. In section \ref{sec.e3.2}, we state the geometric quantities of $E^3$ which are required for the application of Theorem \ref{The first integrals of an autonomous holonomic dynamical system}. It is seen that the number of parameters introduced from the KT components is large. This remark and the fact that the associated system of PDEs is overdetermined have the result that one will find special solutions only by assuming particular values of the geometric parameters. In section \ref{sec.e3.pot.1}, we consider the QFI $I_{(1,1)}$ ($\ell=1$) and the relevant PDEs for this case. We consider various values for the parameters and recover all the existing results together with new ones. For easy reference, the various potentials are grouped in Tables \ref{Table.e3.1} - \ref{Table.e3.3}. In section \ref{sec.e3.pot.2}, we consider the potentials admitting QFIs of the type $I_{(2,0)}$ ($\ell=0$). These results are presented in Tables \ref{Table.e3.4} - \ref{Table.e3.6}. In section \ref{sec.e3.pot.3}, we consider time-dependent LFIs/QFIs of the type $I_{(3)}$ and the results are collected in Tables \ref{Table.e3.7} - \ref{Table.e3.9}. In section \ref{Comparison with other works}, we compare and discuss the results listed in the tables with the existing results of the literature. Finally, in section \ref{Conclusions}, we draw our conclusions.

\subsection*{List of abbreviations and notations/conventions}

For the convenience of the reader, we give a list of abbreviations and notations used throughout the text.

Abbreviations:
\begin{itemize}

\item
FI $=$ first integral

\item
HV $=$ homothetic vector

\item
KT $=$ Killing tensor

\item
KV $=$ Killing vector

\item
LFI $=$ linear first integral

\item
$N$d $=$ $N$-dimensional

\item
ODE $=$ ordinary differential equation

\item
PB $=$ Poisson bracket

\item
PDE $=$ partial differential equation

\item
QFI $=$ quadratic first integral

\end{itemize}

Mathematical notations/conventions:
\begin{itemize}
\item
$E^{n}$ $=$ $n$-dimensional Euclidean space

\item
$r=\sqrt{x^{2}+y^{2}+z^{2}}$, $R=\sqrt{x^{2}+y^{2}}$, $\tan\theta= \frac{y}{x}$, and $w=x+iy=Re^{i\theta}$.

\item
The angular momentum $\mathbf{M}\equiv M_{i}= \left( M_{1}, M_{2}, M_{3}\right)$ $=\left( y\dot{z} -z\dot{y}, z\dot{x} -x\dot{z}, x\dot{y} -y\dot{x} \right)$ with square magnitude $\mathbf{M}^{2}= M_{1}^{2}+M_{2}^{2}+M_{3}^{2}$.

\item
The kinetic metric $\gamma_{ab}(q)$ of the dynamical system is used for lowering and raising the indices.

\item
A comma indicates partial derivative and a semicolon Riemannian covariant derivative.
\end{itemize}

Coordinate systems of $E^{3}$:
\begin{itemize}
\item
Cartesian coordinates: $(x,y,z)$.

\item
Spherical coordinates: $(r,\theta,\phi)$ with  $x=r\sin\theta \cos\phi$, $y= r\sin\theta \sin\phi$ and $z=r\cos\theta$.

\item
Parabolic cylindrical coordinates: $\left( \lambda', \mu', z\right)$ with $\lambda'= R+y$ and $\mu'=R-y$.

\item
Rotational parabolic coordinates: $\left(\zeta, \eta, \phi\right)$ with $\zeta= r+z$, $\eta= r-z$, $\phi= \tan^{-1} \left( \frac{y}{x} \right)$ or, equivalently, $x= \sqrt{\zeta\eta} \cos\phi$, $y=\sqrt{\zeta\eta} \sin\phi$, $z=\frac{1}{2}\left( \zeta -\eta \right)$.

\end{itemize}

\section{Integrable/superintegrable 2+1 separable potentials}

\label{sec.2d3d}

As it has been remarked, the $2+1$ separable integrable/superintegrable potentials in $E^{3}$ are given in terms of the integrable/superintegrable potentials $\Phi(x,y)$ in $E^{2}$. From the latter potentials, the ones that admit LFIs/QFIs are collected in the review papers \cite{MitsTsam sym} and \cite{Hietarinta 1987}. Using these results, the $2+1$ separable potentials in $E^{3}$
\begin{equation}
V(x,y,z)= \Phi(x,y) +F(z) \label{eq.e23.1}
\end{equation}
where $F(z)$ is an arbitrary smooth function, are  integrable/superintegrable due to the additional QFI $I= \frac{1}{2}\dot{z}^{2} +F(z)$ which is in involution with the FIs of $\Phi(x,y)$.

Applying the above procedure to the results of \cite{MitsTsam sym, Hietarinta 1987}, we find the integrable and superintegrable potentials in $E^{3}$ listed in Tables \ref{Table.e23.1} - \ref{Table.e23.3}. The QFI of the Hamiltonian $H$ is not included in the tables. In Tables \ref{Table.e23.2} and \ref{Table.e23.3}, we compare with the results of \cite{Evans 1990}. A similar comparison cannot be done in Table \ref{Table.e23.1} because in \cite{Evans 1990} only superintegrable potentials are considered. Concerning the notation, we set $r=\sqrt{x^{2}+y^{2}+z^{2}}$, $R=\sqrt{x^{2}+y^{2}}$ and the angular momentum $M_{i}= \left( y\dot{z} -z\dot{y}, z\dot{x} -x\dot{z}, x\dot{y} -y\dot{x} \right)$.

\newpage

\begin{longtable}{|l|l|}
\hline
\multicolumn{2}{|c|}{{\large{Integrable $2+1$ separable potentials}}} \\ \hline
{\large Potential} & {\large LFIs and
QFIs} \\ \hline
$V= F_{1}\left(\frac{R^{2}}{2} +b_{1}y-b_{2}x \right) +F_{2}(z)$ & \makecell[l]{$I_{1} =M_{3} -b_{1}\dot{x} -b_{2}\dot{y}$, $I_{2}= \frac{1}{2}\dot{z}^{2} +F_{2}(z)$} \\ \hline
$V= \frac{F_{1}\left( \frac{y}{x}\right) }{R^{2}}%
+F_{2}(R) +F_{3}(z)$ & \makecell[l]{$I_{1}=M_{3}^{2}+2F_{1}\left( \frac{y}{x}\right)$, $I_{2}= \frac{1}{2}\dot{z}^{2} +F_{3}(z)$} \\ \hline
$V= \frac{k}{x^{2}+\ell y^{2}} + F_{1}(R) +F_{2}(z)$ & $I_{1} =M_{3}^{2} + \frac{2k(1-\ell) y^{2}}{x^{2}+\ell y^{2}}$, $I_{2}= \frac{1}{2}\dot{z}^{2} +F_{2}(z)$ \\
\hline
\makecell[l]{$V= \frac{F_{1}(u)-F_{2}(v)}{u^{2}-v^{2}} +F_{3}(z)$ \\
$u^{2}=R^{2}+A+\left[ (R^{2}+A)^{2}-4Ax^{2}\right] ^{1/2}$ and \\
$v^{2}=R^{2}+A-\left[ (R^{2}+A)^{2}-4Ax^{2}\right] ^{1/2}$} & \makecell[l]{$I_{1}=M_{3}^{2} +A\dot{x}^{2} +\frac{v^{2}F_{1}(u) -u^{2}F_{2}(v)}{u^{2}-v^{2}}$ \\ $I_{2}= \frac{1}{2}\dot{z}^{2} +F_{3}(z)$} \\ \hline
\makecell[l]{$V= \frac{F_{1}(u)-F_{2}(v)}{u^{2}-v^{2}} +F_{3}(z)$ \\
$u^{2}=R^{2}+\left[ R^{4}-4A(x\pm iy)^{2}\right] ^{1/2}$ and \\
$v^{2}=R^{2}-\left[ R^{4}-4A(x\pm iy)^{2}\right] ^{1/2}$} & \makecell[l]{$I_{1}=M_{3}^{2}+A(\dot{x}\pm i\dot{y})^{2}+\frac{v^{2}F_{1}(u)-u^{2}F_{2}(v)}{u^{2}-v^{2}}$ \\ $I_{2}= \frac{1}{2}\dot{z}^{2} +F_{3}(z)$} \\ \hline
$V= \frac{F_{1}(R+y)+F_{2}(R-y)}{R} +F_{3}(z)$ & \makecell[l]{$I_{1}= -M_{3}\dot{x} +\frac{(R+y)F_{2}(R-y) -(R-y)F_{1}(R+y)}{R}$ \\ $I_{2}= \frac{1}{2}\dot{z}^{2} +F_{3}(z)$} \\ \hline
\makecell[l]{$V= \bar{w}^{-1/2}\left[
F_{1}(w+\sqrt{\bar{w}})+F_{2}(w-\sqrt{\bar{w}})\right] +F_{3}(z)$ \\ $w=x+iy$ and $\bar{w}=x-iy$} & \makecell[l]{$I_{1}= -M_{3}(\dot{x} +i\dot{y})+\frac{i}{8}(\dot{x} -i\dot{y})^{2} +$ \\ \qquad \enskip $+i\left(
1-\frac{w}{\sqrt{\bar{w}}}\right) F_{1}(w+\sqrt{\bar{w}})+$ \\ \qquad \enskip $+i\left(
-1-\frac{w}{\sqrt{\bar{w}}}\right) F_{2}(w-\sqrt{\bar{w}})$ \\ $I_{2}= \frac{1}{2}\dot{z}^{2} +F_{3}(z)$} \\ \hline
\makecell[l]{$V= \frac{F_{1}(w)}{r}+F_{2}^{\prime }(w) +F_{3}(z)$ \\ $F_{2}^{\prime}=\frac{dF_{2}}{dw}$ and $w=x\pm iy$} & \makecell[l]{$I_{1}= -M_{3}(\dot{x}\pm i\dot{y})-iwV+iF_{2}(w)$ \\ $I_{2}= \frac{1}{2}\dot{z}^{2} +F_{3}(z)$} \\ \hline
$V=F_{1}(x)+F_{2}(y) +F_{3}(z)$ & $I_{1}=\frac{1}{2} \dot{x}^{2}+F_{1}$, $I_{2}=\frac{1}{2}\dot{y}^{2} +F_{2}$, $I_{3}= \frac{1}{2}\dot{z}^{2}+F_{3}$ \\ \hline
\makecell[l]{$V= F_{1}\left(y+b_{0}x+\sqrt{b_{0}^{2}+1}x \right) +$ \\ \qquad \enskip $+ F_{2}\left(y+b_{0}x-\sqrt{b_{0}^{2}+1}x\right) +F_{3}(z)$ \\
where $b_{0}\equiv \frac{A-B}{2C}$} & \makecell[l]{$I_{1}=A\dot{x}^{2}+B\dot{y}^{2}+ 2C\dot{x}\dot{y}+(A+B)(F_{1} +F_{2})+$ \\ \qquad \enskip $+
2C\sqrt{b_{0}^{2}+1}(F_{1}-F_{2})$ \\ $I_{2}= \frac{1}{2}\dot{z}^{2} +F_{3}(z)$ \\ } \\ \hline
$V(b_{0}=0) =F_{1}(y+x)+F_{2}(y-x)+F_{3}(z)$ & $I_{1}=\dot{x}\dot{y}+F_{1}-F_{2}$, $I_{2}= \frac{1}{2}\dot{z}^{2} +F_{3}(z)$ \\ \hline
\caption{\label{Table.e23.1} Integrable potentials $V(x,y,z)= \Phi(x,y) +F(z)$ in $E^{3}$, where $\Phi(x,y)$ are integrable potentials in $E^{2}$.}
\end{longtable}

\newpage

\begin{longtable}{|l|c|l|}
\hline
\multicolumn{3}{|c|}{{\large{Minimally superintegrable $2+1$ separable potentials}}} \\ \hline
{\large Potential} & {\large Ref \cite{Evans 1990}} & {\large LFIs and
QFIs} \\ \hline
\makecell[l]{$V= cx+F_{1}(y-bx)+F_{2}(z)$ \\ $c\neq0$, $\frac{d^{2}F_{1}}{dw^{2}}\neq 0$ and $w\equiv y-bx$} & New & \makecell[l]{$I_{1}=\dot{x}+b\dot{y}+ct$ \\ $I_{2}= (\dot{x}+b\dot{y})^{2}+2c(x+by)$ \\ $I_{3}= \frac{1}{2} \dot{z}^{2} +F_{2}(z)$} \\ \hline
\makecell[l]{$V= F_{1}(y-bx)+F_{2}(z)$ \\ $\frac{d^{2}F_{1}}{dw^{2}}\neq 0$ and $w\equiv y-bx$} & New & \makecell[l]{$I_{1}= \dot{x}+b\dot{y}$ \\ $I_{2}= t(\dot{x}+b\dot{y}) -(x+by)$ \\ $I_{3}= \frac{1}{2} \dot{z}^{2} +F_{2}(z)$} \\ \hline
$V= \frac{k_{1}}{2}(x^{2}+4y^{2}) + \frac{k_{2}}{x^{2}} + k_{3}y +F(z)$ & \makecell[c]{Table II \\ $k_{3}=0$ \\ $x \leftrightarrow y$} & \makecell[l]{$I_{1} =
M_{3}\dot{x} +k_{1}yx^{2} -\frac{2k_{2}y}{x^{2}} +\frac{k_{3}}{2} x^{2}$ \\ $I_{2}= \frac{1}{2}\dot{x}^{2} +\frac{k_{1}}{2}x^{2} + \frac{k_{2}}{x^{2}}$ \\ $I_{3}= \frac{1}{2}\dot{y}^{2} + 2k_{1}y^{2}+ k_{3}y$ \\ $I_{4}= \frac{1}{2}\dot{z}^{2} +F(z)$} \\ \hline
$V = \frac{k_{1}}{x^{2}} + \frac{k_{2}}{R} + \frac{k_{3}y}{Rx^{2}} +F(z)$ & \makecell[c]{Table II \\ $x \leftrightarrow y$} & \makecell[l]{$I_{1}= M_{3}^{2} + 2k_{1}\frac{y^{2}}{x^{2}} + 2k_{3}\frac{Ry}{x^{2}}$ \\ $I_{2}=M_{3}\dot{x} -2k_{1} \frac{y}{x^{2}} -k_{2}\frac{y}{R} -k_{3}\frac{x^{2}+2y^{2}}{Rx^{2}}$ \\ $I_{3}= \frac{1}{2}\dot{z}^{2} +F(z)$} \\ \hline
$V = \frac{k_{1}}{R} + k_{2} \frac{\sqrt{R+y}}{R} + k_{3}\frac{\sqrt{R-y}}{R} +F(z)$ & Table II & \makecell[l]{$I_{1} = M_{3}\dot{x} -\frac{k_{1}y}{R} -\frac{k_{3}(R+y)\sqrt{R-y} -k_{2}(R-y)\sqrt{R+y}}{R}$ \\
$I_{2}= M_{3}\dot{y} + G(x,y)$ \\ $I_{3}= \frac{1}{2}\dot{z}^{2} +F(z)$ \\ $G_{,x}=-yV_{,y}$ and $G_{,y}=2xV_{,y} -yV_{,x}$} \\ \hline
$V =F_{1}(x) +\frac{k}{\left( y +c\right)^{2}} +F_{2}(z)$ & New & \makecell[l]{$I_{1}= \frac{1}{2}\dot{x}^{2} +F_{1}$ \\ $I_{2}= \frac{1}{2}\dot{y}^{2} +\frac{k}{\left( y+c\right)^{2}}$ \\ $I_{3}= \frac{1}{2}\dot{z}^{2} +F_{2}$ \\ $I_{4}=-\frac{t^{2}}{2}\dot{y}^{2}+t(y+c)\dot{y}-
t^{2}\frac{k}{(y+c)^{2}}-\frac{1}{2}y^{2}-cy$} \\ \hline
\makecell[l]{$V= \frac{\lambda}{2}R^{2} +b_{1}y -b_{2}x +F(z)$ \\ $\lambda \neq 0$} & New & \makecell[l]{$I_{1}=\lambda M_{3} -b_{1}\dot{x} -b_{2}\dot{y}$ \\ $I_{2}= \frac{1}{2}\dot{x}^{2}+ \frac{1}{2}\lambda x^{2}-b_{2}x$ \\ $I_{3}=\frac{1}{2}\dot{y}^{2}+\frac{1}{2}\lambda y^{2} + b_{1}y$ \\ $I_{4}=\dot{x}\dot{y}+\lambda xy +b_{1}x -b_{2}y$ \\ $I_{5}= \frac{1}{2}\dot{z}^{2} +F(z)$}
\\ \hline
\caption{\label{Table.e23.2} Minimally superintegrable potentials $V(x,y,z)= \Phi(x,y) +F(z)$ in $E^{3}$, where $\Phi(x,y)$ are superintegrable potentials in $E^{2}$.}
\end{longtable}

\newpage

\begin{longtable}{|l|c|l|}
\hline
\multicolumn{3}{|c|}{{\large{Maximally superintegrable $2+1$ separable potentials}}} \\ \hline
{\large Potential} & {\large Ref \cite{Evans 1990}} & {\large LFIs and QFIs} \\ \hline
$V= cx+\lambda y +F(z)$ & New & %
\makecell[l]{$I_{1}=\dot{x}+ct$, $I_{2}= \dot{y} + \lambda t$, $I_{3}=\frac{1}{2} \dot{x}^{2} + cx$, \\ $I_{4}= \frac{1}{2}\dot{y}^{2} + \lambda y$, $I_{5}= \frac{1}{2}\dot{z}^{2} +F(z)$}
\\ \hline
\makecell[l]{$V= cx-\frac{1}{2}\lambda ^{2}y^{2} +F(z)$ \\ $\lambda \neq 0$} & New & \makecell[l]{$I_{1}= \dot{x} +ct$, $I_{2}=e^{\lambda t}(\dot{y}-\lambda y)$, $I_{3}= \frac{1}{2}\dot{x}^{2} +cx$, \\ $I_{4}= \frac{1}{2}\dot{y}^{2}-\frac{1}{2}\lambda^{2}y^{2}$, $I_{5}= \frac{1}{2}\dot{z}^{2} +F(z)$} \\ \hline
\makecell[l]{$V= -\frac{k^{2}}{2}R^{2}+F(z)$ \\ $k\neq 0$} & New & \makecell[l]{$I_{1}=M_{3}$, $I_{2}= \frac{1}{2}\dot{x}^{2}-\frac{1}{2}k^{2} x^{2}$, $I_{3}= \frac{1}{2}\dot{y}^{2} - \frac{1}{2}k^{2} y^{2}$, \\ $I_{4}=\dot{x}\dot{y} - k^{2}xy$, $I_{5}= \frac{1}{2}\dot{z}^{2} +F(z)$, \\
$I_{6\pm}=e^{\pm kt}(\dot{x}\mp kx)$, $I_{7\pm}=e^{\pm kt}(\dot{y} \mp ky)$, \\ $4I_{2}I_{3}= I_{4}^{2} -k^{2}M_{3}^{2}$} \\ \hline
$V= \frac{k}{2}R^{2} + \frac{b}{x^{2}} + \frac{c}{y^{2}} +F(z)$ & Table II & \makecell[l]{$I_{1}= M_{3}^{2} + 2b\frac{y^{2}}{x^{2}} + 2c \frac{x^{2}}{y^{2}}$, $I_{2}= \frac{1}{2}\dot{z}^{2} +F(z)$ \\ $I_{3} =
\frac{1}{2}\dot{x}^{2} + \frac{k}{2}x^{2} + \frac{b}{x^{2}}$, $I_{4} =\frac{1}{2}\dot{y}^{2} + \frac{k}{2}y^{2} + \frac{c}{y^{2}}$ \\ - For $k=0$: \\ $I_{5}= -\frac{t^{2}}{2}\dot{y}^{2} +ty\dot{y}-
t^{2}\frac{c}{y^{2}} -\frac{1}{2}y^{2}$ \\ $I_{6}= -\frac{t^{2}}{2}\dot{x}^{2} +tx\dot{x}-
t^{2} \frac{b}{x^{2}}-\frac{1}{2}x^{2}$ \\ - For
$k=-\frac{\lambda^{2}}{4}\neq0$: \\ $I_{5}= e^{\lambda t} \left[ -\dot{x}^{2}+\lambda x\dot{x} -\frac{\lambda^{2}}{4}x^{2} -\frac{2b}{x^{2}} \right]$ \\ $I_{6}=e^{\lambda t}\left[ -\dot{y}^{2} +\lambda
y\dot{y} -\frac{\lambda^{2}}{4}y^{2} -\frac{2c}{y^{2}} \right]$} \\ \hline
$V= \frac{k_{1}}{\left( x+c_{1} \right)^{2}}+ \frac{k_{2}}{\left(y+c_{2} \right)^{2}} +F(z)$ & New & \makecell[l]{$I_{1}= \frac{1}{2}\dot{x}^{2} +\frac{k_{1}}{\left( x+c_{1} \right)^{2}}$ \\ $I_{2}= \frac{1}{2}\dot{y}^{2} +\frac{k_{2}}{\left( y+c_{2} \right)^{2}}$ \\ $I_{3}= \frac{1}{2}\dot{z}^{2} +F(z)$ \\
$I_{4}=-\frac{t^{2}}{2}\dot{y}^{2}+t(y+c_{2})\dot{y}-
t^{2}\frac{k_{2}}{(y+c_{2})^{2}}-\frac{1}{2}y^{2}-c_{2}y$ \\
$I_{5}=-\frac{t^{2}}{2}\dot{x}^{2}+t(x+c_{1})\dot{x}-
t^{2}\frac{k_{1}}{(x+c_{1})^{2}}-\frac{1}{2}x^{2}-c_{1}x$} \\ \hline
\makecell[l]{$V =-\frac{\lambda ^{2}}{8}R^{2}-\frac{\lambda
^{2}}{4}\left( c_{1}x+ c_{2}y\right)-$ \\ \qquad \quad
$-\frac{k_{1}}{(x+c_{1})^{2}}-\frac{k_{2}}{(y+c_{2})^{2}} +F(z)$ \\ $\lambda \neq0$} & New & \makecell[l]{$I_{1}= \frac{1}{2}\dot{x}^{2} -\frac{\lambda^{2}}{8} x^{2} -\frac{c_{1}\lambda^{2}}{4}x -\frac{k_{1}}{\left( x+c_{1} \right)^{2}}$ \\ $I_{2}= \frac{1}{2}\dot{y}^{2} -\frac{\lambda^{2}}{8} y^{2} -\frac{c_{2}\lambda^{2}}{4}y -\frac{k_{2}}{\left( y+c_{2} \right)^{2}}$ \\ $I_{3}= \frac{1}{2}\dot{z}^{2} +F(z)$ \\ $I_{4}=e^{\lambda
t}\left[ -\dot{x}^{2}+\lambda (x+c_{1})\dot{x}-\frac{\lambda
^{2}}{4}(x+c_{1})^{2}+\frac{2k_{1}}{(x+c_{1})^{2}}\right] $ \\
$I_{5}=e^{\lambda t}\left[ -\dot{y}^{2}+\lambda
(y+c_{2})\dot{y}-\frac{\lambda
^{2}}{4}(y+c_{2})^{2}+\frac{2k_{2}}{(y+c_{2})^{2}}\right]$} \\ \hline
\caption{\label{Table.e23.3} Maximally superintegrable potentials $V(x,y,z)= \Phi(x,y) +F(z)$ in $E^{3}$, where $\Phi(x,y)$ are superintegrable potentials in $E^{2}$.}
\end{longtable}

\textbf{Note 1:} The results indicated as `New' in Tables \ref{Table.e23.2} and \ref{Table.e23.3} do not appear in \cite{Evans 1990} where only autonomous QFIs are considered.
\bigskip

\textbf{Note 2:} In Table II of \cite{Evans 1990}, the potential (see Table \ref{Table.e23.3})
\begin{equation}
V= \frac{k}{2}R^{2} +\frac{b}{x^{2}} +\frac{c}{y^{2}} +F(z) \label{eq.notev.1}
\end{equation}
where $k,b,c$ are arbitrary constants and $F(z)$ is an arbitrary smooth function, is said to be minimally superintegrable because of the four independent autonomous QFIs:
\[
I_{1}= M_{3}^{2} + 2b\frac{y^{2}}{x^{2}} + 2c \frac{x^{2}}{y^{2}}, \enskip I_{2}= \frac{1}{2}\dot{z}^{2} +F(z), \enskip I_{3}= \frac{1}{2}\dot{x}^{2} + \frac{k}{2}x^{2} + \frac{b}{x^{2}}, \enskip I_{4}= \frac{1}{2}\dot{y}^{2} + \frac{k}{2}y^{2} + \frac{c}{y^{2}}.
\]
However, using in addition the time-dependent QFIs:
\[
\text{For $k=0$:} \quad I_{5}= -\frac{t^{2}}{2}\dot{y}^{2} +ty\dot{y}-t^{2}\frac{c}{y^{2}} -\frac{1}{2}y^{2}, \enskip I_{6}= -\frac{t^{2}}{2}\dot{x}^{2} +tx\dot{x}-
t^{2} \frac{b}{x^{2}}-\frac{1}{2}x^{2}
\]
and
\[
\text{For $k=-\frac{\lambda^{2}}{4}\neq0$:} \quad I_{5}= e^{\lambda t} \left[ -\dot{x}^{2}+\lambda x\dot{x} -\frac{\lambda^{2}}{4}x^{2} -\frac{2b}{x^{2}} \right], \enskip I_{6}=e^{\lambda t}\left[ -\dot{y}^{2} +\lambda
y\dot{y} -\frac{\lambda^{2}}{4}y^{2} -\frac{2c}{y^{2}} \right]
\]
it is seen that the potential (\ref{eq.notev.1}) for these values of $k$ is maximally superintegrable.

Moreover, if we assume the canonical transformation $x \to x+c_{1}$ and $y \to y+c_{2}$ where $c_{1}$ and $c_{2}$ are arbitrary constants, it is shown that the potential (\ref{eq.notev.1}) is transformed canonically into the last two potentials of Table \ref{Table.e23.3}. Indeed, for $k=0$, $b=k_{1}$ and $c=k_{2}$, we get the potential
\[
V = \frac{k_{1}}{(x+c_{1})^{2}} +\frac{k_{2}}{(y+c_{2})^{2}} +F(z)
\]
while for $k=-\frac{\lambda^{2}}{4}$, $b=-k_{1}$ and $c=-k_{2}$, we get the potential
\[
V =-\frac{\lambda ^{2}}{8}R^{2}-\frac{\lambda
^{2}}{4}\left( c_{1}x+ c_{2}y\right)- \frac{k_{1}}{(x+c_{1})^{2}}-\frac{k_{2}}{(y+c_{2})^{2}} -\frac{\lambda^{2}}{8}(c_{1}^{2} +c_{2}^{2}) +F(z).
\]
The constant term $-\frac{\lambda^{2}}{8}(c_{1}^{2} +c_{2}^{2})$ is overlooked because it does not contribute to the dynamical equations.
\bigskip

\textbf{Note 3:} From Table \ref{Table.e23.2}, we observe that the minimally superintegrable potential
\begin{equation}
V = \frac{k_{1}}{R} + k_{2} \frac{\sqrt{R+y}}{R} + k_{3}\frac{\sqrt{R-y}}{R} +F(z) \label{eq.notev.2}
\end{equation}
where $k_{1}, k_{2}, k_{3}$  are arbitrary constants and $F(z)$ is an arbitrary smooth function, admits the two autonomous QFIs:
\begin{eqnarray}
I_{1} &=& M_{3}\dot{x} -\frac{k_{1}y}{R} +\frac{k_{2}(R-y) \sqrt{R+y}}{R} -\frac{k_{3}(R+y)\sqrt{R-y}}{R} \label{eq.notev.3.1} \\
I_{2}&=& M_{3}\dot{y} + G(x,y). \label{eq.notev.3.2}
\end{eqnarray}
The function $G(x,y)$ must satisfy the system of PDEs:
\begin{eqnarray}
G_{,x} +yV_{,y} &=& 0 \label{eq.vs1} \\
G_{,y} +yV_{,x} -2xV_{,y} &=& 0. \label{eq.vs2}
\end{eqnarray}
Using the parabolic cylindrical coordinates $(\lambda',\mu',z)$ (see eqs. (3.19) and (3.51) in \cite{Evans 1990}) with $\lambda'=R +y$ and $\mu'=R-y$,
the QFI (\ref{eq.notev.3.1}) becomes\footnote{We recall that the coordinates $\lambda', \mu'$ are either positive or zero because $\lambda' +\mu'= 2R$, $\lambda' -\mu'= 2y$, and $\lambda'\mu'=x^{2}$.}
\begin{equation}
I_{1}= M_{3}\dot{x} -\frac{2}{\lambda' +\mu'} \left[ \frac{k_{1}}{2}(\lambda' -\mu') -k_{2}\mu'\sqrt{\lambda'} +k_{3}\lambda'\sqrt{\mu'} \right]. \label{eq.vs3}
\end{equation}
The QFI $I_{2}$ in eq. (3.57) of \cite{Evans 1990} is not correct and should be replaced by the QFI (\ref{eq.vs3}).

In the parabolic cylindrical coordinates $(u,v,z)$ with $u=R +x$, $v=R-x$ and\footnote{
For $x,y>0$ we have: $\sqrt{R+x} +\sqrt{R-x}= \sqrt{2} \sqrt{R+y}$ and $\sqrt{R+x} -\sqrt{R-x} =\sqrt{2} \sqrt{R-y}$.} $x,y>0$, the system of PDEs (\ref{eq.vs1}) - (\ref{eq.vs2}) becomes $G_{,v}=uV_{,v}$ and $G_{,u}=-vV_{,u}$. The solution of this system is
\[
G(u,v)= \frac{2}{u+v} \left[ \frac{k_{1}}{2}(u-v) -(k_{2}+k_{3})v\sqrt{\frac{u}{2}} +(k_{2}-k_{3}) u \sqrt{\frac{v}{2}} \right]
\]
or, equivalently, in Cartesian coordinates
\[
G(x,y)= \frac{1}{R} \left[ k_{1}x -(k_{2}+k_{3})(R-x)\sqrt{\frac{R+x}{2}} +(k_{2}-k_{3}) (R+x) \sqrt{\frac{R-x}{2}} \right].
\]
Then, the QFI (\ref{eq.notev.3.2}) is
\begin{equation}
I_{2}= M_{3}\dot{y} +\frac{2}{u+v} \left[ \frac{k_{1}}{2}(u-v) -(k_{2}+k_{3})v\sqrt{\frac{u}{2}} +(k_{2}-k_{3}) u \sqrt{\frac{v}{2}} \right]. \label{eq.vs4}
\end{equation}
There is a misprint in the QFI $I_{3}$ of eq. (3.57) in \cite{Evans 1990}; the correct answer is the QFI (\ref{eq.vs4}).
\bigskip

\textbf{Note 4:} The two superintegrable potentials given in eq. (17) of \cite{Kalnins 2006Pr} are subcases of the potential (see Table \ref{Table.e23.2})
\begin{equation}
V= \frac{\lambda}{2}R^{2} +b_{1}y -b_{2}x +F(z) \label{eq.notev.4}
\end{equation}
for $F(z)= \frac{\lambda}{2}z^{2} +b_{3}z$ and $F(z)= \frac{\lambda}{8}z^{2} +\frac{b_{3}}{z^{2}}$, where $b_{3}$ is an arbitrary constant.
\bigskip

\textbf{Note 5:} The potential (see Table \ref{Table.e23.2})
\begin{equation}
V_{1}= cx +F_{1}(y-bx) +F_{2}(z) \label{eq.pot23.1}
\end{equation}
where $c$ is an arbitrary non-zero constant, $w\equiv y-bx$ and $\frac{d^{2}F_{1}}{dw^{2}}\neq0$, admits the following LFIs/QFIs (apart from the Hamiltonian $H$):
\[
I_{1}= \dot{x} +b\dot{y} +ct, \enskip I_{2}= t(\dot{x}+b\dot{y}) -(x+by) +\frac{c}{2}t^{2}, \enskip I_{3}=(\dot{x}+b\dot{y})^{2} +2c(x+by), \enskip I_{4}= \frac{1}{2}\dot{z}^{2} +F_{2}(z).
\]
We compute the PBs:
\[
\{H, I_{1}\}=c, \enskip \{H, I_{2}\}= I_{1}, \enskip \{I_{1}, I_{2}\}= 1+b^{2}, \enskip \{I_{1}, I_{3}\}= -2c(1+b^{2}), \enskip \{I_{2}, I_{3}\}= -2(1+b^{2})I_{1}.
\]
The three FIs $H, I_{3}, I_{4}$ are (functionally) independent and in involution; therefore, the potential (\ref{eq.pot23.1}) is integrable. The five FIs $H, I_{1}, I_{2}, I_{3}, I_{4}$ are not independent because $I_{1}^{2}= I_{3} +2cI_{2}$. However, the four FIs $H, I_{3}, I_{4}, I_{1}$, or the $H, I_{3}, I_{4}, I_{2}$, are independent and, therefore, the potential (\ref{eq.pot23.1}) is minimally superintegrable.

\section{The Theorem for QFIs}

\label{sec.e3.1}

In order to compute in a systematic way the QFIs of non-decomposable potentials, we need to recall a theorem which is proved in \cite{Tsamparlis 2020B}.

\begin{theorem}
\label{The first integrals of an autonomous holonomic dynamical system} The independent QFIs of the $n$-dimensional autonomous holonomic dynamical system
\begin{equation}
\ddot{q}^{a}= -\Gamma^{a}_{bc}(q)\dot{q}^{b}\dot{q}^{c} -Q^{a}(q) \label{eq.e3.1}
\end{equation}
where $q^{a}$ are the coordinates of the configuration space, $\dot{q}^{a}= \frac{dq^{a}}{dt}$, $t$ is the time variable, $\Gamma^{a}_{bc}(q)$ are the Riemannian connection coefficients of the kinetic metric $\gamma_{ab}(q)$ defined by the kinetic energy of the system and $-Q^{a}(q)$ are the generalized forces, are the following:
\bigskip

\textbf{Integral 1.}
\begin{eqnarray*}
I_{(1,\ell)} &=& \left( - \frac{t^{2\ell}}{2\ell} L_{(2\ell-1)(a;b)} - ... - \frac{t^{4}}{4} L_{(3)(a;b)} - \frac{t^{2}}{2} L_{(1)(a;b)} + C_{ab} \right) \dot{q}^{a} \dot{q}^{b} + t^{2\ell-1} L_{(2\ell-1)a}\dot{q}^{a} + ... +t^{3}L_{(3)a}\dot{q}^{a} + \\
&& + t L_{(1)a}\dot{q}^{a} + \frac{t^{2\ell}}{2\ell} L_{(2\ell-1)a}Q^{a} +... + \frac{t^{4}}{4} L_{(3)a}Q^{a} + \frac{t^{2}}{2} L_{(1)a}Q^{a} + G(q)
\end{eqnarray*}
where\footnote{We note that for $\ell=0$ the conditions for the QFI $I_{(1,0)}$ are given by nullifying all the vectors $L_{(M)a}$.} $C_{ab}(q)$ and $L_{(M)(a;b)}(q)$ for $M=1,3,...,2\ell-1$ are KTs, $\left( L_{(2\ell-1)b} Q^{b} \right)_{,a}= -2L_{(2\ell-1)(a;b)}Q^{b}$, $\left( L_{(k-1)b} Q^{b} \right)_{,a} =-2L_{(k-1)(a;b)}Q^{b} - k(k+1)L_{(k+1)a}$ for $k=2,4,...,2\ell-2$, and $G_{,a}= 2C_{ab}Q^{b} - L_{(1)a}$.

\textbf{Integral 2.}
\begin{eqnarray*}
I_{(2,\ell)} &=& \left( - \frac{t^{2\ell+1}}{2\ell+1} L_{(2\ell)(a;b)} - ... -
\frac{t^{3}}{3} L_{(2)(a;b)} - t L_{(0)(a;b)} \right) \dot{q}^{a} \dot{q}%
^{b} + t^{2\ell} L_{(2\ell)a}\dot{q}^{a} + ... + t^{2}L_{(2)a}\dot{q}^{a} +
\\
&& + L_{(0)a}\dot{q}^{a}+ \frac{t^{2\ell+1}}{2\ell+1} L_{(2\ell)a}Q^{a} +
... + \frac{t^{3}}{3} L_{(2)a}Q^{a} +t L_{(0)a}Q^{a}
\end{eqnarray*}
where $L_{M(a;b)}(q)$ for $M=0,2,...,2\ell$ are KTs, $\left( L_{(2\ell)b} Q^{b}\right)_{,a} = -2L_{(2\ell)(a;b)}Q^{b}$, and \\ $\left( L_{(k-1)b}Q^{b}\right)_{,a} = -2L_{(k-1)(a;b)}Q^{b} - k(k+1)L_{(k+1)a}$ for $k=1,3,...,2\ell-1$.

\textbf{Integral 3.}
\begin{equation*}
I_{(3)} = e^{\lambda t} \left(-L_{(a;b)}\dot{q}^{a}\dot{q}^{b} + \lambda L_{a} \dot{q}^{a} + L_{a}Q^{a} \right)
\end{equation*}
where the vector $L_{a}(q)$ is such that $L_{(a;b)}$ is a KT and $\left(L_{b}Q^{b}\right)_{,a} = -2L_{(a;b)} Q^{b} - \lambda^{2} L_{a}$.
\end{theorem}

Notation: The Einstein summation convention is used, round (square) brackets indicate symmetrization (antisymmetrization) of the enclosed indices, indices enclosed between vertical lines are overlooked by  antisymmetrization or symmetrization symbols, a comma indicates partial derivative and a semicolon Riemannian covariant derivative.

Before we proceed, we recall the geometric quantities of the Euclidean space $E^{3}$ required by Theorem \ref{The first integrals of an autonomous holonomic dynamical system}.

\section{The geometric quantities of $E^{3}$}

\label{sec.e3.2}

- $E^{3}$ admits three gradient Killing vectors (KVs) $\partial_{x}, \partial_{y}, \partial_{z}$ whose generating functions are $x,y,z$, respectively, and three non-gradient KVs $y\partial _{x}-x\partial y$, $z\partial_{y}-y\partial_{z}$, $z\partial_{x} - x\partial _{z}$. These vectors are written collectively as
\begin{equation}
L_{a}=
\left(
  \begin{array}{c}
    b_{1} -b_{4}y +b_{5}z  \\
    b_{2} +b_{4}x -b_{6}z \\
    b_{3} -b_{5}x +b_{6}y \\
  \end{array}
\right) \label{eq.e3.2}
\end{equation}
where $b_{1}, b_{2}, ..., b_{6}$ are arbitrary constants.

- The general second order KT in $E^{3}$ has independent components:
\begin{eqnarray}
C_{11} &=&\frac{a_{6}}{2}y^{2}+\frac{a_{1}}{2}%
z^{2}+a_{4}yz+a_{5}y+a_{2}z+a_{3}  \notag \\
C_{12} &=&\frac{a_{10}}{2}z^{2}-\frac{a_{6}}{2}xy-\frac{a_{4}}{2}xz- \frac{%
a_{14}}{2}yz-\frac{a_{5}}{2}x-\frac{a_{15}}{2}y+a_{16}z+a_{17}  \notag \\
C_{13} &=&\frac{a_{14}}{2}y^{2}-\frac{a_{4}}{2}xy-\frac{a_{1}}{2}xz- \frac{a_{10}}{2}yz-\frac{a_{2}}{2}x+a_{18}y-\frac{a_{11}}{2}z +a_{19}  \label{eq.e3.3}
\\
C_{22} &=&\frac{a_{6}}{2}x^{2}+\frac{a_{7}}{2}%
z^{2}+a_{14}xz+a_{15}x+a_{12}z+a_{13}  \notag \\
C_{23} &=&\frac{a_{4}}{2}x^{2}-\frac{a_{14}}{2}xy-\frac{a_{10}}{2}xz -\frac{%
a_{7}}{2}yz-(a_{16}+a_{18})x-\frac{a_{12}}{2}y-\frac{a_{8}}{2}z+ a_{20}
\notag \\
C_{33} &=&\frac{a_{1}}{2}x^{2}+\frac{a_{7}}{2}%
y^{2}+a_{10}xy+a_{11}x+a_{8}y+a_{9}  \notag
\end{eqnarray}
where $a_{K}$ with $K=1,2,...,20$ are arbitrary constants.

- The vector $L_{a}$ generating the reducible KT $C_{ab}= L_{(a;b)}$ is
\begin{equation}
L_{a}=\left(
\begin{array}{c}
-a_{15}y^{2}-a_{11}z^{2}+a_{5}xy+a_{2}xz+2(a_{16}+a_{18})yz+
a_{3}x+2a_{4}y+2a_{1}z+a_{6} \\
-a_{5}x^{2}-a_{8}z^{2}+a_{15}xy-2a_{18}xz+a_{12}yz+
2(a_{17}-a_{4})x+a_{13}y+2a_{7}z+a_{14} \\
-a_{2}x^{2}-a_{12}y^{2}-2a_{16}xy+a_{11}xz+a_{8}yz+
2(a_{19}-a_{1})x+2(a_{20}-a_{7})y+a_{9}z+a_{10}%
\end{array}%
\right)  \label{eq.e3.4}
\end{equation}
and the generated KT is
\begin{equation}
C_{ab}=\left(
\begin{array}{ccc}
a_{5}y+a_{2}z+a_{3} & -\frac{a_{5}}{2}x-\frac{a_{15}}{2}y+a_{16}z+a_{17} & -%
\frac{a_{2}}{2}x+a_{18}y-\frac{a_{11}}{2}z+a_{19} \\
-\frac{a_{5}}{2}x-\frac{a_{15}}{2}y+a_{16}z+a_{17} & a_{15}x+a_{12}z+a_{13}
& -(a_{16}+a_{18})x-\frac{a_{12}}{2}y-\frac{a_{8}}{2}z+a_{20} \\
-\frac{a_{2}}{2}x+a_{18}y-\frac{a_{11}}{2}z+a_{19} & -(a_{16}+a_{18})x-\frac{%
a_{12}}{2}y-\frac{a_{8}}{2}z+a_{20} & a_{11}x+a_{8}y+a_{9}%
\end{array}%
\right)  \label{eq.e3.5}
\end{equation}
which is a subcase of the general KT (\ref{eq.e3.3}) for $a_{1}=a_{4}=a_{6}=a_{7}=a_{10}=a_{14}=0$.

\section{The QFI $I_{(1,1)}$ where $\ell=1$}

\label{sec.e3.pot.1}

We set $L_{(1)a}=L_{a}$ and the QFI $I_{(1,\ell)}$ for $\ell=1$ becomes
\begin{equation}
I_{(1,1)}= \left( -\frac{t^{2}}{2} L_{(a;b)} + C_{ab} \right) \dot{q}^{a} \dot{q}^{b} +tL_{a}\dot{q}^{a} + \frac{t^{2}}{2} L_{a}V^{,a} + G(x,y,z) \label{eq.e3pot.1}
\end{equation}
where $C_{ab}$ is a second order KT given by (\ref{eq.e3.3}), the vector $L_{a}$ is given by (\ref{eq.e3.4}), the generated KT $L_{(a;b)}$ is the (\ref{eq.e3.5}) and the following conditions must be satisfied:
\begin{eqnarray}
\left( L_{b} V^{,b} \right)_{,a} &=& -2L_{(a;b)}V^{,b} \label{eq.e3pot.2a} \\
G_{,a}&=& 2C_{ab}V^{,b} - L_{a}. \label{eq.e3pot.2b}
\end{eqnarray}
Equations (\ref{eq.e3pot.2a}) and (\ref{eq.e3pot.2b}) must be supplemented with the three integrability conditions for the function $G$ and the three integrability conditions for the function $L_{a}V^{,a}$.

Finally, we have an overdetermined system of twelve PDEs with unknowns the two functions $G(x,y,z)$ and $V(x,y,z)$, and forty free parameters. Obviously, the general solution is not possible, and we have to look for special solutions which are achieved by introducing simplifying assumptions.

\subsection{Case $L_{a}=0$}

\label{sec.e3.pot.1.1}

In this case, the QFI (\ref{eq.e3pot.1}) is the well-known autonomous QFI
\begin{equation}
I_{(1,1)}(L_{a}=0)= C_{ab} \dot{q}^{a} \dot{q}^{b} + G(x,y,z) \label{eq.e3pot.3}
\end{equation}
where the second order KT $C_{ab}$ has independent components
\begin{eqnarray}
C_{11} &=& \frac{a_{6}}{2}y^{2} +\frac{a_{1}}{2}z^{2} +a_{4}yz +a_{5}y +a_{2}z +a_{3} \notag \\
C_{12} &=& \frac{a_{10}}{2}z^{2} -\frac{a_{6}}{2}xy -\frac{a_{4}}{2}xz- \frac{a_{14}}{2}yz -\frac{a_{5}}{2}x -\frac{a_{15}}{2}y +a_{16}z +a_{17}  \notag \\
C_{13} &=&\frac{a_{14}}{2}y^{2}-\frac{a_{4}}{2}xy-\frac{a_{1}}{2}xz- \frac{a_{10}}{2}yz-\frac{a_{2}}{2}x+a_{18}y-\frac{a_{11}}{2}z +a_{19}  \label{eq.e3pot.4a}
\\
C_{22}&=& \frac{a_{6}}{2}x^{2} +\frac{a_{7}}{2}z^{2} +a_{14}xz +a_{15}x +a_{12}z +a_{13}  \notag \\
C_{23} &=& \frac{a_{4}}{2}x^{2} -\frac{a_{14}}{2}xy -\frac{a_{10}}{2}xz -\frac{a_{7}}{2}yz -(a_{16}+a_{18})x -\frac{a_{12}}{2}y -\frac{a_{8}}{2}z +a_{20}
\notag \\
C_{33} &=&\frac{a_{1}}{2}x^{2}+\frac{a_{7}}{2}%
y^{2}+a_{10}xy+a_{11}x+a_{8}y+a_{9}  \notag
\end{eqnarray}
the parameters $a_{1}, ..., a_{20}$ are arbitrary constants and the function $G(x,y,z)$ satisfies the condition
\begin{equation}
G_{,a}= 2C_{ab}V^{,b}. \label{eq.e3pot.4b}
\end{equation}
The integrability condition $G_{,[ab]}=0$ gives:
\begin{eqnarray}
0&=& C_{12}\left( V_{,yy} -V_{,xx} \right) + \left[ \frac{a_{6}(y^{2} -x^{2})}{2} + \frac{(a_{1}-a_{7})z^{2}}{2} -(a_{14}x -a_{4}y)z -a_{15}x +a_{5}y +(a_{2} -a_{12})z + \right. \notag \\
&& \left. + a_{3} -a_{13}  \right] V_{,xy} + C_{13}V_{,yz} -C_{23}V_{,xz} + \frac{3}{2} (a_{6}y +a_{4}z +a_{5}) V_{,x} -\frac{3}{2} (a_{6}x +a_{14}z +a_{15}) V_{,y}+ \notag \\
&& +\left( \frac{3a_{14}}{2}y -\frac{3a_{4}}{2}x +2a_{18} +a_{16} \right) V_{,z} \label{eq.e3pot.5a} \\
0&=& C_{13} \left( V_{,zz} - V_{,xx} \right) +\left[ \frac{a_{1}(z^{2}-x^{2})}{2} +\frac{(a_{6}-a_{7})y^{2}}{2} -(a_{10}x -a_{4}z)y -a_{11}x +(a_{5}-a_{8})y +a_{2}z + \right. \notag \\
&& \left. +a_{3} -a_{9} \right] V_{,xz}+C_{12}V_{,yz} -C_{23}V_{,xy} +\frac{3}{2}(a_{4}y +a_{1}z +a_{2})V_{,x} +\left( \frac{3a_{10}}{2}z -\frac{3a_{4}}{2}x +2a_{16} +a_{18} \right) V_{,y} -\notag \\
&& -\frac{3}{2} (a_{1}x +a_{10}y +a_{11}) V_{,z} \label{eq.e3pot.5b} \\
0&=& C_{23} \left( V_{,zz} -V_{,yy} \right) +\left[ \frac{a_{7}(z^{2}-y^{2})}{2} +\frac{(a_{6}-a_{1})x^{2}}{2} -(a_{10}y -a_{14}z)x +(a_{15}-a_{11})x -a_{8}y +a_{12}z + \right. \notag \\
&& \left. +a_{13} -a_{9} \right] V_{,yz} +C_{12}V_{,xz} -C_{13}V_{,xy} +\left( \frac{3a_{10}}{2}z -\frac{3a_{14}}{2}y +a_{16} -a_{18} \right) V_{,x} +\frac{3}{2}(a_{14}x +a_{7}z +a_{12})V_{,y}- \notag \\
&& -\frac{3}{2}(a_{10}x +a_{7}y +a_{8})V_{,z}. \label{eq.e3pot.5c}
\end{eqnarray}
We note that in the case of 2d Newtonian potentials the integrability condition $G_{,[ab]}=0$ leads to just one equation, which is the well-known Bertrand-Darboux PDE (see e.g. eq. (28) of \cite{MitsTsam sym} and eq. (3.2.5) of \cite{Hietarinta 1987}).

The system of PDEs (\ref{eq.e3pot.5a}) - (\ref{eq.e3pot.5c}) has to be solved in order to find potentials $V(x,y,z)$ that admit autonomous QFIs of the form (\ref{eq.e3pot.3}). Replacing these potentials in the remaining condition (\ref{eq.e3pot.4b}), we find the functions $G(x,y,x)$ which determine the associated QFIs (\ref{eq.e3pot.3}). Since the general solution $V(x,y,z)$ is not possible, we consider again several cases for various values of the twenty free parameters $a_{1}, a_{2}, ..., a_{20}$.

\subsubsection{The components of the KT $C_{ab}$ are constants}

\label{sec.e3.pot.1.1.1}

In this case, the possibly non-zero parameters are the $a_{3}, a_{9}, a_{13}, a_{17}, a_{19}$, and $a_{20}$. A detailed study leads to the following five cases (only non-vanishing parameters are listed).

1) $a_{3}=a$, $a_{17}=\frac{b}{2}$, and $a_{19}=\frac{c}{2}$, where $a,b,c$ are arbitrary constants\footnote{If instead of the triplet $a_{3}, a_{17}, a_{19}$ we take as non-vanishing parameters the triplets $a_{13}, a_{17}, a_{20}$ or $a_{9}, a_{19}, a_{20}$, the resulting potentials are symmetric up to a cyclic permutation of the coordinates $x,y,z$.}.

The potential is
\begin{equation}
V(x,y,z)= F_{1}\left( cz +by +(\sqrt{a^{2} +b^{2} +c^{2}} +a)x \right) +F_{2}\left( cz +by -(\sqrt{a^{2} +b^{2} +c^{2}} -a)x \right) +F_{3}(bz -cy) \label{eq.e3pot.8d}
\end{equation}
where $F_{1}, F_{2}$, and $F_{3}$ are arbitrary smooth functions of their arguments.

The associated QFI (\ref{eq.e3pot.3}) is
\begin{equation}
I_{(1,1)}= \left( a\dot{x} +b\dot{y} +c\dot{z} \right)\dot{x} +a(F_{1} +F_{2}) +\sqrt{a^{2}+b^{2}+c^{2}}(F_{1} -F_{2}). \label{eq.e3pot.8e}
\end{equation}
We note that the constants $a, b, c$ are parameters of the potential (\ref{eq.e3pot.8d}); therefore, they cannot generate three distinct QFIs.

2)$a_{3}=a$, $a_{13}=-a$, and $a_{17}=ia$, where $a$ is an arbitrary constant.

The potential is
\begin{equation}
V(x,y,z)= F_{1}(w,z) +F_{2}(w)\bar{w} \label{eq.kal25}
\end{equation}
where $w=x+iy$, $\bar{w}= x-iy$ and $F_{1}, F_{2}$ are arbitrary smooth functions of their arguments.

The associated autonomous QFI (\ref{eq.e3pot.3}) is
\begin{equation}
I_{(1,1)}= \left( \dot{x} +i\dot{y} \right)^{2} +4\int F_{2}(w)dw. \label{eq.kal27}
\end{equation}

If $F_{1}(w,z)= F_{3}(w) +F_{4}(z)$, then
\begin{equation}
V(x,y,z)= F_{2}(w)\bar{w} +F_{3}(w) +F_{4}(z) \label{eq.kal28}
\end{equation}
is a new integrable potential due to the additional QFI $I= \frac{1}{2}\dot{z}^{2} +F_{4}(z)$.

3) $a_{3}=a$, $a_{13}=-a$, and $a_{17}=-ia$, where $a$ is an arbitrary constant.

The potential
\begin{equation}
V(x,y,z)= F_{1}(\bar{w},z) +F_{2}(\bar{w})w \label{eq.kal30}
\end{equation}
and the associated QFI
\begin{equation}
I_{(1,1)}= \left( \dot{x} -i\dot{y} \right)^{2} +4\int F_{2}(\bar{w})d\bar{w}. \label{eq.kal32}
\end{equation}

If $F_{1}(\bar{w},z)= F_{3}(\bar{w}) +F_{4}(z)$, then
\begin{equation}
V(x,y,z)= F_{2}(\bar{w})w +F_{3}(\bar{w}) +F_{4}(z) \label{eq.kal33}
\end{equation}
is a new integrable potential due to the additional QFI $I= \frac{1}{2}\dot{z}^{2} +F_{4}(z)$.

4) $a_{19}\neq0$ and $a_{20}= ia_{19}$.

The potential is
\begin{equation}
V(x,y,z)= F_{2}'z^{2} +F_{3}(w)z +F_{4}(w) +F_{2}(w)\bar{w} \label{eq.kal35}
\end{equation}
where $w=x+iy$, $\bar{w}= x-iy$, $F_{2}, F_{3}, F_{4}$ are arbitrary smooth functions of their arguments, and $F_{2}'\equiv \frac{dF_{2}}{dw}$.

The associated autonomous QFI (\ref{eq.e3pot.3}) is
\begin{equation}
I_{(1,1)}= \frac{1}{2}\dot{z}\left( \dot{x} +i\dot{y} \right) +F_{2}(w)z +\frac{1}{2}\int F_{3}(w)dw. \label{eq.kal37}
\end{equation}

Because the potential (\ref{eq.kal35}) is of the general form (\ref{eq.kal25}) for $F_{1}(w,z)= F_{2}'z^{2} +F_{3}(w)z +F_{4}(w)$, it admits the additional QFI (\ref{eq.kal27}). Therefore, it is integrable because the independent QFIs $H$, (\ref{eq.kal27}) and (\ref{eq.kal37}) are in involution\footnote{
The PB of the QFIs (\ref{eq.kal27}) and (\ref{eq.kal37}) vanishes because for an integrable function of the form $M(w)= \int F(w) dw$ with $w=x+iy$, it holds that: $M'\equiv \frac{dM}{dw}= F$, $M_{,x} =F$ and $M_{,y}= iF$.}.

For $F_{2}= k_{1}w +k_{2}$ and $F_{3}=k_{3}$ where $k_{1}, k_{2}, k_{3}$ are arbitrary constants, the potential (\ref{eq.kal35}) becomes
\begin{equation}
V(x,y,z)= k_{1}r^{2} +k_{2}\bar{w} +k_{3}z +F_{4}(w). \label{eq.kal35.1}
\end{equation}
This is a new minimally superintegrable potential because it is separable in the coordinate $z$.

5) $a_{19}\neq0$ and $a_{20}= -ia_{19}$.

The potential is
\begin{equation}
V(x,y,z)= F_{2}'z^{2} +F_{3}(\bar{w})z +F_{4}(\bar{w}) +F_{2}(\bar{w})w. \label{eq.kal39}
\end{equation}
where $w=x+iy$ and $F_{2}'\equiv \frac{dF_{2}}{d\bar{w}}$.

The associated autonomous QFI (\ref{eq.e3pot.3}) is
\begin{equation}
I_{(1,1)}= \frac{1}{2}\dot{z}\left( \dot{x} -i\dot{y} \right) +F_{2}(\bar{w})z +\frac{1}{2}\int F_{3}(\bar{w})d\bar{w}. \label{eq.kal40}
\end{equation}

The potential (\ref{eq.kal39}) is of the general form (\ref{eq.kal30}) for $F_{1}(\bar{w},z)= F_{2}'z^{2} +F_{3}(\bar{w})z +F_{4}(\bar{w})$; therefore, it is integrable due to the additional QFI (\ref{eq.kal32}).

Moreover, for $F_{2}= k_{1}\bar{w} +k_{2}$ and $F_{3}=k_{3}$, the potential (\ref{eq.kal39}) becomes
\begin{equation}
V(x,y,z)= k_{1}r^{2} +k_{2}w +k_{3}z +F_{4}(\bar{w}). \label{eq.kal35.2}
\end{equation}
This is a new minimally superintegrable potential because it is separable in the coordinate $z$.

\subsubsection{The components of the KT $C_{ab}$ are linear functions of $x,y,z$}

\label{sec.e3.pot.1.1.2}

The possibly non-zero parameters are the $a_{2}, a_{5}, a_{8}, a_{11}, a_{12}, a_{15}, a_{16}$, and $a_{18}$. In this case, there are six different combinations which lead to new results.

1) $a_{2}=a$ and $a_{5}=b$, where $a, b$ are arbitrary constants.

The potential is
\begin{equation}
V(x,y,z)= (a^{2}+b^{2})x^{2} +4(az+by)^{2} +\frac{k_{1}}{x^{2}} +k_{2}(az +by) +F(ay -bz) \label{eq.e3pot.12d}
\end{equation}
where $F$ is an arbitrary smooth function of its argument and $k_{1}, k_{2}$ are arbitrary constants. For $a=0$, the potential (\ref{eq.e3pot.12d}) reduces to the minimally superintegrable potential of the form (see Table \ref{Table.e23.2})
\begin{equation}
V(x,y,z)= \frac{k_{1}}{2}(x^{2} +4y^{2}) +\frac{k_{2}}{x^{2}} +k_{3}y +F(z). \label{eq.e3pot.11}
\end{equation}

The associated QFI (\ref{eq.e3pot.3}) is
\begin{equation}
I_{(1,1)}= (aM_{2} -bM_{3})\dot{x} -\frac{k_{2}}{2}(a^{2} +b^{2})x^{2} -2(a^{2}+b^{2})(az+by)x^{2} +\frac{2k_{1}(az+by)}{x^{2}} \label{eq.e3pot.12e}
\end{equation}
where $M_{i}=(y\dot{z}-z\dot{y}, z\dot{x}-x\dot{z}, x\dot{y} -y\dot{x})$ is the angular momentum.

Since the potential (\ref{eq.e3pot.12d}) is separable in the coordinate $x$, it admits the additional QFI
\[
I= \frac{1}{2}\dot{x}^{2}+ (a^{2}+b^{2})x^{2} +\frac{k_{1}}{x^{2}}.
\]
However, it is not integrable because the PB $\{I_{(1,1)}, I\} \neq0$.

2) $a_{2}=a$ and $a_{12}=b$, where $a, b$ are arbitrary constants.

We find the fully separable potential
\begin{equation}
V(x,y,z)= k_{1}(x^{2} +y^{2} +4z^{2}) +\frac{k_{2}}{x^{2}} +\frac{k_{3}}{y^{2}} +k_{4}z \label{eq.e3pot.16d}
\end{equation}
where $k_{1}, k_{2}, k_{3}$, and $k_{4}$ are arbitrary constants. We note that the potential in Table I of \cite{Evans 1990} is a subcase of the potential (\ref{eq.e3pot.16d}) for $k_{4}=0$.

The associated QFI (\ref{eq.e3pot.3}) consists of the following two independent QFIs:
\begin{eqnarray}
J_{1}&=& M_{2}\dot{x} +2z\left( \frac{k_{2}}{x^{2}} -k_{1}x^{2} \right) -\frac{k_{4}}{2}x^{2} \label{eq.e3pot.17a} \\
J_{2}&=& -M_{1}\dot{y} +2z\left( \frac{k_{3}}{y^{2}} -k_{1}y^{2} \right) -\frac{k_{4}}{2}y^{2}. \label{eq.e3pot.17b}
\end{eqnarray}

Moreover, the potential (\ref{eq.e3pot.16d}) is of the integrable form $V= \frac{F_{1}\left( \frac{y}{x} \right)}{R^{2}} +F_{2}(R) +F_{3}(z)$ (see Table \ref{Table.e23.1}) for
\[
F_{1}\left(\frac{y}{x}\right)= k_{2} \left[ 1 +\left(\frac{y}{x}\right)^{2} \right] + k_{3} \left[ 1+ \left(\frac{x}{y}\right)^{2} \right], \enskip F_{2}(R)= k_{1}(x^{2} +y^{2}), \enskip F_{3}(z)= 4k_{1}z^{2} +k_{4}z.
\]
Therefore, it admits the additional QFI
\begin{equation}
J_{3}= \frac{1}{2}M_{3}^{2} + k_{2}\left(\frac{y}{x}\right)^{2} +k_{3} \left(\frac{x}{y}\right)^{2}. \label{eq.e3pot.17c}
\end{equation}

In order to compare the QFIs (\ref{eq.e3pot.17a}), (\ref{eq.e3pot.17b}), (\ref{eq.e3pot.17c}) with the QFIs $I_{3}, I_{4}$ of eq. (3.43) in \cite{Evans 1990}, we set $k_{4}=0$ and we use the rotational parabolic coordinates (see eqs. (3.8) and (3.9) in \cite{Evans 1990}):
\[
\zeta= r+z, \enskip \eta= r-z, \enskip \phi= \tan^{-1} \left( \frac{y}{x} \right)
\]
or equivalently
\[
x= \sqrt{\zeta\eta} \cos\phi, \enskip y=\sqrt{\zeta\eta} \sin\phi, \enskip z=\frac{1}{2}\left( \zeta -\eta \right).
\]
We compute (for $k_{4}=0$):
\begin{eqnarray}
J_{1}&=& M_{2}\dot{x} +(\zeta -\eta) \left( \frac{k_{2}}{\zeta \eta \cos^{2}\phi} -k_{1}\zeta\eta \cos^{2}\phi \right) \label{eq.e3pot.17.1} \\
J_{2}&=& -M_{1}\dot{y} +(\zeta -\eta)\left( \frac{k_{3}}{\zeta \eta\sin^{2}\phi} -k_{1}\zeta\eta \sin^{2}\phi \right) \label{eq.e3pot.17.2} \\
J_{3}&=& \frac{1}{2}M_{3}^{2} + \frac{k_{2}}{\cos^{2}\phi} +\frac{k_{3}}{\sin^{2}\phi} -k_{2}-k_{3}. \label{eq.e3pot.17.3}
\end{eqnarray}
Then, $J_{3}= I_{3}-k_{2}-k_{3}$ and
\[
J_{1}+J_{2}= M_{2}\dot{x} -M_{1}\dot{y} -(\zeta-\eta) \left( k_{1}\zeta\eta -\frac{k_{2}}{\zeta\eta\cos^{2}\phi} -\frac{k_{3}}{\zeta\eta\sin^{2}\phi} \right)= I_{4}.
\]
There is a misprint in eq. (3.43) of \cite{Evans 1990} concerning the leading term of the QFI $I_{4}$. It must be $L_{2}P_{1} -L_{1}P_{2}$.

We conclude that the potential (\ref{eq.e3pot.16d}) is maximally superintegrable. However, from the seven QFIs (the QFIs $H, J_{1}, J_{2}, J_{3}$ plus the three QFIs arising from the separability of $x,y,z$) only five are functionally independent.

3) $a_{2}=a$ and $a_{8}=b$, where $a, b$ are arbitrary constants.

We find the separable potential
\begin{equation}
V(x,y,z)= \frac{k_{1}}{4}(x^{2} +16y^{2} +4z^{2}) +\frac{k_{2}}{x^{2}} +k_{3}y \label{eq.e3pot.18d}
\end{equation}
where $k_{1}, k_{2}$, and $k_{3}$ are arbitrary constants.

The associated QFI (\ref{eq.e3pot.3}) consists of the following two independent QFIs:
\begin{eqnarray}
J_{1}&=& M_{2}\dot{x} +z\left( \frac{2k_{2}}{x^{2}} -\frac{k_{1}}{2}x^{2} \right) \label{eq.e3pot.18e} \\
J_{2}&=& M_{1}\dot{z} -z^{2} \left( 2k_{1}y +\frac{k_{3}}{2} \right). \label{eq.e3pot.18f}
\end{eqnarray}
Therefore, the separable potential (\ref{eq.e3pot.18d}) is a new maximally superintegrable potential.

4) $a_{2}=a$ and $a_{16}=-a_{18}=\frac{b}{2}$, where $a, b$ are arbitrary constants.

The potential is
\begin{equation}
V(x,y,z)= \frac{k}{\sqrt{(ax+by)^{2} +(a^{2}+b^{2})z^{2}}} \label{eq.e3pot.20d}
\end{equation}
where $k$ is an arbitrary constant.

The associated QFI (\ref{eq.e3pot.3}) is
\begin{equation}
I_{(1,1)}= aM_{2}\dot{x} -bM_{1}\dot{x} +azV. \label{eq.e3pot.20e}
\end{equation}
In order to show that the potential (\ref{eq.e3pot.20d}) is integrable, we need one more independent FI in involution.

5) $a_{2}=a_{12}\neq0$.

In this case, we find the following three potentials\footnote{We note that any linear combination of these potentials is a solution of the system of PDEs (\ref{eq.e3pot.5a}) - (\ref{eq.e3pot.5c}) for $a_{2}=a_{12}\neq0$.}:
\begin{eqnarray}
V_{1}(x,y,z)&=& \frac{F_{1}\left(\frac{y}{x}\right)}{R^{2}} +k(R^{2}+4z^{2}) \label{eq.e3pot.27a} \\
V_{2}(x,y,z)&=& \frac{F_{1}\left(\frac{y}{x}\right)}{R^{2}} +\frac{k_{1}z}{rR^{2}} -\frac{k_{2}}{r} \label{eq.e3pot.27b} \\
V_{3}(x,y,z)&=& \frac{F_{1}\left(\frac{y}{x}\right)}{R^{2}} +kz \label{eq.e3pot.27c}
\end{eqnarray}
where $k, k_{1}, k_{2}$ are arbitrary constants, $R=\sqrt{x^{2} +y^{2}}$ and $r=\sqrt{x^{2} +y^{2} +z^{2}}$. The first two potentials, i.e. $V_{1}$ and $V_{2}$, are included in Table II of \cite{Evans 1990}, whereas the third potential $V_{3}$ is not included.

The associated QFIs (\ref{eq.e3pot.3}) are:\newline
- For the potential\footnote{In Table II of \cite{Evans 1990}, there is a misprint in the QFIs $I_{3}$ associated with the potentials (\ref{eq.e3pot.27a}) and (\ref{eq.e3pot.27b}). The leading part of $I_{3}$ should be $L_{2}P_{1} -P_{2}L_{1}$.} (\ref{eq.e3pot.27a}):
\begin{eqnarray}
I_{(1,1)}&=& M_{2}\dot{x} -M_{1}\dot{y} +\frac{2z F_{1}\left(\frac{y}{x}\right)}{x^{2} +y^{2}} -2kz(x^{2} +y^{2}) \notag \\
&=& M_{2}\dot{x} -M_{1}\dot{y} +(\zeta -\eta) \left( \frac{F_{1}(\tan\phi)}{\zeta\eta} -k\zeta\eta \right). \label{eq.e3pot.28a}
\end{eqnarray}
- For the potential (\ref{eq.e3pot.27b}):
\begin{eqnarray}
I_{(1,1)}&=& M_{2}\dot{x} -M_{1}\dot{y} +\frac{2z F_{1}\left( \frac{y}{x} \right)}{x^{2} +y^{2}} +\frac{2k_{1} z^{2}}{r(x^{2} +y^{2})} +\frac{k_{1}}{r} -\frac{k_{2}z}{r} \notag \\
&=& M_{2}\dot{x} -M_{1}\dot{y} +\frac{(\zeta-\eta)F_{1}(\tan\phi)}{\zeta\eta} +\frac{k_{1}(\zeta^{2} +\eta^{2})}{\zeta\eta(\zeta +\eta)} -\frac{k_{2}(\zeta-\eta)}{\zeta+\eta}. \label{eq.e3pot.28b}
\end{eqnarray}
- For the potential (\ref{eq.e3pot.27c}):
\begin{eqnarray}
I_{(1,1)}&=& M_{2}\dot{x} -M_{1}\dot{y} +\frac{2z F_{1}\left( \frac{y}{x} \right)}{x^{2} +y^{2}} -k\frac{R^{2}}{2}. \label{eq.e3pot.28c}
\end{eqnarray}

We note that both the potentials (\ref{eq.e3pot.27a}) and (\ref{eq.e3pot.27c}) are minimally superintegrable, because they are of the form $V= \frac{F_{1}\left( \frac{y}{x}\right)}{R^{2}} +F_{2}(R) +F_{3}(z)$ (see Table \ref{Table.e23.1}).

Moreover, from cases 2) and 3) of the following subsection \ref{sec.e3.pot.1.1.3}, the potential (\ref{eq.e3pot.27b}) becomes minimally superintegrable because it admits the additional QFIs (\ref{eq.e3pot.21d}) and (\ref{eq.e3pot.25b}) which are also in involution.

6) $a_{2}\neq0$, $a_{12}= -a_{2}$, $a_{16}= ia_{2}$ and $a_{18}= -i\frac{a_{2}}{2}$.

The potential is
\begin{equation}
V(x,y,z)= k_{1}(R^{2} +4z^{2}) +k_{2}z +\frac{k_{3}}{w^{2}} +k_{4}\frac{\bar{w}}{w^{3}} \label{eq.kal9}
\end{equation}
where $k_{1}, k_{2}, k_{3}, k_{4}$ are arbitrary constants, $w=x+iy$, and $\bar{w}= x-iy$. This result coincides with the potential given in eq. (14) of \cite{Kalnins 2006Pr} if we apply the canonical transformation $x \rightarrow y$, $y \rightarrow z$ and $z \rightarrow x$.

The associated autonomous QFI (\ref{eq.e3pot.3}) is
\begin{equation}
I_{1}= \frac{1}{2}(\dot{x} +i\dot{y}) \left( M_{2} -iM_{1} \right) -k_{1}zw^{2} -\frac{k_{2}}{4}w^{2} -k_{4}\frac{z}{w^{2}}. \label{eq.kal11.2}
\end{equation}

Moreover, the potential (\ref{eq.kal9}) admits the additional QFIs:
\begin{eqnarray}
I_{2}&=& \frac{1}{2} \left( \dot{x} +i\dot{y} \right)^{2} +k_{1}w^{2} -\frac{k_{4}}{w^{2}} \label{eq.kal11.1} \\
I_{3}&=& \frac{1}{2}\dot{z}^{2} +4k_{1}z^{2} +k_{2}z \label{eq.ka11.3} \\
I_{4}&=& \frac{1}{2}M_{3}^{2} +k_{3}e^{-2i\theta} +k_{4}e^{-4i\theta} \label{eq.kal11.4} \\
I_{5}&=& \frac{1}{2} \left(M_{2}\dot{x} -M_{1}\dot{y}\right) +k_{3}\frac{z}{w^{2}} +k_{4}\frac{z\bar{w}}{w^{3}} -k_{1}zR^{2} -k_{2}\frac{R^{2}}{4} \label{eq.kal11.5}
\end{eqnarray}
because it is of the form (\ref{eq.kal25}) for $F_{1}= 4k_{1}z^{2} +k_{2}z +\frac{k_{3}}{w^{2}}$ and $F_{2}= k_{1}w +\frac{k_{4}}{w^{3}}$, and of the form (see subsection \ref{sec.e3.pot.1.1.2} and Table \ref{Table.e3.2})
\begin{equation}
V(x,y,z)= \frac{F_{1}\left(\frac{y}{x}\right)}{R^{2}} +k_{1}(R^{2} +4z^{2}) +k_{2}z \label{eq.kal12}
\end{equation}
for $F_{1}\left(\frac{y}{x}\right)= k_{3}e^{-2i\theta} +k_{4}e^{-4i\theta}$, where $\tan\theta= \frac{y}{x}$. Therefore, the potential (\ref{eq.kal9}) is maximally superintegrable.

\subsubsection{The components of the KT $C_{ab}$ depend on $xy, xz, yz,$ $x^{2}$, $y^{2}, z^{2}$}

\label{sec.e3.pot.1.1.3}

In this case, the possibly non-zero parameters are the $a_{1}, a_{4}, a_{6}, a_{7}, a_{10},$ and $a_{14}$. New results are produced for the following six cases.

1) $a_{1}=a, a_{6}=b$, and $a_{7}=c$, where $a,b,c$ are arbitrary constants.

The potential is (see Table II in \cite{Evans 1990})
\begin{equation}
V(x,y,z)= \frac{k_{1}}{x^{2}} +\frac{k_{2}}{y^{2}} +\frac{k_{3}}{z^{2}} +F(r) \label{eq.e3pot.13d}
\end{equation}
where $k_{1},k_{2},k_{3}$ are arbitrary constants, $r=\sqrt{x^{2} +y^{2} +z^{2}}$ and $F$ is an arbitrary smooth function of $r$.

The associated QFI (\ref{eq.e3pot.3}) consists of the following three independent FIs (one for each parameter $a,b,c$):
\begin{eqnarray}
I_{1}&=& \frac{1}{2}M_{1}^{2} + k_{2}\frac{z^{2}}{y^{2}} +k_{3} \frac{y^{2}}{z^{2}} \label{eq.e3pot.14a} \\
I_{2}&=& \frac{1}{2}M_{2}^{2} + k_{1}\frac{z^{2}}{x^{2}} +k_{3} \frac{x^{2}}{z^{2}} \label{eq.e3pot.14b} \\
I_{3}&=& \frac{1}{2}M_{3}^{2} + k_{1}\frac{y^{2}}{x^{2}} +k_{2} \frac{x^{2}}{y^{2}}. \label{eq.e3pot.14c}
\end{eqnarray}

Using spherical coordinates $x=r\sin\theta \cos\phi$, $y= r\sin\theta \sin\phi$ and $z=r\cos\theta$, the QFIs (\ref{eq.e3pot.14a}) - (\ref{eq.e3pot.14c}) coincide with those found in Table II of \cite{Evans 1990}. Moreover, by adding the above QFIs, we find the QFI
\begin{equation}
I_{4}= \frac{1}{2} \mathbf{M}^{2} +\frac{k_{1}}{\sin^{2}\theta \cos^{2}\phi} +\frac{k_{2}}{\sin^{2}\theta\sin^{2}\phi} +\frac{k_{3}}{\cos^{2}\theta} \label{eq.e3pot.15}
\end{equation}
where $\mathbf{M}^{2}= M_{1}^{2}+M_{2}^{2}+M_{3}^{2}$ is the square magnitude of the angular momentum.

Even though the potential (\ref{eq.e3pot.13d}) admits the four independent QFIs $H, I_{1}, I_{2}$, and $I_{3}$, it is not integrable because the PBs $\{I_{i},I_{j}\}\neq0$.

For $F(r)=kr^{2}$, where $k$ is an arbitrary constant, the potential (\ref{eq.e3pot.13d}) becomes (see Table I in \cite{Evans 1990})
\begin{equation}
V(x,y,z)= k \left(x^{2}+y^{2}+z^{2}\right) +\frac{k_{1}}{x^{2}} +\frac{k_{2}}{y^{2}} +\frac{k_{3}}{z^{2}} \label{eq.e3.16}
\end{equation}
which is maximally superintegrable (see Tables \ref{Table.e23.1} and \ref{Table.e23.3}).

2) $a_{6}\neq0$.

The potential is
\begin{equation}
V(x,y,z)= \frac{F_{1}\left(\frac{y}{x}\right)}{R^{2}} +F_{2}(R,z) \label{eq.e3pot.21c}
\end{equation}
where $F_{1}$ and $F_{2}$ are arbitrary smooth functions of their arguments.

The associated QFI (\ref{eq.e3pot.3}) is
\begin{equation}
I_{(1,1)}= \frac{1}{2}M_{3}^{2} +F_{1}\left(\frac{y}{x}\right). \label{eq.e3pot.21d}
\end{equation}

If $F_{2}(R,z)= F_{3}(R) +F_{4}(z)$ where $F_{3}$ and $F_{4}$ are arbitrary smooth functions, then the potential (\ref{eq.e3pot.21c}) is integrable (see Table \ref{Table.e23.1}).

3) $a_{1}=a_{6}=a_{7} \neq0$.

The potential is
\begin{equation}
V(x,y,z; m)= \sum_{j=1}^{m} \frac{F_{j}\left(\frac{y}{x}\right)}{R^{2}} N_{j}\left(\frac{z}{R}\right) +F(r) \label{eq.e3pot.23.1}
\end{equation}
where $F_{j}, N_{j}$ and $F$ with $j=1,2,...,m$ are smooth functions of their arguments.

The associated QFI (\ref{eq.e3pot.3}) is
\begin{equation}
I_{(1,1)}= \frac{1}{2}\mathbf{M}^{2} +\sum_{j=1}^{m} \frac{r^{2}F_{j}\left(\frac{y}{x}\right)}{R^{2}} N_{j}\left(\frac{z}{R}\right). \label{eq.e3pot.23.3}
\end{equation}

We note that for \newline
a. $m=2$, $N_{1}=F_{2}=1$, $N_{2}= k_{2}\frac{R^{2}}{z^{2}}$, $F(r)= k_{1}r^{2}$; and\newline
b. $m=2$, $N_{1}=F_{2}=1$, $N_{2}= \frac{k_{1}\frac{z}{R}}{\sqrt{1+ \frac{z^{2}}{R^{2}}}}$, $F(r)= -\frac{k_{2}}{r}$ \newline
the potential (\ref{eq.e3pot.23.1}) reduces, respectively, to the following potentials (see Table II in \cite{Evans 1990}):
\begin{eqnarray}
V_{1}(x,y,z)&=& \frac{F_{1}\left(\frac{y}{x}\right)}{x^{2} +y^{2}} +k_{1}(x^{2}+y^{2}+z^{2}) +\frac{k_{2}}{z^{2}} \label{eq.e3pot.24a} \\
V_{2}(x,y,z)&=& \frac{F_{1}\left(\frac{y}{x}\right)}{x^{2} +y^{2}} +\frac{k_{1}z}{r(x^{2}+y^{2})} -\frac{k_{2}}{r} \label{eq.e3pot.24b}
\end{eqnarray}
where $k_{1}$ and $k_{2}$ are arbitrary constants. Both the above potentials are also of the general form (\ref{eq.e3pot.21c}).

The associated QFIs (\ref{eq.e3pot.23.3}) are as follows:\newline
- For the potential (\ref{eq.e3pot.24a}):
\begin{equation}
I_{(1,1)}=\frac{1}{2}\mathbf{M}^{2} +\frac{r^{2} F_{1}\left(\frac{y}{x}\right)}{x^{2} +y^{2}} + \frac{k_{2}(x^{2}+y^{2})}{z^{2}}. \label{eq.e3pot.25a}
\end{equation}
-For the potential (\ref{eq.e3pot.24b}):
\begin{equation}
I_{(1,1)}=\frac{1}{2}\mathbf{M}^{2} +\frac{r^{2} F_{1}\left(\frac{y}{x}\right)}{x^{2} +y^{2}} + \frac{k_{1}zr}{x^{2}+y^{2}}.  \label{eq.e3pot.25b}
\end{equation}

We note that the potential (\ref{eq.e3pot.24a}) is minimally superintegrable because it is separable in the coordinate $z$ and is also of the form (\ref{eq.e3pot.21c}).

4) $a_{1}\neq0$, $a_{7}= -a_{1}$ and $a_{10}= ia_{1}$.

The potential is
\begin{equation}
V(x,y,z)= F_{3}(w) +\frac{z^{2}}{w}F_{2}(w) +\frac{F_{4}\left( \frac{z}{w} \right)}{w^{2}} +F_{2}(w)\bar{w} \label{eq.kal43}
\end{equation}
where $w=x+iy$, $\bar{w}= x-iy$, and $F_{2}$, $F_{3}$, $F_{4}$ are arbitrary smooth functions of their arguments.

The associated QFI (\ref{eq.e3pot.3}) is
\begin{equation}
I_{1}= \frac{1}{2} \left( M_{2} -iM_{1} \right)^{2} +F_{4}\left( \frac{z}{w} \right). \label{eq.kal44}
\end{equation}

Moreover, the potential (\ref{eq.kal43}) admits the additional QFI
\begin{equation}
I_{2}= \left( \dot{x} +i\dot{y} \right)^{2} +4\int F_{2}(w)dw  \label{eq.kal45}
\end{equation}
because it is of the form (\ref{eq.kal25}) with $F_{1}= F_{3}(w) +\frac{z^{2}}{w}F_{2}(w) +\frac{F_{4}\left( \frac{z}{w} \right)}{w^{2}}$. Since the PB $\{I_{1},I_{2}\}=0$, the potential (\ref{eq.kal43}) is (Liouville) integrable.

Finally, for $F_{2}=k_{1}w$ and $F_{4}= k_{2}\frac{w^{2}}{z^{2}}$, the potential (\ref{eq.kal43}) becomes
\begin{equation}
V(x,y,z)= k_{1}r^{2} +\frac{k_{2}}{z^{2}} +F_{3}(w)
\label{eq.kal43.1}
\end{equation}
which is a new minimally superintegrable potential due to the separability in the $z$-coordinate.

5) $a_{1}\neq0$, $a_{7}= -a_{1}$ and $a_{10}= -ia_{1}$.

Similarly to the previous case 4), we find the integrable potential
\begin{equation}
V(x,y,z)= F_{3}(\bar{w}) +\frac{z^{2}}{\bar{w}}F_{2}(\bar{w}) +\frac{F_{4}\left( \frac{z}{\bar{w}} \right)}{\bar{w}^{2}} +F_{2}(\bar{w})w \label{eq.kal46}
\end{equation}
and the associated QFI
\begin{equation}
I_{1}= \frac{1}{2} \left( M_{2} +iM_{1} \right)^{2} +F_{4}\left( \frac{z}{\bar{w}} \right). \label{eq.kal47}
\end{equation}

Moreover, the potential (\ref{eq.kal46}) admits the additional QFI
\begin{equation}
I_{2}= \left( \dot{x} -i\dot{y} \right)^{2} +4\int F_{2}(\bar{w})d\bar{w}  \label{eq.kal48}
\end{equation}
because it is of the form (\ref{eq.kal30}) for $F_{1}= F_{3}(\bar{w}) +\frac{z^{2}}{\bar{w}}F_{2}(\bar{w}) +\frac{F_{4}\left( \frac{z}{\bar{w}} \right)}{\bar{w}^{2}}$.

Finally, for $F_{2}=k_{1}\bar{w}$ and $F_{4}= k_{2}\frac{\bar{w}^{2}}{z^{2}}$, the potential (\ref{eq.kal46}) becomes
\begin{equation}
V(x,y,z)= k_{1}r^{2} +\frac{k_{2}}{z^{2}} +F_{3}(\bar{w})
\label{eq.kal43.2}
\end{equation}
which is a new minimally superintegrable potential due to the separability in the $z$-coordinate.

6) $a_{4}\neq0$ and $a_{14}= -ia_{4}$.

The potential is (see eq. (12) of \cite{Kalnins 2006Pr})
\begin{equation}
V(x,y,z)= k_{1}r^{2} +\frac{k_{2}}{w^{2}} +k_{3}\frac{z}{w^{3}} +k_{4}\frac{R^{2} -3z^{2}}{w^{4}} \label{eq.kal5}
\end{equation}
where $k_{1}, k_{2}, k_{3}, k_{4}$ are arbitrary constants and $w=x+iy$.

The associated QFI (\ref{eq.e3pot.3}) is
\begin{equation}
I_{1}= M_{3}\left( iM_{1} -M_{2} \right) +k_{3}\frac{y}{w} -2ik_{2}\frac{z}{w} -\frac{3ik_{3}}{2}\frac{z^{2}}{w^{2}} -4ik_{4}\frac{z(x^{2}+y^{2}-z^{2})}{w^{3}}. \label{eq.kal7.2}
\end{equation}

Moreover, the potential (\ref{eq.kal5}) admits the additional QFIs:
\begin{eqnarray}
I_{2}&=& \frac{1}{2} \left( M_{2} -iM_{1} \right)^{2} +k_{3}\frac{z}{w} -4k_{4}\frac{z^{2}}{w^{2}} \label{eq.kal7.1} \\
I_{3}&=& \frac{1}{2}\dot{z} \left( \dot{x} +i\dot{y} \right) +k_{1}zw -\frac{k_{3}}{4w^{2}} +k_{4}\frac{z}{w^{3}}. \label{eq.kal7.4} \\
I_{4}&=& \frac{1}{2} \left( \dot{x} +i\dot{y} \right)^{2} +k_{1}w^{2} -\frac{k_{4}}{w^{2}} \label{eq.kal7.3} \\
I_{5}&=& \frac{1}{2}\mathbf{M}^{2} +k_{2}\frac{r^{2}}{w^{2}} +k_{3}\frac{zr^{2}}{w^{3}} +k_{4}\frac{r^{2}(x^{2} +y^{2} -3z^{2})}{w^{4}}. \label{eq.kal7.5}
\end{eqnarray}
Specifically, we have the following: \newline
1) It admits the QFI (\ref{eq.kal7.1}) because it is of the form (\ref{eq.kal43}) for $F_{2}=k_{1}w+\frac{k_{4}}{%
w^{3}}$, $F_{3}=\frac{k_{2}}{w^{2}}$ and $F_{4}=k_{3}\frac{z}{w}-4k_{4}\frac{%
z^{2}}{w^{2}}$. \newline
2) It admits the QFI (\ref{eq.kal7.4}) because it is of the form (\ref{eq.kal35}) for $F_{2}=k_{1}w+\frac{k_{4}}{%
w^{3}}$, $F_{3}=\frac{k_{3}}{w^{3}}$ and $F_{4}=\frac{k_{2}}{w^{2}}$. \newline
3) It admits the QFI (\ref{eq.kal7.3}) because it is of the
form (\ref{eq.kal25}) for $F_{1}=k_{1}z^{2}+\frac{k_{2}}{w^{2}}+k_{3}\frac{z%
}{w^{3}}-3k_{4}\frac{z^{2}}{w^{4}}$ and $F_{2}=k_{1}w+\frac{k_{4}}{w^{3}}$. \newline
4) It admits the QFI (\ref{eq.kal7.5}) because it is of the form (\ref{eq.e3pot.23.1}) for $m=4$, $N_{1}=1$, $%
F_{1}=k_{2}e^{-2i\theta }$, $N_{2}=\frac{z}{R}$, $F_{2}=k_{3}e^{-3i\theta }$, $N_{3}=1$, $F_{3}=k_{4}e^{-4i\theta }$, $N_{4}=\frac{z^{2}}{R^{2}}$, $%
F_{4}=-3k_{4}e^{-4i\theta }$ and $F(r)=k_{1}r^{2}$.

We note that the variable $\theta=\tan ^{-1}\left( \frac{y}{x}\right)$; hence, $w=x+iy=Re^{i\theta }$ and $R^{2}=w\bar{w}$.

We conclude that the potential (\ref{eq.kal5}) is maximally superintegrable. Specifically, it is integrable due to the triplet $H, I_{2}, I_{4}$ and superintegrable because it admits the five independent QFIs $H, I_{1}, I_{2}, I_{3}, I_{4}$.

\subsubsection{The components of the KT $C_{ab}$ depend on products of $x,y,z$ of mixed degree}

\label{sec.e3.pot.1.1.4}

In this subsection, we continue by considering mixed combinations of the twenty parameters $a_{1}, ..., a_{20}$ so that the components of the KT $C_{ab}$ contain products of $x,y,z$ of mixed degree. We note that we do not exhaust all possible cases; therefore, other authors could consider other cases and determine new non-decomposable integrable/superintegrable potentials in $E^{3}$.

1) The only non-vanishing parameters are the $a_{3}=\frac{iB}{4}$, $a_{5}= B$, $a_{13}= -\frac{iB}{4}$, $a_{17}= -\frac{B}{4}$ and $a_{15}= iB$, where $B$ is an arbitrary constant.

The KT (\ref{eq.e3pot.4a}) is
\begin{equation}
C_{ab}=
\left(
  \begin{array}{ccc}
   By +\frac{iB}{4} & -\frac{B}{2}x -\frac{iB}{2}y -\frac{B}{4} & 0 \\
   -\frac{B}{2}x -\frac{iB}{2}y -\frac{B}{4} & iBx -\frac{iB}{4} & 0 \\
   0 & 0 & 0 \\
  \end{array}
\right). \label{eq.kal13}
\end{equation}

For the KT (\ref{eq.kal13}) the system of PDEs (\ref{eq.e3pot.5a}) - (\ref{eq.e3pot.5c}) gives the potential
\begin{equation}
V(x,y,z)= 4k_{1} \left( R^{2} -\frac{\bar{w}^{3}}{2} \right) +k_{2} \left( 2w -3\bar{w}^{2} \right) +k_{3}\bar{w} +F(z) \label{eq.kal14}
\end{equation}
where $k_{1}, k_{2}, k_{3}$ are arbitrary constants and $F(z)$ is an arbitrary smooth function.

The associated QFI (\ref{eq.e3pot.3}) is
\begin{eqnarray}
I_{1}&=& \frac{1}{4} \left( \dot{x} +i\dot{y} \right)^{2} +iM_{3} \left( \dot{x} -i\dot{y} \right) -2k_{1} \left( \frac{3}{4}\bar{w}^{4} -w^{2} +R^{2}\bar{w} \right)- \notag \\
&& -2k_{2} \left( \bar{w}^{3} +2R^{2} \right) +k_{3} \left( \frac{\bar{w}^{2}}{2} +w \right). \label{eq.kal16.2}
\end{eqnarray}

Moreover, the potential (\ref{eq.kal14}) admits the additional QFIs:
\begin{eqnarray}
I_{2}&=& \frac{1}{8} \left( \dot{x} -i\dot{y} \right)^{2} +k_{1}\bar{w}^{2} +k_{2}\bar{w} \label{eq.kal16.1} \\
I_{3}&=& \frac{1}{2}\dot{z}^{2} +F(z) \label{eq.kal16.3}
\end{eqnarray}
because it is of the form (\ref{eq.kal30}) for $F_{1}= -2k_{1}\bar{w}^{3} -3k_{2}\bar{w}^{2} +k_{3}\bar{w} +F(z)$ and $F_{2}= 4k_{1}\bar{w} +2k_{2}$, and it is separable on the $z$-coordinate. Therefore, it is minimally superintegrable due to the four independent QFIs $H, I_{1}, I_{2}, I_{3}$.

We note that the potential given in eq. (15) of \cite{Kalnins 2006Pr} is a subcase of (\ref{eq.kal14}) for $F(z)= k_{1}z^{2} +\frac{k_{4}}{z^{2}}$, where $k_{4}$ is an arbitrary constant. As it will be shown below, in this special case, the resulting potential admits additional QFIs which promote it to a maximally superintegrable potential.

2) The only non-vanishing parameters are the $a_{1}= C$, $a_{7}= -C$, $a_{8}= -iD +2iC$, $a_{10}= -iC$ and $a_{11}= D$, where $C,D$ are arbitrary constants.

The KT (\ref{eq.e3pot.4a}) is
\begin{equation}
C_{ab}=
\left(
  \begin{array}{ccc}
   \frac{C}{2}z^{2} & -\frac{iC}{2}z^{2} & -\frac{C}{2}xz +\frac{iC}{2}yz -\frac{D}{2}z \\
   -\frac{iC}{2}z^{2} & -\frac{C}{2}z^{2} & \frac{iC}{2}xz +\frac{C}{2}yz +i \left(\frac{D}{2} -C\right)z  \\
   -\frac{C}{2}xz +\frac{iC}{2}yz -\frac{D}{2}z & \frac{iC}{2}xz +\frac{C}{2}yz +i \left(\frac{D}{2} -C\right)z & \frac{C}{2}(x^{2} -y^{2}) +iCy(2 -x) +D(x -iy) \\
  \end{array}
\right). \label{eq.kal17}
\end{equation}

For the KT (\ref{eq.kal17}) the system of PDEs (\ref{eq.e3pot.5a}) - (\ref{eq.e3pot.5c}) gives the potential (see eq. (15) of \cite{Kalnins 2006Pr})
\begin{equation}
V(x,y,z)= 4k_{1} \left( R^{2} -\frac{\bar{w}^{3}}{2} +\frac{z^{2}}{4} \right) +k_{2} \left( 2w -3\bar{w}^{2} \right) +k_{3}\bar{w} +\frac{k_{4}}{z^{2}} \label{eq.kal18}
\end{equation}
where $k_{1}, k_{2}, k_{3}$, and $k_{4}$ are arbitrary constants.

The associated QFI (\ref{eq.e3pot.3}) consists of the following independent QFIs:
\begin{eqnarray}
I_{1}&=& \frac{1}{2} \left( M_{2} +iM_{1} \right)^{2} +2i\dot{z}M_{1} +k_{1}z^{2}\left( 3\bar{w}^{2} -4iy \right) +2k_{2}z^{2} \left( 2\bar{w} +1\right) -\notag \\
&& -k_{3}z^{2} +\frac{k_{4}}{z^{2}} \left( \bar{w}^{2} +4iy \right) \label{eq.kal19.1} \\
I_{2}&=& \frac{1}{2}\dot{z} \left( M_{2} +iM_{1} \right) +k_{1}z^{2}\bar{w} +k_{2}z^{2} -k_{4}\frac{\bar{w}}{z^{2}}. \label{eq.kal19.2}
\end{eqnarray}

Moreover, the potential (\ref{eq.kal18}) admits the three additional QFIs (\ref{eq.kal16.2}) - (\ref{eq.kal16.3}) because it is of the form (\ref{eq.kal14}) for $F(z)= k_{1}z^{2} +\frac{k_{4}}{z^{2}}$. Therefore, it is maximally superintegrable.

3) The only non-vanishing parameters are the:
\[
a_{1}=-2C, \enskip a_{2}=iB -C, \enskip a_{5}=a_{8}= iA, \enskip  a_{7}= 2C, \enskip a_{9}= \frac{iB}{4} -\frac{C}{4}, \enskip a_{10}= -2iC, \enskip a_{11}=a_{15}=A,
\]
\[
a_{12}= -iB +2C, \enskip a_{13}= \frac{C}{4}, \enskip a_{16}= -B -\frac{3iC}{2}, \enskip a_{18}= \frac{B}{2} +\frac{3iC}{2}, \enskip a_{20}= \frac{iA}{4}
\]
where $A, B$, and $C$ are arbitrary constants.

The KT (\ref{eq.e3pot.4a}) has independent components:
\begin{eqnarray}
C_{11} &=& -Cz^{2} +iAy +(iB-C)z \notag \\
C_{12} &=& -iCz^{2} -\frac{iA}{2}\bar{w} -\left( B +\frac{3iC}{2} \right)z \notag \\
C_{13} &=& Cxz- -iCyz -\frac{iB-C}{2}x +\frac{1}{2}(B +3iC)y -\frac{A}{2}z \label{eq.kal20}
\\
C_{22}&=& Cz^{2} +Ax -(iB -2C)z -\frac{C}{4}  \notag \\
C_{23} &=& iCxz -Cyz +\frac{B}{2}x +\frac{iB-2C}{2}y -\frac{iA}{2}z +\frac{iA}{4}
\notag \\
C_{33} &=& -Cx^{2}+Cy^{2} -2iCxy +Aw +\frac{iB-C}{4}  \notag
\end{eqnarray}
where $w=x+iy$.

For the KT (\ref{eq.kal20}) the system of PDEs (\ref{eq.e3pot.5a}) - (\ref{eq.e3pot.5c}) gives the potential (see eq. (16) of \cite{Kalnins 2006Pr})
\begin{equation}
V(x,y,z)= k_{1}w +k_{2}\left( 3w^{2} +z \right) +k_{3}\left( 4w^{3} +3wz +\frac{\bar{w}}{4} \right) +k_{4} \left( \frac{5}{2}w^{4} +\frac{r^{2}}{2} +3w^{2}z \right) \label{eq.kal21}
\end{equation}
where $k_{1}, k_{2}, k_{3}$, and $k_{4}$ are arbitrary constants.

The associated QFI (\ref{eq.e3pot.3}) consists of the following independent QFIs:
\begin{eqnarray}
I_{1}&=& M_{3}\dot{w} -\left( M_{1} +iM_{2} \right)\dot{z} -\frac{1}{2}\dot{y}\dot{z} +\frac{ik_{1}}{2}\left( w^{2}-z \right) +ik_{2}\left( 2w^{3} -zw +\frac{i}{2}y \right) -\frac{ik_{3}}{8}\left( w^{2}-z \right)+ \notag \\
&& +ik_{3}\left( 3w^{4} -z^{2} +iyw \right) -\frac{k_{4}}{2}y \left( w^{2}+z \right) +ik_{4}w \left( 2w^{4} +zw^{2} -z^{2} \right) \label{eq.kal22.1} \\
I_{2}&=& M_{1}\left( 2\dot{x} +i\dot{y} \right) +iM_{2}\dot{x} +M_{3}\dot{z} +\frac{i}{4}\dot{z}^{2} +\frac{ik_{2}}{2} \left( z -w^{2} \right) +ik_{3}w \left( z -w^{2} \right) + \notag \\
&& + \frac{ik_{4}}{4} \left(z^{2} +2zw^{2} -3w^{4}\right) \label{eq.kal22.2} \\
I_{3}&=& \left( M_{1} +iM_{2} \right)^{2} +\left( 2iM_{1} -M_{2} \right) \dot{w} -iM_{3}\dot{z} +\frac{1}{4}\left( \dot{y}^{2} -\dot{z}^{2} \right) +k_{1}zw +\frac{ik_{1}}{2}y + \notag \\
&& +2k_{2}w \left( 2zw +iy \right) +\frac{k_{2}}{2} \left( w^{2} -z \right) +k_{3}w^{3}(6z+1) +k_{3}zw\left(2z -\frac{1}{4}\right) + \notag \\
&& +ik_{3}y\left( 3w^{2} -\frac{1}{8} \right) -k_{3}xz +k_{4}w^{4} \left( 4z +\frac{3}{4} \right) +2ik_{4}yw^{3} + \notag \\
&& +k_{4}z^{2} \left( 3w^{2} -\frac{1}{4}\right) +\frac{k_{4}}{2} \left( \frac{y^{2}}{2} -zR^{2} \right). \label{eq.kal22.3}
\end{eqnarray}
We note that the parameter $A$ produces the QFI $I_{1}$, $B$ the $I_{2}$, and $C$ the $I_{3}$.

Moreover, the potential (\ref{eq.kal21}) admits the additional QFIs:
\begin{eqnarray}
I_{4}&=& \dot{w}^{2} +k_{3}w +k_{4}w^{2} \label{eq.kal22.4} \\
I_{5}&=& \dot{z}\dot{w} +\left(k_{4}w +\frac{k_{3}}{2}\right)z +k_{4}w^{3} +\frac{3k_{3}}{2}w^{2} +k_{2}w \label{eq.kal22.5}
\end{eqnarray}
because it is of the form (\ref{eq.kal25}) for $F_{1}= k_{1}w +k_{2}(3w^{2}+z) +k_{3}w(4w^{2} +3z) +k_{4}\left( \frac{5}{2}w^{4} +3w^{2}z +\frac{z^{2}}{2} \right)$ and $F_{2}= \frac{k_{4}}{2}w +\frac{k_{3}}{4}$; and of the form (\ref{eq.kal35}) for $F_{2}= \frac{k_{4}}{2}w +\frac{k_{3}}{4}$, $F_{3}= 3k_{4}w^{2} +3k_{3}w +k_{2}$ and $F_{4}= \frac{5k_{4}}{2}w^{4} +4k_{3}w^{3} +3k_{2}w^{2} +k_{1}w$.

We compute the PB $\{I_{2},I_{4}\}=0$; therefore, the potential (\ref{eq.kal21}) is maximally superintegrable.

\subsubsection{Special superintegrable potentials}

\label{sec.e3.pot.1.1.5}

In this subsection, we construct potentials whose form belongs to two or more of the previous general results. We have the following cases:

1) Consider the potential (see Table I in \cite{Evans 1990})
\begin{equation}
V(x,y,z)= -\frac{c_{1}}{r} +\frac{c_{2}}{x^{2}} +\frac{c_{3}}{y^{2}} \label{eq.e3pot.29a}
\end{equation}
where $c_{1}, c_{2}$, and $c_{3}$ are arbitrary constants. This potential is of the general form (\ref{eq.e3pot.13d}) for $F(r)=-\frac{c_{1}}{r}$, $k_{1}=c_{2}$, $k_{2}=c_{3}$ and $k_{3}=0$; and of the form (\ref{eq.e3pot.27b}) for $k_{1}=0$, $k_{2}=c_{1}$ and $F_{1}\left(\frac{y}{x}\right)= c_{2} \left[ 1 +\left(\frac{y}{x}\right)^{2} \right] +c_{3} \left[ 1 +\left(\frac{x}{y}\right)^{2} \right]$. Therefore, it admits the additional QFIs:
\begin{eqnarray}
I_{1}&=& \frac{1}{2}M_{1}^{2} + c_{3}\frac{z^{2}}{y^{2}} \label{eq.e3pot.30a} \\
I_{2}&=& \frac{1}{2}M_{2}^{2} + c_{2}\frac{z^{2}}{x^{2}} \label{eq.e3pot.30b} \\
I_{3}&=& \frac{1}{2}M_{3}^{2} +c_{2}\frac{y^{2}}{x^{2}} +c_{3} \frac{x^{2}}{y^{2}} \label{eq.e3pot.30c} \\
I_{4}&=& M_{2}\dot{x} -M_{1}\dot{y} -2z \left( \frac{c_{1}}{2r} -\frac{c_{2}}{x^{2}} -\frac{c_{3}}{y^{2}} \right). \label{eq.e3pot.30d}
\end{eqnarray}
We conclude that the potential (\ref{eq.e3pot.29a}) is maximally superintegrable because the QFIs $H, I_{3}, I_{4}$ are in involution and the five QFIs $H, I_{1}, I_{2}, I_{3}, I_{4}$ are functionally independent.

2) Consider the potential (see Table I in \cite{Evans 1990})
\begin{equation}
V(x,y,z)= \frac{c_{1}y}{x^{2}R} +\frac{c_{2}}{x^{2}} +\frac{c_{3}}{z^{2}}  \label{eq.e3pot.31}
\end{equation}
where $c_{1}, c_{2}$, and $c_{3}$ are arbitrary constants. This potential is of the form $V= \frac{k_{1}}{x^{2}} +\frac{k_{2}}{R} +\frac{k_{3}y}{Rx^{2}} +F(z)$ (see Table \ref{Table.e23.2}) for $k_{1}=c_{2}$, $k_{2}=0$, $k_{3}=c_{1}$, and $F(z)= \frac{c_{3}}{z^{2}}$; and of the form (\ref{eq.e3pot.24a}) for $k_{1}=0$, $k_{2}=c_{3}$, and $F_{1}\left( \frac{y}{x} \right)= \left( \frac{c_{1}}{\sqrt{1 +\frac{x^{2}}{y^{2}}}} +c_{2} \right) \left( 1 +\frac{y^{2}}{x^{2}} \right)$. Therefore, it admits the additional QFIs:
\begin{eqnarray}
I_{1}&=& \frac{1}{2}\dot{z}^{2} +\frac{c_{3}}{z^{2}} \label{eq.e3pot.32a} \\
I_{2}&=& \frac{1}{2} M_{3}^{2} +c_{2}\frac{y^{2}}{x^{2}} +c_{1}\frac{yR}{x^{2}} \label{eq.e3pot.32b} \\
I_{3}&=& M_{3}\dot{x} -2c_{2} \frac{y}{x^{2}} -c_{1} \frac{x^{2} +2y^{2}}{x^{2}R} \label{eq.e3pot.32c} \\
I_{4}&=& \frac{1}{2} \mathbf{M}^{2} +c_{1} \frac{yr^{2}}{Rx^{2}} +c_{2}\frac{r^{2}}{x^{2}} +c_{3}\frac{R^{2}}{z^{2}} \label{eq.e3pot.32d}
\end{eqnarray}
and it is maximally superintegrable.

3) Another maximally superintegrable potential is the (see Table I in \cite{Evans 1990})
\begin{equation}
V(x,y,z)= \frac{c_{1}y}{x^{2}R} +\frac{c_{2}}{x^{2}} +c_{3}z \label{eq.e3pot.33}
\end{equation}
where $c_{1}, c_{2}$, and $c_{3}$ are arbitrary constants. This potential is of the form $V= \frac{k_{1}}{x^{2}} +\frac{k_{2}}{R} +\frac{k_{3}y}{Rx^{2}} +F(z)$ (see Table \ref{Table.e23.2}) for $k_{1}=c_{2}$, $k_{2}=0$, $k_{3}=c_{1}$, and $F(z)= c_{3}z$; and of the form (\ref{eq.e3pot.27c}) for $k=c_{3}$, and $F_{1}\left( \frac{y}{x} \right)= \left( \frac{c_{1}}{\sqrt{1 +\frac{x^{2}}{y^{2}}}} +c_{2} \right) \left( 1 +\frac{y^{2}}{x^{2}} \right)$. Therefore, it admits the additional QFIs:
\begin{eqnarray}
I_{1}&=& \frac{1}{2}\dot{z}^{2} +c_{3}z \label{eq.e3pot.34a} \\
I_{2}&=& \frac{1}{2} M_{3}^{2} +c_{2}\frac{y^{2}}{x^{2}} +c_{1}\frac{yR}{x^{2}} \label{eq.e3pot.34b} \\
I_{3}&=& M_{3}\dot{x} -2c_{2} \frac{y}{x^{2}} -c_{1} \frac{x^{2} +2y^{2}}{x^{2}R} \label{eq.e3pot.34c} \\
I_{4}&=& M_{2}\dot{x} -M_{1}\dot{y} +c_{1}\frac{2yz}{x^{2}R} +c_{2}\frac{2z}{x^{2}} -c_{3}\frac{R^{2}}{2}. \label{eq.e3pot.34d}
\end{eqnarray}

4) Consider the potential (see eq. (11) of \cite{Kalnins 2006Pr})
\begin{equation}
V(x,y,z)= k_{1}r^{2} +k_{2}\frac{\bar{w}}{w^{3}} +\frac{k_{3}}{w^{2}} +\frac{k_{4}}{z^{2}} \label{eq.kal2}
\end{equation}
where $k_{1}, k_{2}, k_{3}, k_{4}$ are arbitrary constants, $w=x+iy$ and $\bar{w}= x-iy$.

This potential admits the additional QFIs:
\begin{eqnarray}
I_{1}&=& \frac{1}{2} \left( M_{2} -iM_{1} \right)^{2} +k_{4}\frac{w^{2}}{z^{2}} -k_{2}\frac{z^{2}}{w^{2}}. \label{eq.kal3.1} \\
I_{2}&=& \frac{1}{2} \left( \dot{x} +i\dot{y} \right)^{2} +k_{1}w^{2} -\frac{k_{2}}{w^{2}} \label{eq.kal3.2} \\
I_{3}&=& \frac{1}{2}M_{3}^{2} +k_{2}e^{-4i\theta} +k_{3}e^{-2i\theta} = \frac{1}{2}M_{3}^{2} +k_{2}\left( \frac{\bar{w}}{w} \right)^{2} +k_{3}\frac{\bar{w}}{w} \label{eq.kal3.3} \\
I_{4}&=& \frac{1}{2}\dot{z}^{2} +k_{1}z^{2} +\frac{k_{4}}{z^{2}} \label{eq.kal3.4} \\
I_{5}&=& \frac{1}{2} \mathbf{M}^{2} +k_{2}\frac{r^{2}\bar{w}}{w^{3}} +k_{3}\frac{r^{2}}{w^{2}} +k_{4} \frac{r^{2}}{z^{2}} \label{eq.kal3.5}
\end{eqnarray}
because it is of the form (\ref{eq.kal43}) for $F_{2}= k_{1}w +\frac{k_{2}}{w^{3}}$, $F_{3}= \frac{k_{3}}{w^{2}}$ and $F_{4}= -k_{2}\frac{z^{2}}{w^{2}} +k_{4}\frac{w^{2}}{z^{2}}$; of the form (\ref{eq.kal25}) for $F_{1}= k_{1}z^{2} +\frac{k_{3}}{w^{2}} +\frac{k_{4}}{z^{2}}$ and $F_{2}= k_{1}w +\frac{k_{2}}{w^{3}}$; of the form (\ref{eq.e3pot.21c}) for $F_{1}= k_{2}e^{-4i\theta} +k_{3}e^{-2i\theta}$ and $F_{2}= k_{1}r^{2} +\frac{k_{4}}{z^{2}}$; separable on the $z$-coordinate; and of the form (\ref{eq.e3pot.23.1}) for $m=2$, $F_{1}=N_{2}=1$, $N_{1}=k_{4}\frac{R^{2}}{z^{2}}$, $F_{2}= k_{2}e^{-4i\theta} +k_{3}e^{-2i\theta}$ and $F(r)= k_{1}r^{2}$.

The variable $\theta= \tan^{-1}\left( \frac{y}{x} \right)$ and, hence, $w= x+iy =Re^{i\theta}$. We recall that
\[
w= Re^{i\theta} \implies e^{in\theta}= \left( \frac{w}{R} \right)^{n} \implies e^{in\theta}= \left( \frac{1 +i\frac{y}{x}}{\sqrt{1+ \left(\frac{y}{x}\right)^{2}}} \right)^{n}
\]
where $n$ is an arbitrary real constant. If $n=2k \in \mathbb{R}$, then $e^{2ik\theta}= \left( \frac{w}{\bar{w}} \right)^{k}$ because $R^{2}= w\bar{w}$.

We conclude that the potential (\ref{eq.kal2}) is maximally superintegrable.
\bigskip

We collect the results of this section in Tables \ref{Table.e3.1} - \ref{Table.e3.3}.

\begin{longtable}{|l|l|}
\hline
{\large Potential} & {\large LFIs and
QFIs} \\ \hline
\makecell[l]{$V= F_{1}\left( cz +by +(\sqrt{a^{2} +b^{2} +c^{2}} +a)x \right) +$ \\ \qquad $+F_{2}\left( cz +by -(\sqrt{a^{2} +b^{2} +c^{2}} -a)x \right)+$ \\ \qquad $+F_{3}(bz -cy)$} & \makecell[l]{$I_{1}= \left( a\dot{x} +b\dot{y} +c\dot{z} \right)\dot{x} +a(F_{1} +F_{2}) +$ \\ \qquad $+\sqrt{a^{2}+b^{2}+c^{2}}(F_{1} -F_{2})$} \\ \hline
\makecell[l]{$V= (a^{2}+b^{2})x^{2} +4(az+by)^{2} +\frac{k_{1}}{x^{2}}+$ \\ \qquad $+k_{2}(az +by) +F(ay -bz)$} & \makecell[l]{$I_{1} = aM_{2}\dot{x} -bM_{3}\dot{x} -\frac{k_{2}}{2}(a^{2} +b^{2})x^{2} -$ \\ \qquad $-2(a^{2}+b^{2})(az+by)x^{2} +\frac{2k_{1}(az+by)}{x^{2}}$ \\
$I_{2}= \frac{1}{2}\dot{x}^{2}+ (a^{2}+b^{2})x^{2} +\frac{k_{1}}{x^{2}}$} \\ \hline
$V= \frac{k}{\sqrt{(ax+by)^{2} +(a^{2}+b^{2})z^{2}}}$ & $I_{1}= aM_{2}\dot{x} -bM_{1}\dot{x} +azV$ \\ \hline
$V= \frac{k_{1}}{x^{2}} +\frac{k_{2}}{y^{2}} +\frac{k_{3}}{z^{2}} +F(r)$ & \makecell[l]{$I_{1}= \frac{1}{2}M_{1}^{2} + k_{2}\frac{z^{2}}{y^{2}} +k_{3} \frac{y^{2}}{z^{2}}$ \\ $I_{2}= \frac{1}{2}M_{2}^{2} + k_{1}\frac{z^{2}}{x^{2}} +k_{3} \frac{x^{2}}{z^{2}}$ \\ $I_{3}= \frac{1}{2}M_{3}^{2} + k_{1}\frac{y^{2}}{x^{2}} +k_{2} \frac{x^{2}}{y^{2}}$} \\ \hline
$V= \frac{F_{1}\left(\frac{y}{x}\right)}{x^{2}+y^{2}} +F_{2}(x^{2}+y^{2},z)$ & $I_{1}=\frac{1}{2}M_{3}^{2} +F_{1}\left(\frac{y}{x}\right)$ \\ \hline
$V=\sum_{j=1}^{m} \frac{F_{j}\left(\frac{y}{x}\right)}{R^{2}} N_{j}\left(\frac{z}{R}\right) +F(r)$ & $I=\frac{1}{2}\mathbf{M}^{2} +\sum_{j=1}^{m} \frac{r^{2}F_{j}\left(\frac{y}{x}\right)}{R^{2}} N_{j}\left(\frac{z}{R}\right)$ \\ \hline
$V= F_{1}(w,z) +F_{2}(w)\bar{w}$ & $I_{1}= \left( \dot{x} +i\dot{y} \right)^{2} +4\int F_{2}(w)dw$ \\ \hline
$V= F_{1}(\bar{w},z) +F_{2}(\bar{w})w$ & $I_{1}= \left( \dot{x} -i\dot{y} \right)^{2} +4\int F_{2}(\bar{w})d\bar{w}$ \\ \hline
\caption{\label{Table.e3.1} Possibly non-integrable potentials $V(x,y,z)$ in $E^{3}$ that admit one or more QFIs of the type $I_{(1,1)}$ which are not in involution.}
\end{longtable}

\begin{longtable}{|l|l|}
\hline
\multicolumn{2}{|c|}{{\large{Integrable potentials}}} \\ \hline
{\large Potential} &  {\large LFIs and QFIs} \\ \hline
$V= F_{2}(w)\bar{w} +F_{3}(w) +F_{4}(z)$ & \makecell[l]{$I_{1}= \left( \dot{x} +i\dot{y} \right)^{2} +4\int F_{2}(w)dw$ \\ $I_{2}= \frac{1}{2}\dot{z}^{2} +F_{4}(z)$} \\ \hline
$V= F_{2}(\bar{w})w +F_{3}(\bar{w}) +F_{4}(z)$ & \makecell[l]{$I_{1}= \left( \dot{x} -i\dot{y} \right)^{2} +4\int F_{2}(\bar{w})d\bar{w}$ \\ $I_{2}= \frac{1}{2}\dot{z}^{2} +F_{4}(z)$} \\ \hline
$V= F_{2}'z^{2} +F_{3}(w)z +F_{4}(w) +F_{2}(w)\bar{w}$ & \makecell[l]{$I_{1}= \frac{1}{2}\dot{z}\left( \dot{x} +i\dot{y} \right) +F_{2}(w)z +\frac{1}{2}\int F_{3}(w)dw$ \\ $I_{2}= \left( \dot{x} +i\dot{y} \right)^{2} +4\int F_{2}(w)dw$} \\ \hline
$V= F_{2}'z^{2} +F_{3}(\bar{w})z +F_{4}(\bar{w}) +F_{2}(\bar{w})w$ & \makecell[l]{$I_{1}= \frac{1}{2}\dot{z}\left( \dot{x} -i\dot{y} \right) +F_{2}(\bar{w})z +\frac{1}{2}\int F_{3}(\bar{w})d\bar{w}$ \\ $I_{2}= \left( \dot{x} -i\dot{y} \right)^{2} +4\int F_{2}(\bar{w})d\bar{w}$} \\ \hline
$V= F_{3}(w) +\frac{z^{2}}{w}F_{2}(w) +\frac{F_{4}\left( \frac{z}{w} \right)}{w^{2}} +F_{2}(w)\bar{w}$ & \makecell[l]{$I_{1}= \frac{1}{2} \left( M_{2} -iM_{1} \right)^{2} +F_{4}\left( \frac{z}{w} \right)$ \\ $I_{2}= \left( \dot{x} +i\dot{y} \right)^{2} +4\int F_{2}(w)dw$} \\ \hline
$V= F_{3}(\bar{w}) +\frac{z^{2}}{\bar{w}}F_{2}(\bar{w}) +\frac{F_{4}\left( \frac{z}{\bar{w}} \right)}{\bar{w}^{2}} +F_{2}(\bar{w})w$ & \makecell[l]{$I_{1}= \frac{1}{2} \left( M_{2} +iM_{1} \right)^{2} +F_{4}\left( \frac{z}{\bar{w}} \right)$ \\ $I_{2}= \left( \dot{x} -i\dot{y} \right)^{2} +4\int F_{2}(\bar{w})d\bar{w}$} \\ \hline
\caption{\label{Table.e3.1.1} Integrable potentials $V(x,y,z)$ in $E^{3}$ that admit QFIs of the type $I_{(1,1)}$.}
\end{longtable}

\newpage

\begin{longtable}{|l|c|l|}
\hline
\multicolumn{3}{|c|}{{\large{Minimally superintegrable potentials}}} \\ \hline
{\large Potential} & {\large Ref \cite{Evans 1990}} & {\large LFIs and QFIs} \\ \hline
$V= \frac{F_{1}\left(\frac{y}{x}\right)}{R^{2}} +k_{1}(R^{2}+4z^{2}) +k_{2}z$ & \makecell[l]{Table II \\ $k_{2}=0$} & \makecell[l]{$I_{1}= \frac{1}{2}\dot{z}^{2} +4k_{1}z^{2} +k_{2}z$ \\ $I_{2}= \frac{1}{2}M_{3}^{2} +F_{1}\left(\frac{y}{x}\right)$ \\ $I_{3}= M_{2}\dot{x} -M_{1}\dot{y} +\frac{2zF_{1}\left(\frac{y}{x} \right)}{R^{2}} -2k_{1}zR^{2} -k_{2}\frac{R^{2}}{2}$} \\ \hline
$V= \frac{F_{1}\left(\frac{y}{x}\right)}{R^{2}} +\frac{k_{1}z}{rR^{2}} -\frac{k_{2}}{r}$ & Table II & \makecell[l]{$I_{1}=\frac{1}{2}M_{3}^{2} +F_{1}\left(\frac{y}{x} \right)$ \\ $I_{2}=\frac{1}{2}\mathbf{M}^{2} +\frac{r^{2} F_{1}\left(\frac{y}{x}\right)}{R^{2}} + \frac{k_{1}zr}{R^{2}}$ \\ $I_{3}= M_{2}\dot{x} -M_{1}\dot{y} +\frac{2z F_{1}\left( \frac{y}{x} \right)}{R^{2}} +\frac{2k_{1} z^{2}}{rR^{2}} +\frac{k_{1}}{r} -\frac{k_{2}z}{r}$} \\ \hline
$V= \frac{F_{1}\left(\frac{y}{x}\right)}{R^{2}} +k_{1}r^{2} +\frac{k_{2}}{z^{2}}$ & Table II & \makecell[l]{$I_{1}= \frac{1}{2}\dot{z}^{2} +k_{1}z^{2} +\frac{k_{2}}{z^{2}}$ \\ $I_{2}= \frac{1}{2}M_{3}^{2} +F_{1}\left(\frac{y}{x}\right)$ \\ $I_{3}= \frac{1}{2}\mathbf{M}^{2} +\frac{r^{2} F_{1}\left(\frac{y}{x}\right)}{R^{2}} + \frac{k_{2}R^{2}}{z^{2}}$} \\ \hline
\makecell[l]{$V= 4k_{1} \left( R^{2} -\frac{\bar{w}^{3}}{2} \right) +k_{2} \left( 2w -3\bar{w}^{2} \right) +$ \\ \qquad  $+k_{3}\bar{w} +F(z)$} & New & \makecell[l]{$I_{1}= \frac{1}{8} \left( \dot{x} -i\dot{y} \right)^{2} +k_{1}\bar{w}^{2} +k_{2}\bar{w}$ \\ $I_{2}= \frac{1}{4} \dot{w}^{2} +iM_{3} \left( \dot{x} -i\dot{y} \right) -2k_{1} \left( \frac{3}{4}\bar{w}^{4} -w^{2} +R^{2}\bar{w} \right)-$ \\ \qquad $-2k_{2} \left( \bar{w}^{3} +2R^{2} \right) +k_{3} \left( \frac{\bar{w}^{2}}{2} +w \right)$ \\ $I_{3}= \frac{1}{2}\dot{z}^{2} +F(z)$} \\ \hline
$V= k_{1}r^{2} +k_{2}\bar{w} +k_{3}z +F_{4}(w)$ & New & \makecell[l]{$I_{1}= \frac{1}{2}\dot{z}\left( \dot{x} +i\dot{y} \right) +k_{1}wz +k_{2}z +\frac{k_{3}}{2}w$ \\ $I_{2}= \left( \dot{x} +i\dot{y} \right)^{2} +2k_{1}w^{2} +4k_{2}w$ \\ $I_{3}= \frac{1}{2}\dot{z}^{2} +k_{1}z^{2} +k_{3}z$} \\ \hline
$V= k_{1}r^{2} +k_{2}w +k_{3}z +F_{4}(\bar{w})$ & New & \makecell[l]{$I_{1}= \frac{1}{2}\dot{z}\left( \dot{x} -i\dot{y} \right) +k_{1}\bar{w}z +k_{2}z +\frac{k_{3}}{2}\bar{w}$ \\ $I_{2}= \left( \dot{x} -i\dot{y} \right)^{2} +2k_{1}\bar{w}^{2} +4k_{2}\bar{w}$ \\ $I_{3}= \frac{1}{2}\dot{z}^{2} +k_{1}z^{2} +k_{3}z$} \\ \hline
$V= k_{1}r^{2} +\frac{k_{2}}{z^{2}} +F_{3}(w)$ & New & \makecell[l]{$I_{1}= \frac{1}{2} \left( M_{2} -iM_{1} \right)^{2} +k_{2}\frac{w^{2}}{z^{2}}$ \\ $I_{2}= \left( \dot{x} +i\dot{y} \right)^{2} +2k_{1}w^{2}$ \\ $I_{3}= \frac{1}{2}\dot{z}^{2} +k_{1}z^{2} +\frac{k_{2}}{z^{2}}$} \\ \hline
$V= k_{1}r^{2} +\frac{k_{2}}{z^{2}} +F_{3}(\bar{w})$ & New & \makecell[l]{$I_{1}= \frac{1}{2} \left( M_{2} +iM_{1} \right)^{2} +k_{2}\frac{\bar{w}^{2}}{z^{2}}$ \\ $I_{2}= \left( \dot{x} -i\dot{y} \right)^{2} +2k_{1}\bar{w}^{2}$ \\ $I_{3}= \frac{1}{2}\dot{z}^{2} +k_{1}z^{2} +\frac{k_{2}}{z^{2}}$} \\ \hline
\caption{\label{Table.e3.2} Minimally superintegrable potentials $V(x,y,z)$ in $E^{3}$ that admit QFIs of the type $I_{(1,1)}$.}
\end{longtable}

\newpage

\begin{longtable}{|l|c|c|l|}
\hline
\multicolumn{4}{|c|}{{\large{Maximally superintegrable potentials}}} \\ \hline
{\large Potential} & {\large Ref \cite{Evans 1990}} & {\large Ref \cite{Kalnins 2006Pr}} & {\large LFIs and
QFIs} \\ \hline
\makecell[l]{$V= k_{1}(R^{2} +4z^{2}) +\frac{k_{2}}{x^{2}} +\frac{k_{3}}{y^{2}} +k_{4}z$} & \makecell[l]{Table I \\ $k_{4}=0$} & \makecell[l]{eq. (13) \\ $z \leftrightarrow x$} & \makecell[l]{$I_{1}= \frac{1}{2}\dot{x}^{2} +k_{1}x^{2} +\frac{k_{2}}{x^{2}}$ \\ $I_{2}= \frac{1}{2}\dot{y}^{2} +k_{1}y^{2} +\frac{k_{3}}{y^{2}}$ \\ $I_{3}= \frac{1}{2}\dot{z}^{2} +4k_{1}z^{2} +k_{4}z$ \\ $I_{4}=M_{2}\dot{x} +2z\left( \frac{k_{2}}{x^{2}} -k_{1}x^{2} \right) -\frac{k_{4}}{2}x^{2}$ \\ $I_{5}= -M_{1}\dot{y} +2z\left( \frac{k_{3}}{y^{2}} -k_{1}y^{2} \right) -\frac{k_{4}}{2}y^{2}$ \\ $I_{6}= \frac{1}{2}M_{3}^{2} + k_{2}\left(\frac{y}{x}\right)^{2} +k_{3} \left(\frac{x}{y}\right)^{2}$} \\ \hline
\makecell[l]{$V= \frac{k_{1}}{4} \left( x^{2} +16y^{2} +4z^{2} \right) +\frac{k_{2}}{x^{2}} +$ \\ \qquad $+k_{3}y$} & New & New & \makecell[l]{$I_{1}= \frac{1}{2}\dot{x}^{2} +\frac{k_{1}}{4}x^{2} +\frac{k_{2}}{x^{2}}$ \\ $I_{2}= \frac{1}{2}\dot{y}^{2} +4k_{1}y^{2} +k_{3}y$ \\ $I_{3}= \frac{1}{2}\dot{z}^{2} +k_{1}z^{2}$ \\ $I_{4}= M_{2}\dot{x} +z\left( \frac{2k_{2}}{x^{2}} -\frac{k_{1}}{2}x^{2} \right)$ \\ $I_{5}= M_{1}\dot{z} -z^{2} \left( 2k_{1}y +\frac{k_{3}}{2} \right)$} \\ \hline
$V= kr^{2} + \frac{k_{1}}{x^{2}} +\frac{k_{2}}{y^{2}} +\frac{k_{3}}{z^{2}}$ & Table I & eq. (10) & \makecell[l]{$I_{1}= \frac{1}{2}\dot{x}^{2} +kx^{2} +\frac{k_{1}}{x^{2}}$ \\ $I_{2}= \frac{1}{2}\dot{y}^{2} +ky^{2} +\frac{k_{2}}{y^{2}}$ \\ $I_{3}= \frac{1}{2}\dot{z}^{2} +kz^{2} +\frac{k_{3}}{z^{2}}$ \\ $I_{4}= \frac{1}{2}M_{1}^{2} + k_{2}\frac{z^{2}}{y^{2}} +k_{3} \frac{y^{2}}{z^{2}}$ \\ $I_{5}= \frac{1}{2}M_{2}^{2} + k_{1}\frac{z^{2}}{x^{2}} +k_{3} \frac{x^{2}}{z^{2}}$ \\ $I_{6}= \frac{1}{2}M_{3}^{2} + k_{1}\frac{y^{2}}{x^{2}} +k_{2} \frac{x^{2}}{y^{2}}$} \\ \hline
$V= -\frac{c_{1}}{r} +\frac{c_{2}}{x^{2}} +\frac{c_{3}}{y^{2}}$ & Table I & \makecell[c]{not \\ included} & \makecell[l]{$I_{1}= \frac{1}{2}M_{1}^{2} + c_{3}\frac{z^{2}}{y^{2}}$ \\
$I_{2}= \frac{1}{2}M_{2}^{2} + c_{2}\frac{z^{2}}{x^{2}}$ \\ $I_{3}= \frac{1}{2}M_{3}^{2} +c_{2}\frac{y^{2}}{x^{2}} +c_{3} \frac{x^{2}}{y^{2}}$ \\ $I_{4}= M_{2}\dot{x} -M_{1}\dot{y} -2z \left(\frac{c_{1}}{2r} -\frac{c_{2}}{x^{2}} -\frac{c_{3}}{y^{2}} \right)$} \\ \hline
$V= \frac{c_{1}y}{x^{2}R} +\frac{c_{2}}{x^{2}} +\frac{c_{3}}{z^{2}}$ & \makecell[l]{Table I \\ $x \leftrightarrow y$} & \makecell[c]{not \\ included} & \makecell[l]{$I_{1}= \frac{1}{2}\dot{z}^{2} +\frac{c_{3}}{z^{2}}$ \\ $I_{2}= \frac{1}{2} M_{3}^{2} +c_{2}\frac{y^{2}}{x^{2}} +c_{1}\frac{yR}{x^{2}}$ \\ $I_{3}= M_{3}\dot{x} -2c_{2} \frac{y}{x^{2}} -c_{1} \frac{x^{2} +2y^{2}}{x^{2}R}$ \\ $I_{4}= \frac{1}{2} \mathbf{M}^{2} +c_{1} \frac{yr^{2}}{Rx^{2}} +c_{2}\frac{r^{2}}{x^{2}} +c_{3}\frac{R^{2}}{z^{2}}$} \\ \hline
$V= \frac{c_{1}y}{x^{2}R} +\frac{c_{2}}{x^{2}} +c_{3}z$ & \makecell[l]{Table I \\ $x \leftrightarrow y$} & \makecell[c]{not \\ included} & \makecell[l]{$I_{1}= \frac{1}{2}\dot{z}^{2} +c_{3}z$ \\ $I_{2}= \frac{1}{2} M_{3}^{2} +c_{2}\frac{y^{2}}{x^{2}} +c_{1}\frac{yR}{x^{2}}$ \\ $I_{3}= M_{3}\dot{x} -2c_{2} \frac{y}{x^{2}} -c_{1} \frac{x^{2} +2y^{2}}{x^{2}R}$ \\ $I_{4}= M_{2}\dot{x} -M_{1}\dot{y} +c_{1}\frac{2yz}{x^{2}R} +c_{2}\frac{2z}{x^{2}} -c_{3}\frac{R^{2}}{2}$} \\ \hline
\makecell[l]{$V= k_{1}r^{2} +k_{2}\frac{\bar{w}}{w^{3}} +\frac{k_{3}}{w^{2}} +\frac{k_{4}}{z^{2}}$ \\ $w=x+iy =Re^{i\theta}$ \\ $\bar{w}= x-iy$} & \makecell[c]{not \\ included} & eq. (11) & \makecell[l]{$I_{1}= \frac{1}{2} \left( M_{2} -iM_{1} \right)^{2} +k_{4}\frac{w^{2}}{z^{2}} -k_{2}\frac{z^{2}}{w^{2}}$ \\
$I_{2}= \frac{1}{2} \left( \dot{x} +i\dot{y} \right)^{2} +k_{1}w^{2} -\frac{k_{2}}{w^{2}}$ \\ $I_{3}= \frac{1}{2}M_{3}^{2} +k_{2}e^{-4i\theta} +k_{3}e^{-2i\theta}$ \\ \enskip \enskip $= \frac{1}{2}M_{3}^{2} +k_{2}\left( \frac{\bar{w}}{w} \right)^{2} +k_{3}\frac{\bar{w}}{w}$ \\ $I_{4}= \frac{1}{2}\dot{z}^{2} +k_{1}z^{2} +\frac{k_{4}}{z^{2}}$ \\ $I_{5}= \frac{1}{2} \mathbf{M}^{2} +k_{2}\frac{r^{2}\bar{w}}{w^{3}} +k_{3}\frac{r^{2}}{w^{2}} +k_{4} \frac{r^{2}}{z^{2}}$} \\ \hline
\makecell[l]{$V= k_{1}r^{2} +\frac{k_{2}}{w^{2}} +k_{3}\frac{z}{w^{3}}+$ \\ \qquad $+k_{4}\frac{R^{2} -3z^{2}}{w^{4}}$} & \makecell[c]{not \\ included} & eq. (12) & \makecell[l]{$I_{1}= \frac{1}{2} \left( M_{2} -iM_{1} \right)^{2} +k_{3}\frac{z}{w} -4k_{4}\frac{z^{2}}{w^{2}}$ \\ $I_{2}= M_{3}\left( iM_{1} -M_{2} \right) +k_{3}\frac{y}{w} -2ik_{2}\frac{z}{w} -$ \\ \qquad $-\frac{3ik_{3}}{2}\frac{z^{2}}{w^{2}} -4ik_{4}\frac{z(x^{2}+y^{2}-z^{2})}{w^{3}}$ \\ $I_{3}= \frac{1}{2} \left( \dot{x} +i\dot{y} \right)^{2} +k_{1}w^{2} -\frac{k_{4}}{w^{2}}$ \\ $I_{4}= \frac{1}{2}\dot{z} \left( \dot{x} +i\dot{y} \right) +k_{1}zw -\frac{k_{3}}{4w^{2}} +k_{4}\frac{z}{w^{3}}$ \\ $I_{5}= \frac{1}{2}\mathbf{M}^{2} +k_{2}\frac{r^{2}}{w^{2}} +k_{3}\frac{zr^{2}}{w^{3}} +k_{4}\frac{r^{2}(x^{2} +y^{2} -3z^{2})}{w^{4}}$} \\ \hline
\makecell[l]{$V= k_{1}(R^{2} +4z^{2}) +k_{2}z +\frac{k_{3}}{w^{2}}+$ \\ \qquad  $+k_{4}\frac{\bar{w}}{w^{3}}$} & \makecell[c]{not \\ included} & eq. (14) & \makecell[l]{$I_{1}= \frac{1}{2} \left( \dot{x} +i\dot{y} \right)^{2} +k_{1}w^{2} -\frac{k_{4}}{w^{2}}$ \\ $I_{2}= \frac{1}{2}(\dot{x} +i\dot{y}) \left( M_{2} -iM_{1} \right) -k_{1}zw^{2} -$ \\ \qquad  $-\frac{k_{2}}{4}w^{2} -k_{4}\frac{z}{w^{2}}$ \\
$I_{3}= \frac{1}{2}\dot{z}^{2} +4k_{1}z^{2} +k_{2}z$ \\ $I_{4}= \frac{1}{2}M_{3}^{2} +k_{3}e^{-2i\theta} +k_{4}e^{-4i\theta}$ \\ $I_{5}= \frac{1}{2} \left(M_{2}\dot{x} -M_{1}\dot{y}\right) +k_{3}\frac{z}{w^{2}} +k_{4}\frac{z\bar{w}}{w^{3}}-$ \\ \qquad $-k_{1}zR^{2} -k_{2}\frac{R^{2}}{4}$} \\ \hline
\makecell[l]{$V= 4k_{1} \left( R^{2} -\frac{\bar{w}^{3}}{2} +\frac{z^{2}}{4} \right)+$ \\ \qquad $+k_{2} \left( 2w -3\bar{w}^{2} \right) +k_{3}\bar{w}+$ \\ \qquad $+\frac{k_{4}}{z^{2}}$} & \makecell[c]{not \\ included} & eq. (15) & \makecell[l]{$I_{1}= \frac{1}{8} \left( \dot{x} -i\dot{y} \right)^{2} +k_{1}\bar{w}^{2} +k_{2}\bar{w}$ \\ $I_{2}= \frac{1}{4}\dot{w}^{2} +iM_{3}\dot{\bar{w}} -2k_{1} \left( \frac{3}{4}\bar{w}^{4} -w^{2} +R^{2}\bar{w} \right)-$ \\ \qquad $-2k_{2} \left( \bar{w}^{3} +2R^{2} \right) +k_{3} \left( \frac{\bar{w}^{2}}{2} +w \right)$ \\ $I_{3}= \frac{1}{2}\dot{z}^{2} +k_{1}z^{2} +\frac{k_{4}}{z^{2}}$ \\ $I_{4}= \frac{1}{2} \left( M_{2} +iM_{1} \right)^{2} +2i\dot{z}M_{1} +$ \\ \qquad $+k_{1}z^{2}\left( 3\bar{w}^{2} -4iy \right) +2k_{2}z^{2} \left( 2\bar{w} +1\right) -$ \\ \qquad $-k_{3}z^{2} +\frac{k_{4}}{z^{2}} \left( \bar{w}^{2} +4iy \right)$ \\ $I_{5}= \frac{1}{2}\dot{z} \left( M_{2} +iM_{1} \right) +k_{1}z^{2}\bar{w} +k_{2}z^{2} -k_{4}\frac{\bar{w}}{z^{2}}$} \\ \hline
\makecell[l]{$V= k_{1}w +k_{2}\left( 3w^{2} +z \right)+$ \\ \qquad $+k_{3}\left( 4w^{3} +3wz +\frac{\bar{w}}{4} \right) +$ \\ \qquad $+k_{4} \left( \frac{5}{2}w^{4} +\frac{r^{2}}{2} +3w^{2}z \right)$} & \makecell[c]{not \\ included} & eq. (16) & \makecell[l]{$I_{1}= M_{3}\dot{w} -\left( M_{1} +iM_{2} \right)\dot{z} -\frac{1}{2}\dot{y}\dot{z}+$ \\ \qquad $+\frac{ik_{1}}{2}\left( w^{2}-z \right) +ik_{2}\left( 2w^{3} -zw +\frac{i}{2}y \right) -$ \\ \qquad $-\frac{ik_{3}}{8}\left( w^{2}-z \right)+ik_{3}\left( 3w^{4} -z^{2} +iyw \right)-$ \\ \qquad $ -\frac{k_{4}}{2}y \left( w^{2}+z \right)+ik_{4}w \left( 2w^{4} +zw^{2} -z^{2} \right)$ \\ $I_{2}= M_{1}\left( 2\dot{x} +i\dot{y} \right) +iM_{2}\dot{x} +M_{3}\dot{z} +\frac{i}{4}\dot{z}^{2}+$ \\ \qquad $+\frac{ik_{2}}{2} \left( z -w^{2} \right) +ik_{3}w \left( z -w^{2} \right)+$ \\
\qquad $+ \frac{ik_{4}}{4} \left(z^{2} +2zw^{2} -3w^{4}\right)$ \\ $I_{3}= \left( M_{1} +iM_{2} \right)^{2} +\left( 2iM_{1} -M_{2} \right) \dot{w}-$ \\ \qquad $-iM_{3}\dot{z} +\frac{1}{4}\left( \dot{y}^{2} -\dot{z}^{2} \right) +k_{1}zw +$ \\ \qquad  $ +\frac{ik_{1}}{2}y +2k_{2}w \left( 2zw +iy \right)+$ \\ \qquad $+\frac{k_{2}}{2} \left( w^{2} -z \right) +k_{3}w^{3}(6z+1)+$ \\ \qquad $+k_{3}zw\left(2z -\frac{1}{4}\right)+ik_{3}y\left( 3w^{2} -\frac{1}{8} \right)-$ \\ \qquad $ -k_{3}xz +k_{4}w^{4} \left( 4z +\frac{3}{4} \right) +2ik_{4}yw^{3} +$ \\ \qquad $+k_{4}z^{2} \left( 3w^{2} -\frac{1}{4}\right) +\frac{k_{4}}{2} \left( \frac{y^{2}}{2} -zR^{2} \right)$ \\ $I_{4}= \dot{w}^{2} +k_{3}w +k_{4}w^{2}$ \\ $I_{5}= \dot{z}\dot{w} +\left(k_{4}w +\frac{k_{3}}{2}\right)z +k_{4}w^{3}+$ \\ \qquad $+\frac{3k_{3}}{2}w^{2} +k_{2}w$} \\ \hline
\caption{\label{Table.e3.3} Maximally superintegrable potentials $V(x,y,z)$ in $E^{3}$ that admit QFIs of the type $I_{(1,1)}$.}
\end{longtable}

\textbf{Remark:} The potential (\ref{eq.e3pot.13d}) admits the four independent QFIs $H, I_{1}, I_{2}$ and $I_{3}$ (see Table \ref{Table.e3.1}); however, it is not second order integrable because the PBs $\{I_{i}, I_{j}\} \neq 0$ for $i\neq j$. In Table II of \cite{Evans 1990}, it is claimed that this potential is minimally superintegrable because in that paper superintegrability is defined without the requirement of the integrability (i.e. the vanishing of the PBs). Indeed, we have:
\begin{equation}
I_{4} \equiv \{I_{1},I_{3}\}= \{I_{2},I_{1}\}= \{I_{3},I_{2}\}= M_{1} \left( x^{2}\dot{y}\dot{z} +2k_{1}\frac{yz}{x^{2}} \right) +M_{2} \left( y^{2}\dot{x}\dot{z} +2k_{2}\frac{xz}{y^{2}} \right) +M_{3} \left( z^{2}\dot{x}\dot{y} +2k_{3}\frac{xy}{z^{2}} \right). \label{eq.newp.1}
\end{equation}
The third order (cubic) FI $I_{4}$ cannot be used for establishing integrability because the PBs $\{I_{i}, I_{4}\}\neq 0$, where $i,j= 1,2,3$.

\section{The QFI $I_{(2,0)}$ where $\ell=0$}

\label{sec.e3.pot.2}

We set $L_{(0)a}=L_{a}$ and the QFI $I_{(2,\ell)}$ for $\ell=0$ becomes
\begin{equation}
I_{(2,0)}= -t L_{(a;b)}\dot{q}^{a} \dot{q}^{b} +L_{a} \dot{q}^{a} +t L_{a}V^{,a} \label{eq.e3p2.0}
\end{equation}
where the vector $L_{a}$ is given by (\ref{eq.e3.4}), the generated KT $L_{(a;b)}$ is given by (\ref{eq.e3.5}) and the following condition is satisfied
\begin{equation}
\left( L_{b} V^{,b}\right)_{,a} = -2L_{(a;b)}V^{,b}. \label{eq.e3p2.1}
\end{equation}

Condition (\ref{eq.e3p2.1}) is a subcase of the general condition (\ref{eq.e3pot.4b}) in the case that the function $G= -L_{a}V^{,a}$ and the general second order KT $C_{ab}= L_{(a;b)}$ is reducible. In section \ref{sec.e3.pot.1.1}, we have computed (not all) pairs of functions $(G, V)$ which satisfy the condition (\ref{eq.e3pot.4b}). Therefore, in order to find potentials $V(x,y,z)$ that admit QFIs of the form (\ref{eq.e3p2.0}), it is sufficient to solve the constraint
\begin{equation}
G= -L_{a}V^{,a} \label{eq.e3p2.2}
\end{equation}
for all pairs $(G, V)$ for which the KT $C_{ab}$ is given by the reducible form (\ref{eq.e3.5}). If the constraint (\ref{eq.e3p2.2}) is not satisfied for some pairs $(G, V)$, then the corresponding potentials $V$ of these pairs do not admit QFIs of the type $I_{(2,0)}$.

Moreover, the QFI (\ref{eq.e3p2.0}) is written as
\[
I_{(2,0)}= -Jt +L_{a}\dot{q}^{a}
\]
where $J$ is the associated autonomous QFI (\ref{eq.e3pot.3}). The PB $\{ H, I_{(2,0)}\}= \frac{\partial I_{(2,0)}}{\partial t}= -J$. Therefore: \newline
\emph{The time-dependent QFI $I_{(2,0)}$ generates an autonomous QFI of the type $I_{(1,0)}$.} \newline
This is an interesting connection between (first degree) time-dependent and autonomous QFIs.

We consider the following cases.

1) In section \ref{sec.e3.pot.1.1.2}, we determined the functions:
\begin{eqnarray}
V(x,y,z)&=& (a^{2}+b^{2})x^{2} +4(az+by)^{2} +\frac{k_{1}}{x^{2}} +k_{2}(az +by) +F(ay -bz) \label{eq.e3p2.3a} \\
G(x,y,z)&=& -\frac{k_{2}}{2}(a^{2}+b^{2})x^{2} -2ab(ay +bz) x^{2} -2(a^{3}z +b^{3}y)x^{2} +\frac{2k_{1}(by+az)}{x^{2}}. \label{eq.e3p2.3b}
\end{eqnarray}
Then, the vector
\begin{equation}
L_{a}=
\left(
  \begin{array}{c}
    bxy +axz +2b_{2}y +2b_{1}z +b_{3} \\
    -bx^{2} -2b_{2}x +2b_{4}z +b_{6} \\
    -ax^{2} -2b_{1}x -2b_{4}y +b_{5} \\
  \end{array}
\right) \label{eq.e3p2.4}
\end{equation}
where $b_{1}, b_{2}, ..., b_{6}$ are arbitrary constants. Replacing (\ref{eq.e3p2.3a}), (\ref{eq.e3p2.3b}) and (\ref{eq.e3p2.4}) in the condition (\ref{eq.e3p2.2}), we find:
\[
b_{1}=b_{2}=b_{3}=b_{4}=0, \enskip a=\pm ib, \enskip b_{5}=\pm b_{6}.
\]
Therefore, the potential (\ref{eq.e3p2.3a}) becomes\footnote{
The function $F(iz \pm y)$ is either the $F(y+iz)$ or the $F(y -iz)$. Therefore, we can write $F(y \pm iz)$.
} (see the potential $V= F_{1}(y-bx) +F_{2}(z)$ in Table \ref{Table.e23.2})
\begin{equation}
V(x,y,z)= \frac{k_{1}}{x^{2}} +\underbrace{4b(y \pm iz)^{2} +bk_{2}(y \pm iz) +F(y \pm iz)}_{= F_{1}(y \pm iz)} = \frac{k_{1}}{x^{2}} +F_{1}(y \pm iz)  \label{eq.e3p2.5}
\end{equation}
and the vector
\begin{equation}
L_{a}=
\left(
  \begin{array}{c}
    bx(y \pm iz) \\
    -bx^{2} +b_{6} \\
    \pm i (-bx^{2} +b_{6}) \\
  \end{array}
\right). \label{eq.e3p2.6}
\end{equation}

The associated time-dependent QFI (\ref{eq.e3p2.0}) is
\begin{eqnarray}
I_{(2,0)}&=& -bt(y \pm iz)\dot{x}^{2} +btx\dot{x}(\dot{y} \pm i\dot{z}) +b(y \pm iz)x\dot{x} -(bx^{2} -b_{6})\dot{y} \mp i (bx^{2} -b_{6})\dot{z} -\frac{2k_{1}b t(y \pm iz)}{x^{2}} \notag \\
&=& b_{6}J_{1} -bJ_{2} \label{eq.e3p2.7}
\end{eqnarray}
which contains the independent FIs:
\[
J_{1}= \dot{y} \pm i\dot{z}, \enskip J_{2}= t\left( \dot{x}^{2} +\frac{2k_{1}}{x^{2}} \right) (y\pm iz) -x\dot{x} (y\pm iz)  - J_{1}x( t\dot{x} -x).
\]

From section \ref{sec.e3.pot.1.1.2} we have that the potential (\ref{eq.e3p2.5}) admits also the autonomous QFIs:
\begin{equation*}
J_{3}= (\pm iM_{2} -M_{3})\dot{x} +\frac{2k_{1}(y \pm iz)}{x^{2}}, \enskip J_{4}= \frac{1}{2}\dot{x}^{2} +\frac{k_{1}}{x^{2}}.
\end{equation*}
We note that $J_{2}= J_{3}t -x\dot{x}(y\pm iz) +J_{1}x^{2}$.

The potential (\ref{eq.e3p2.5}) is maximally superintegrable due to the five linearly independent FIs $H, J_{1}, J_{2}, J_{3}$, and $J_{4}$. The autonomous FIs $H, J_{1}, J_{4}$ are in involution. This is a new result which could not be found in \cite{Evans 1990} because of the additional time-dependent QFI $J_{2}$.

The PBs are:
\[
\{H, J_{2}\} =\frac{\partial J_{2}}{\partial t}= J_{3}, \enskip \{J_{1}, J_{2}\}=\{J_{1}, J_{3}\}=\{J_{1}, J_{4}\}=0, \enskip \{J_{3}, J_{4}\}= -J_{1}  \left( \dot{x}^{2} +\frac{2k_{1}}{x^{2}} \right),
\]
\[
\{J_{2}, J_{3}\}= -2(M_{3} \mp iM_{2})^{2} -\frac{4k_{1}}{x^{2}} (y \pm iz)^{2}, \enskip \{J_{2}, J_{4}\}= -\left( J_{1}t +y \pm iz \right) \left( \dot{x}^{2} +\frac{2k_{1}}{x^{2}} \right) +2J_{1}x\dot{x}.
\]

2) In section \ref{sec.e3.pot.1.1.2}, we determined the functions:
\begin{eqnarray}
V(x,y,z)&=& \frac{F_{1}\left(\frac{y}{x}\right)}{R^{2}} +\frac{k_{1}z}{rR^{2}} -\frac{k_{2}}{r} \label{eq.e3p2.8a} \\
G(x,y,z)&=& a_{2} \frac{2z F_{1}\left( \frac{y}{x} \right)}{R^{2}} +a_{2} \frac{2k_{1} z^{2}}{rR^{2}} +a_{2}\frac{k_{1}}{r} -a_{2}\frac{k_{2}z}{r}. \label{eq.e3p2.8b}
\end{eqnarray}
Then, the vector
\begin{equation}
L_{a}=
\left(
  \begin{array}{c}
    axz +2b_{2}y +2b_{1}z +b_{3} \\
    ayz -2b_{2}x +2b_{4}z +b_{6} \\
    -aR^{2} -2b_{1}x -2b_{4}y +b_{5} \\
  \end{array}
\right). \label{eq.e3p2.9}
\end{equation}
Replacing (\ref{eq.e3p2.8a}), (\ref{eq.e3p2.8b}) and (\ref{eq.e3p2.9}) in the condition (\ref{eq.e3p2.2}), we get:
\[
b_{1}=b_{2}=b_{3}=b_{4}=b_{5}=b_{6}=0, \enskip k_{2}=0.
\]
Therefore, the potential (\ref{eq.e3p2.8a}) becomes
\begin{equation}
V(x,y,z)= \frac{F_{1}\left(\frac{y}{x}\right)}{R^{2}} +\frac{k_{1}z}{rR^{2}} = R^{-2} \left[ F_{1}\left(\frac{y}{x}\right) +\frac{k_{1}z}{r} \right] \label{eq.e3p2.10}
\end{equation}
and the vector $L_{b}= a \left( xz, yz, -R^{2} \right)$.

The associated time-dependent QFI (\ref{eq.e3p2.0}) is
\begin{equation}
I_{(2,0)}= -J_{1} t + z (x\dot{x} +y\dot{y}) -(x^{2} +y^{2}) \dot{z} \label{eq.e3p2.11}
\end{equation}
where $J_{1}$ is the autonomous QFI
\begin{equation*}
J_{1}= M_{2}\dot{x} -M_{1}\dot{y} +\frac{2z F_{1}\left( \frac{y}{x} \right)}{x^{2} +y^{2}} +\frac{2k_{1} z^{2}}{r(x^{2} +y^{2})} +\frac{k_{1}}{r}.
\end{equation*}

From Table \ref{Table.e3.2} we have that the potential (\ref{eq.e3p2.10}) admits the additional autonomous QFIs:
\[
J_{2}= \frac{1}{2}M_{3}^{2} +F_{1}\left( \frac{y}{x} \right), \enskip J_{3}= \frac{1}{2}\mathbf{M}^{2} +\frac{r^{2} F_{1}\left( \frac{y}{x} \right)}{x^{2} +y^{2}} +\frac{k_{1}zr}{x^{2} +y^{2}}.
\]
Therefore, (\ref{eq.e3p2.10}) is a new maximally superintegrable potential due to the five independent QFIs $H, J_{1}, J_{2}, J_{3}$, and (\ref{eq.e3p2.11}). We note that this potential was considered to be minimally superintegrable (see e.g. \cite{Evans 1990}) because only autonomous QFIs were considered.

The PBs are $\{H, I_{(2,0)}\}= -J_{1}$ and $\{I_{(2,0)}, J_{2}\}=0$.

3) In section \ref{sec.e3.pot.1.1.1}, we determined the functions:
\begin{eqnarray}
V(x,y,z)&=& F_{1}(x) +F_{2}(y) +F_{3}(z) \label{eq.e3p2.12a} \\
G(x,y,z)&=& 2a_{3}F_{1}(x) +2a_{13}F_{2}(y) +2a_{9}F_{3}(z). \label{eq.e3p2.12b}
\end{eqnarray}
Then, the vector
\begin{equation}
L_{a}=
\left(
  \begin{array}{c}
    a_{3}x +2b_{2}y +2b_{1}z +b_{3} \\
    a_{13}y -2b_{2}x +2b_{4}z +b_{6} \\
    a_{9}z -2b_{1}x -2b_{4}y +b_{5} \\
  \end{array}
\right). \label{eq.e3p2.13}
\end{equation}
Replacing (\ref{eq.e3p2.12a}), (\ref{eq.e3p2.12b}) and (\ref{eq.e3p2.13}) in the condition (\ref{eq.e3p2.2}), we obtain the following ordinary differential equation (ODE):
\begin{eqnarray}
0&=& a_{3} \left[ xF_{1}' + 2F_{1}(x) \right] +b_{3}F_{1}' +a_{13} \left[ yF_{2}' + 2F_{2}(y) \right] +b_{6}F_{2}' +a_{9} \left[ zF_{3}' + 2F_{3}(z) \right] +b_{5}F_{3}' + \notag \\
&& +2b_{2} \left( F_{1}'y -F_{2}'x \right) +2b_{1} \left( F_{1}'z -F_{3}'x \right) +2b_{4} \left( F_{2}'z -F_{3}'y \right) \label{eq.e3p2.14}
\end{eqnarray}
where $F_{1}'= \frac{dF_{1}}{dx}$, $F_{2}'= \frac{dF_{2}}{dy}$ and $F_{3}'= \frac{dF_{3}}{dz}$.

We consider the following subcases:

3.1. Subcase $b_{1}=b_{2}=b_{4}=0$ and the pairs $(a_{3}, b_{3})$, $(a_{13}, b_{6})$, $(a_{9}, b_{5})$ are not the origin $(0,0)$.

Then, the ODE (\ref{eq.e3p2.14}) gives:
\begin{eqnarray}
a_{3} \left[ xF_{1}' + 2F_{1}(x) \right] +b_{3}F_{1}' &=& \lambda_{1} \label{eq.e3p2.15a} \\
a_{13} \left[ yF_{2}' + 2F_{2}(y) \right] +b_{6}F_{2}' &=& \lambda_{2} \label{eq.e3p2.15b} \\
a_{9} \left[ zF_{3}' + 2F_{3}(z) \right] +b_{5}F_{3}' &=& -\lambda_{1} -\lambda_{2} \label{eq.e3p2.15c}
\end{eqnarray}
where $\lambda_{1}$ and $\lambda_{2}$ are arbitrary constants, and the vector
$L_{a}=
\left(
  \begin{array}{c}
    a_{3}x +b_{3} \\
    a_{13}y +b_{6} \\
    a_{9}z +b_{5} \\
  \end{array}
\right)$.

Solving the system of ODEs (\ref{eq.e3p2.15a}) - (\ref{eq.e3p2.15c}), we find the functions:
\[
F_{1}(x)= \frac{\lambda_{1} \left( \frac{a_{3}}{2}x^{2} +b_{3}x \right) +c_{1}}{(a_{3}x +b_{3})^{2}}, \enskip F_{2}(y)= \frac{\lambda_{2} \left( \frac{a_{13}}{2}y^{2} +b_{6}y \right) +c_{2}}{(a_{13}y +b_{6})^{2}}, \enskip F_{3}(z)= -\frac{(\lambda_{1} +\lambda_{2})\left( \frac{a_{9}}{2}z^{2} +b_{5}z \right) +c_{3}}{(a_{9}z +b_{5})^{2}}
\]
where $c_{1}, c_{2}$, and $c_{3}$ are arbitrary constants.

Then, the potential (\ref{eq.e3p2.12a}) becomes
\begin{equation}
V(x,y,z)= \frac{\lambda_{1} \left( \frac{a_{3}}{2}x^{2} +b_{3}x \right) +c_{1}}{(a_{3}x +b_{3})^{2}}+ \frac{\lambda_{2} \left( \frac{a_{13}}{2}y^{2} +b_{6}y \right) +c_{2}}{(a_{13}y +b_{6})^{2}} -\frac{(\lambda_{1} +\lambda_{2})\left( \frac{a_{9}}{2}z^{2} +b_{5}z \right) +c_{3}}{(a_{9}z +b_{5})^{2}}. \label{eq.e3p2.16}
\end{equation}

The associated time-dependent QFI (\ref{eq.e3p2.0}) is
\begin{equation}
I_{(2,0)}= -Jt +(a_{3}x\dot{x} +a_{13}y\dot{y} +a_{9}z\dot{z}) +b_{3}\dot{x} +b_{6}\dot{y} +b_{5}\dot{z} \label{eq.e3p2.17}
\end{equation}
where $J=2a_{3}I_{1}+ 2a_{13}I_{2} +2a_{9}I_{3}$ is the sum of the three separated QFIs:
\begin{equation}
I_{1}= \frac{1}{2}\dot{x}^{2} +F_{1}(x), \enskip I_{2}= \frac{1}{2}\dot{y}^{2} +F_{2}(y), \enskip I_{3}= \frac{1}{2}\dot{z}^{2} +F_{3}(z). \label{eq.e3p2.18}
\end{equation}
Therefore, (\ref{eq.e3p2.16}) is a new minimally superintegrable potential due to the four independent QFIs $I_{1}, I_{2}, I_{3}$, and (\ref{eq.e3p2.17}). We note that (\ref{eq.e3p2.16}) depends on the eleven parameters $a_{3}, a_{9}, a_{13}, b_{3}, b_{5}, b_{6}, c_{1}, c_{2}, c_{3}, \lambda_{1}$ and $\lambda_{2}$; hence, the time-dependent QFI (\ref{eq.e3p2.17}) is irreducible.

- For $\lambda_{1}=\lambda_{2}=0$ and $a_{3}a_{13}a_{9} \neq0$ we obtain the potential\footnote{Since $a_{3}a_{13}a_{9} \neq0$, we can set $b_{3}= m_{1}a_{3}$, $b_{6}= m_{2}a_{13}$, $b_{5}= m_{3}a_{9}$, $k_{1}= \frac{c_{1}}{a_{3}^{2}}$, $k_{2}= \frac{c_{2}}{a_{13}^{2}}$ and $k_{3}= \frac{c_{3}}{a_{9}^{2}}$.}
\begin{equation}
V(x,y,z)= \frac{k_{1}}{(x +m_{1})^{2}}+ \frac{k_{2}}{(y +m_{2})^{2}} +\frac{k_{3}}{(z +m_{3})^{2}} \label{eq.e3p2.18.1}
\end{equation}
where $k_{1}, k_{2}, k_{3}, m_{1}, m_{2}$, and $m_{3}$ are new arbitrary constants.

Then, the associated time-dependent QFI (\ref{eq.e3p2.17}) consists of the independent QFIs:
\[
I_{4}= -2I_{1}t +(x + m_{1}) \dot{x}, \enskip I_{5}= -2I_{2}t +(y + m_{2}) \dot{y}, \enskip I_{6}= -2I_{3}t +(z + m_{3}) \dot{z}.
\]
Therefore, the potential (\ref{eq.e3p2.18.1}) is maximally superintegrable due to the independent QFIs $I_{1}, I_{2}, ..., I_{6}$. Because time-dependent FIs are considered, the maximum number of independent FIs is six (i.e. greater than five).

3.2. Subcase $b_{1}=b_{2}=b_{4}=0$, $a_{3}\neq0$ and $a_{9}=a_{13}=b_{5}=b_{6}=0$ (i.e. two pairs of parameters from subcase 3.1 vanish).

From the system of ODEs (\ref{eq.e3p2.15a}) - (\ref{eq.e3p2.15c}), we find that $\lambda_{1}= \lambda_{2}=0$ and the potential (\ref{eq.e3p2.12a}) becomes
\begin{equation}
V(x,y,z)= \frac{k_{1}}{(x +m_{1})^{2}} +F_{2}(y) +F_{3}(z) \label{eq.e3p2.20}
\end{equation}
where $k_{1}, m_{1}$ are arbitrary constants and $F_{2}(y)$, $F_{3}(z)$ are arbitrary smooth functions.

The associated time-dependent QFI (\ref{eq.e3p2.0}) is
\begin{equation}
I_{(2,0)}= -2I_{1}t +(x +m_{1})\dot{x} \label{eq.e3p2.21}
\end{equation}
where the QFI $I_{1}= \frac{1}{2}\dot{x}^{2} +\frac{k_{1}}{(x +m_{1})^{2}}$. Therefore, the potential (\ref{eq.e3p2.20}) is minimally superintegrable (see Table \ref{Table.e23.2}).

3.3. Subcase $b_{1}=b_{2}=b_{4}=0$, $a_{9}=b_{5}=0$ and the pairs $(a_{3}, b_{3})$, $(a_{13}, b_{6})$ are not the origin $(0,0)$.

From the system of ODEs (\ref{eq.e3p2.15a}) - (\ref{eq.e3p2.15c}), we find that $\lambda_{2}= -\lambda_{1}$ and the potential (\ref{eq.e3p2.12a}) becomes
\begin{equation}
V(x,y,z)= \frac{\lambda_{1} \left( \frac{a_{3}}{2}x^{2} +b_{3}x \right) +c_{1}}{(a_{3}x +b_{3})^{2}}- \frac{\lambda_{1} \left( \frac{a_{13}}{2}y^{2} +b_{6}y \right) +c_{2}}{(a_{13}y +b_{6})^{2}} +F_{3}(z) \label{eq.e3p2.22}
\end{equation}
where $F_{3}(z)$ is an arbitrary smooth function.

The associated time-dependent QFI (\ref{eq.e3p2.0}) is
\begin{equation}
I_{(2,0)}= -2(a_{3}I_{1} +a_{13}I_{2})t +(a_{3}x +b_{3})\dot{x} +(a_{13}y +b_{6})\dot{y} \label{eq.e3p2.23}
\end{equation}
where the QFIs $I_{1}$ and $I_{2}$ are given by (\ref{eq.e3p2.18}). We note that the potential (\ref{eq.e3p2.22}) is a minimally superintegrable potential.

- For $\lambda_{1}=0$ and $a_{3}a_{13}\neq0$ we obtain the maximally superintegrable potential (see Table \ref{Table.e23.3})
\begin{equation}
V(x,y,z)= \frac{k_{1}}{(x +m_{1})^{2}} +\frac{k_{2}}{(y +m_{2})^{2}} +F_{3}(z) \label{eq.e3p2.23.1}
\end{equation}
which admits the additional time-dependent QFIs:
\[
I_{4}= -2I_{1}t +(x+m_{1})\dot{x}, \enskip I_{5}= -2I_{2}t +(y +m_{2})\dot{y}.
\]

3.4. Subcase $a_{3}=a_{9}=a_{13}=0$ (autonomous LFIs, $L_{(a;b)}=0$ and $G=0$).

The ODE (\ref{eq.e3p2.14}) becomes
\begin{equation}
2b_{2} \left( F_{1}'y -F_{2}'x \right) +2b_{1} \left( F_{1}'z -F_{3}'x \right) +2b_{4} \left( F_{2}'z -F_{3}'y \right) +b_{3}F_{1}' +b_{6}F_{2}' +b_{5}F_{3}' =0 \label{eq.e3p2.24}
\end{equation}
and the vector
\begin{equation}
L_{a}=
\left(
  \begin{array}{c}
    2b_{2}y +2b_{1}z +b_{3} \\
   -2b_{2}x +2b_{4}z +b_{6} \\
   -2b_{1}x -2b_{4}y +b_{5} \\
  \end{array}
\right). \label{eq.e3p2.24.1}
\end{equation}

The ODE (\ref{eq.e3p2.24}) admits solutions of the form:
\begin{equation}
F_{1}(x)= kx^{2} +k_{1}x, \enskip F_{2}(y)= ky^{2} +k_{2}y, \enskip F_{3}(z)= kz^{2} +k_{3}z \label{eq.e3p2.25}
\end{equation}
where $k, k_{1}, k_{2}$, and $k_{3}$ are arbitrary constants. Then, we get the separable potential
\begin{equation}
V(x,y,z)= kr^{2} +k_{1}x +k_{2}y +k_{3}z. \label{eq.e3p2.26}
\end{equation}

Replacing (\ref{eq.e3p2.25}) in (\ref{eq.e3p2.24}), we find the following system of equations:
\begin{eqnarray}
k_{1}b_{3} +k_{2}b_{6} +k_{3}b_{5} &=& 0 \label{eq.e3p2.27.1} \\
kb_{3} -k_{2}b_{2} -k_{3}b_{1} &=& 0 \label{eq.e3p2.27.2} \\
kb_{6} +k_{1}b_{2} -k_{3}b_{4} &=& 0 \label{eq.e3p2.27.3} \\
kb_{5} +k_{1}b_{1} +k_{2}b_{4} &=& 0 \label{eq.e3p2.27.4}.
\end{eqnarray}
We consider the following cases.

- Case $k=0$.

The potential (\ref{eq.e3p2.26}) becomes
\begin{equation}
V(x,y,z)=k_{1}x +k_{2}y +k_{3}z \label{eq.e3p2.28}
\end{equation}
where $k_{1}k_{2}k_{3}\neq0$ in order to have a 3d potential.

Solving the system of equations (\ref{eq.e3p2.27.1}) - (\ref{eq.e3p2.27.4}) for $k=0$, we find:
\[
b_{1}=-\frac{k_{2}}{k_{1}}b_{4}, \enskip b_{2}= \frac{k_{3}}{k_{1}}b_{4}, \enskip b_{3}= -\frac{k_{2}}{k_{1}}b_{6} -\frac{k_{3}}{k_{1}}b_{5}.
\]

The associated QFI (\ref{eq.e3p2.0}) reduces to the LFI
\begin{equation*}
I= L_{a}\dot{q}^{a} = -2b_{4} \sum^{3}_{i=1} k_{i}M_{i} -b_{5} (k_{3}\dot{x} -k_{1}\dot{z}) -b_{6}(k_{2}\dot{x} -k_{1}\dot{y})
\end{equation*}
which consists of the LFIs:
\[
J_{1}= \sum^{3}_{i=1} k_{i}M_{i}, \enskip J_{2}= k_{3}\dot{x} -k_{1}\dot{z}, \enskip J_{3}= k_{2}\dot{x} -k_{1}\dot{y}.
\]
Therefore, the separable potential (\ref{eq.e3p2.28}) is maximally superintegrable.

- Case $k\neq0$.

The system of equations (\ref{eq.e3p2.27.1}) - (\ref{eq.e3p2.27.4}) implies that $b_{3}= \frac{k_{2}}{k}b_{2} +\frac{k_{3}}{k}b_{1}$, $b_{5}= -\frac{k_{1}}{k}b_{1} -\frac{k_{2}}{k}b_{4}$, and $b_{6}= \frac{k_{3}}{k}b_{4} -\frac{k_{1}}{k}b_{2}$.

Similarly, we find the LFIs:
\[
J_{1}= 2kM_{1} +k_{2}\dot{z} -k_{3}\dot{y}, \enskip J_{2}= 2kM_{2} +k_{3}\dot{x} -k_{1}\dot{z}, \enskip J_{3}= 2kM_{3} +k_{1}\dot{y} -k_{2}\dot{x}.
\]
Therefore, the separable potential (\ref{eq.e3p2.26}) is maximally superintegrable. We note that the $k\neq 0$ introduces the term $kr^{2}$ which is the oscillator; therefore, the corresponding change in the FIs is the addition of the components of the angular momentum.

\subsection{Case $L_{a}$ is a KV}

\label{sec.e3.pot.2.1}

We consider that $L_{a}$ is a KV in $E^{3}$. Then, $L_{(a;b)}=0$ and the time-dependent QFI (\ref{eq.e3p2.0}) becomes the time-dependent LFI
\begin{equation}
I= L_{a}\dot{q}^{a} +st \label{eq.e3p2.29}
\end{equation}
where the arbitrary constant $s$ satisfies the condition
\begin{equation}
L_{a}V^{,a}= s. \label{eq.e3p2.30}
\end{equation}

Replacing the general KV $L_{a}$ given by (\ref{eq.e3.2}) in (\ref{eq.e3p2.30}), we find the PDE
\begin{equation}
\left( b_{1} -b_{4}y +b_{5}z \right) \frac{\partial V}{\partial x} +\left( b_{2} +b_{4}x -b_{6}z \right) \frac{\partial V}{\partial y} +\left( b_{3} -b_{5}x +b_{6}y \right) \frac{\partial V}{\partial z} =s \label{eq.e3p2.31}
\end{equation}
where $b_{1}, ..., b_{6}$ are arbitrary constants.

We consider the following cases.

1) Case $b_{1}\neq0$ and $b_{4}=b_{5}=b_{6}=0$.

Then, the PDE (\ref{eq.e3p2.31}) gives the potential
\begin{equation}
V= c_{1}x +F(y -c_{2}x, z -c_{3}x) \label{eq.e3p2.32}
\end{equation}
where $c_{1}= \frac{s}{b_{1}}$, $c_{2}= \frac{b_{2}}{b_{1}}$, $c_{3}= \frac{b_{3}}{b_{1}}$ and $F$ is an arbitrary smooth function of its arguments.

The associated LFI (\ref{eq.e3p2.29}) is
\begin{equation}
I= \dot{x} +c_{2}\dot{y} +c_{3}\dot{z} +c_{1}t. \label{eq.e3p2.33}
\end{equation}

2) Case $b_{2}\neq0$, $b_{1}=0$ and $b_{4}=b_{5}=b_{6}=0$.

We find a subcase of the potential (\ref{eq.e3p2.32}) for $x \leftrightarrow y$ and $c_{2}=0$.

3) Case $b_{3}\neq0$, $b_{1}=b_{2}=0$ and $b_{4}=b_{5}=b_{6}=0$.

We find a subcase of the potential (\ref{eq.e3p2.32}) for $c_{1}=c$, $c_{2}=c_{3}=0$ and $x \leftrightarrow z$.

4) Case $b_{4}\neq0$ and $b_{5}=b_{6}=0$.

Then, the PDE (\ref{eq.e3p2.31}) gives the potential
\begin{equation}
V= c_{0} \tan^{-1} \left( \frac{x+c_{2}}{y+c_{1}} \right) + F \left[ \frac{1}{2}(x^{2} +y^{2}) +c_{2}x +c_{1}y, z +c_{3}\tan^{-1} \left( \frac{x+c_{2}}{y+c_{1}} \right) \right] \label{eq.e3p2.38}
\end{equation}
where $c_{0}= \frac{s}{b_{4}}$, $c_{1}= -\frac{b_{1}}{b_{4}}$, $c_{2}= \frac{b_{2}}{b_{4}}$, $c_{3}= -\frac{b_{3}}{b_{4}}$ and $F$ is an arbitrary function of its arguments.

The associated LFI (\ref{eq.e3p2.29}) is
\begin{equation}
I= M_{3} -c_{1}\dot{x} +c_{2}\dot{y} -c_{3}\dot{z} +c_{0}t. \label{eq.e3p2.39}
\end{equation}

5) Case $b_{4}\neq0$, $b_{6}=0$ and $b_{2}=b_{3}=0$.

Then, the PDE (\ref{eq.e3p2.31}) gives the potential
\begin{equation}
V= \frac{c_{0}}{\sqrt{1+c_{1}^{2}}} \tan^{-1} \left( \frac{y +c_{1}z +c_{2}}{\sqrt{1+c_{1}^{2}} x} \right) + F \left( z -c_{1}y, x^{2} +(1-c_{1}^{2})y^{2} +2c_{2}y +2c_{1}yz \right) \label{eq.e3p2.40}
\end{equation}
where $c_{0}= \frac{s}{b_{4}}$, $c_{1}= -\frac{b_{5}}{b_{4}}$, $c_{2}= -\frac{b_{1}}{b_{4}}$ and $F$ is an arbitrary function of its arguments.

The associated LFI (\ref{eq.e3p2.29}) is
\begin{equation}
I= M_{3} -c_{1}M_{2} -c_{2}\dot{x} +c_{0}t. \label{eq.e3p2.41}
\end{equation}

6) Case $b_{1}=b_{2}=b_{3}=s=0$ and $b_{6}\neq0$.

Then, the PDE (\ref{eq.e3p2.31}) gives the potential
\begin{equation}
V= F(r, x -c_{1}y -c_{2}z) \label{eq.e3p2.42}
\end{equation}
where $c_{1}= -\frac{b_{5}}{b_{6}}$, $c_{2}= -\frac{b_{4}}{b_{6}}$ and $F$ is an arbitrary function of its arguments.

The associated LFI (\ref{eq.e3p2.29}) is
\begin{equation}
I= M_{1} -c_{1}M_{2} -c_{2}M_{3}. \label{eq.e3p2.43}
\end{equation}
\bigskip

We collect the results of section \ref{sec.e3.pot.2} in Tables \ref{Table.e3.4} - \ref{Table.e3.6}.

\newpage

\begin{longtable}{|l|l|l|}
\hline
\multicolumn{3}{|c|}{{\large{Minimally superintegrable potentials}}} \\ \hline
{\large Potential} & {\large Ref \cite{Evans 1990}} & {\large LFIs and
QFIs} \\ \hline
\makecell[l]{$V= \frac{\lambda_{1} \left( \frac{a_{3}}{2}x^{2} +b_{3}x \right) +c_{1}}{(a_{3}x +b_{3})^{2}} +\frac{\lambda_{2} \left( \frac{a_{13}}{2}y^{2} +b_{6}y \right) +c_{2}}{(a_{13}y +b_{6})^{2}} -$ \\ \quad \enskip $-\frac{(\lambda_{1} +\lambda_{2})\left( \frac{a_{9}}{2}z^{2} +b_{5}z \right) +c_{3}}{(a_{9}z +b_{5})^{2}}$} & New & \makecell[l]{$I = -Jt +(a_{3}x\dot{x} +a_{13}y\dot{y} +a_{9}z\dot{z}) +$ \\ \quad \enskip $+b_{3}\dot{x} +b_{6}\dot{y} +b_{5}\dot{z}$  \\
$J=2a_{3}I_{1}+ 2a_{13}I_{2} +2a_{9}I_{3}$ \\
$I_{1}= \frac{1}{2}\dot{x}^{2} +\frac{\lambda_{1} \left( \frac{a_{3}}{2}x^{2} +b_{3}x \right) +c_{1}}{(a_{3}x +b_{3})^{2}}$ \\ $I_{2}= \frac{1}{2}\dot{y}^{2} +\frac{\lambda_{2} \left( \frac{a_{13}}{2}y^{2} +b_{6}y \right) +c_{2}}{(a_{13}y +b_{6})^{2}}$ \\  $I_{3}= \frac{1}{2}\dot{z}^{2} -\frac{(\lambda_{1} +\lambda_{2})\left( \frac{a_{9}}{2}z^{2} +b_{5}z \right) +c_{3}}{(a_{9}z +b_{5})^{2}}$} \\ \hline
$V= \frac{k_{1}}{(x +m_{1})^{2}} +F_{2}(y) +F_{3}(z)$ & New & \makecell[l]{$I_{1}= \frac{1}{2}\dot{x}^{2} +\frac{k_{1}}{(x +m_{1})^{2}}$ \\ $I_{2}= \frac{1}{2}\dot{y}^{2} +F_{2}(y)$ \\ $I_{3}= \frac{1}{2}\dot{z}^{2} +F_{3}(z)$ \\ $I_{4}= -2I_{1}t +(x +m_{1})\dot{x}$} \\ \hline
\makecell[l]{$V= \frac{\lambda_{1} \left( \frac{a_{3}}{2}x^{2} +b_{3}x \right) +c_{1}}{(a_{3}x +b_{3})^{2}}-$ \\ \quad \enskip $-\frac{\lambda_{1} \left( \frac{a_{13}}{2}y^{2} +b_{6}y \right) +c_{2}}{(a_{13}y +b_{6})^{2}} +F_{3}(z)$} & New & \makecell[l]{$I_{1}= \frac{1}{2}\dot{x}^{2} +\frac{\lambda_{1} \left( \frac{a_{3}}{2}x^{2} +b_{3}x \right) +c_{1}}{(a_{3}x +b_{3})^{2}}$ \\ $I_{2}= \frac{1}{2}\dot{y}^{2} -\frac{\lambda_{1} \left( \frac{a_{13}}{2}y^{2} +b_{6}y \right) +c_{2}}{(a_{13}y +b_{6})^{2}}$ \\ $I_{3}= \frac{1}{2}\dot{z}^{2} +F_{3}(z)$ \\ $I_{4}= -2(a_{3}I_{1} +a_{13}I_{2})t +(a_{3}x +b_{3})\dot{x}+$ \\ \qquad $+(a_{13}y +b_{6})\dot{y}$} \\ \hline
\caption{\label{Table.e3.4} Minimally superintegrable potentials $V(x,y,z)$ in $E^{3}$ that admit time-dependent QFIs of the form $I_{(2,0)}$.}
\end{longtable}

\newpage

\begin{longtable}{|l|l|l|}
\hline
\multicolumn{3}{|c|}{{\large{Maximally superintegrable potentials}}} \\ \hline
{\large Potential} & {\large Ref \cite{Evans 1990}} & {\large LFIs and
QFIs} \\ \hline
$V= \frac{k_{1}}{x^{2}} +F_{1}(y \pm iz)$ & New & \makecell[l]{$J_{1}= \dot{y} \pm i\dot{z}$ \\ $J_{2}= J_{3}t -x\dot{x}(y\pm iz) +J_{1}x^{2}$ \\ $J_{3}= (\pm iM_{2} -M_{3})\dot{x} +\frac{2k_{1}(y \pm iz)}{x^{2}}$ \\ $J_{4}= \frac{1}{2}\dot{x}^{2} +\frac{k_{1}}{x^{2}}$} \\ \hline
$V= R^{-2} \left[ F_{1}\left(\frac{y}{x}\right) +\frac{k_{1}z}{r} \right]$ & New & \makecell[l]{$J_{1}= M_{2}\dot{x} -M_{1}\dot{y} +\frac{2z F_{1}\left( \frac{y}{x} \right)}{x^{2} +y^{2}} +\frac{2k_{1} z^{2}}{r(x^{2} +y^{2})} +\frac{k_{1}}{r}$ \\ $J_{2}= \frac{1}{2}M_{3}^{2} +F_{1}\left( \frac{y}{x} \right)$ \\ $J_{3}= \frac{1}{2}\mathbf{M}^{2} +\frac{r^{2} F_{1}\left( \frac{y}{x} \right)}{x^{2} +y^{2}} +\frac{k_{1}zr}{x^{2} +y^{2}}$ \\ $J_{4}= -J_{1} t + z (x\dot{x} +y\dot{y}) -(x^{2} +y^{2}) \dot{z}$} \\ \hline
$V= \frac{k_{1}}{(x +m_{1})^{2}}+ \frac{k_{2}}{(y +m_{2})^{2}} +\frac{k_{3}}{(z +m_{3})^{2}}$ & New & \makecell[l]{$I_{1}= \frac{1}{2}\dot{x}^{2} +\frac{k_{1}}{(x +m_{1})^{2}}$ \\ $I_{2}= \frac{1}{2}\dot{y}^{2} +\frac{k_{2}}{(y +m_{2})^{2}}$ \\ $I_{3}= \frac{1}{2}\dot{z}^{2} +\frac{k_{3}}{(z +m_{3})^{2}}$ \\ $I_{4}= -2I_{1}t +(x + m_{1}) \dot{x}$ \\ $I_{5}= -2I_{2}t +(y + m_{2}) \dot{y}$ \\ $I_{6}= -2I_{3}t +(z + m_{3}) \dot{z}$} \\ \hline
$V= \frac{k_{1}}{(x +m_{1})^{2}} +\frac{k_{2}}{(y +m_{2})^{2}} +F_{3}(z)$ & New & \makecell[l]{$I_{1}= \frac{1}{2}\dot{x}^{2} +\frac{k_{1}}{(x +m_{1})^{2}}$ \\ $I_{2}= \frac{1}{2}\dot{y}^{2} +\frac{k_{2}}{(y +m_{2})^{2}}$ \\ $I_{3}= \frac{1}{2}\dot{z}^{2} +F_{3}(z)$ \\ $I_{4}= -2I_{1}t +(x+m_{1})\dot{x}$ \\ $I_{5}= -2I_{2}t +(y +m_{2})\dot{y}$} \\ \hline
$V=k_{1}x +k_{2}y +k_{3}z$ & New & \makecell[l]{$I_{1}= \frac{1}{2}\dot{x}^{2} +k_{1}x$ \\ $I_{2}= \frac{1}{2}\dot{y}^{2} +k_{2}y$ \\ $I_{3}= \frac{1}{2}\dot{z}^{2} +k_{3}z$ \\ $I_{4}= \sum^{3}_{i=1} k_{i}M_{i}$ \\ $I_{5}= k_{3}\dot{x} -k_{1}\dot{z}$ \\ $I_{6}= k_{2}\dot{x} -k_{1}\dot{y}$} \\ \hline
$V= kr^{2} +k_{1}x +k_{2}y +k_{3}z$ & New & \makecell[l]{$I_{1}= \frac{1}{2}\dot{x}^{2} +kx^{2} +k_{1}x$ \\ $I_{2}= \frac{1}{2}\dot{y}^{2} +ky^{2} +k_{2}y$ \\ $I_{3}= \frac{1}{2}\dot{z}^{2} +kz^{2} +k_{3}z$ \\ $I_{4}= 2kM_{1} +k_{2}\dot{z} -k_{3}\dot{y}$ \\ $I_{5}= 2kM_{2} +k_{3}\dot{x} -k_{1}\dot{z}$ \\ $I_{6}= 2kM_{3} +k_{1}\dot{y} -k_{2}\dot{x}$} \\ \hline
\caption{\label{Table.e3.5} Maximally superintegrable potentials $V(x,y,z)$ in $E^{3}$ that admit QFIs of the form $I_{(2,0)}$.}
\end{longtable}

\begin{longtable}{|l|l|}
\hline
{\large Potential} & {\large LFIs and
QFIs} \\ \hline
$V= c_{1}x +F(y -c_{2}x, z -c_{3}x)$ &
$I= \dot{x} +c_{2}\dot{y} +c_{3}\dot{z} +c_{1}t$ \\ \hline
\makecell[l]{$V= c_{0} \tan^{-1} \left( \frac{x+c_{2}}{y+c_{1}} \right) +$ \\ \quad \enskip $+ F \left[ \frac{1}{2}(x^{2} +y^{2}) +c_{2}x +c_{1}y, z +c_{3}\tan^{-1} \left( \frac{x+c_{2}}{y+c_{1}} \right) \right]$} & $I= M_{3} -c_{1}\dot{x} +c_{2}\dot{y} -c_{3}\dot{z} +c_{0}t$ \\ \hline
\makecell[l]{$V= \frac{c_{0}}{\sqrt{1+c_{1}^{2}}} \tan^{-1} \left( \frac{y +c_{1}z +c_{2}}{\sqrt{1+c_{1}^{2}} x} \right)+$ \\ \quad \enskip $+ F \left( z -c_{1}y, x^{2} +(1-c_{1}^{2})y^{2} +2c_{2}y +2c_{1}yz \right)$} & $I= M_{3} -c_{1}M_{2} -c_{2}\dot{x} +c_{0}t$ \\ \hline
$V= F(r, x -c_{1}y -c_{2}z)$ & $I= M_{1} -c_{1}M_{2} -c_{2}M_{3}$ \\ \hline
\caption{\label{Table.e3.6} Possibly non-integrable potentials $V(x,y,z)$ in $E^{3}$ that admit LFIs of the form $I= L_{a}\dot{q}^{a} +st$.}
\end{longtable}

\section{The QFI $I_{(3)}$}

\label{sec.e3.pot.3}

In this section, we consider the QFI
\begin{equation*}
I_{(3)} = e^{\lambda t} \left(-L_{(a;b)}\dot{q}^{a}\dot{q}^{b} + \lambda L_{a} \dot{q}^{a} + L_{a}V^{,a} \right)
\end{equation*}
where the vector $L_{a}$ is given by (\ref{eq.e3.4}), the generated KT $L_{(a;b)}$ is given by (\ref{eq.e3.5}) and the following condition is satisfied
\begin{equation}
\left( L_{b} V^{,b}\right)_{,a} = -2L_{(a;b)}V^{,b} -\lambda^{2} L_{a}. \label{eq.e3p3.1}
\end{equation}

We consider several cases concerning the parameters $a_{1}, a_{2}, ..., a_{20}$ which define the vector $L_{a}$ given in (\ref{eq.e3.4}).

\subsection{Case containing KVs and the HV: parameters $a_{1}, a_{3}, a_{4}, a_{6}, a_{7}, a_{9}, a_{10}, a_{13}, a_{14}$}

\label{sec.e3.pot.3.1}

In this case, the vector $L_{a}$ given in (\ref{eq.e3.4}) has the general form
\begin{equation}
L_{a}=
\left(
  \begin{array}{c}
    k_{1}x \\
    k_{2}y \\
    k_{3}z \\
  \end{array}
\right) +
\left(
  \begin{array}{c}
    b_{1} -b_{4}y +b_{5}z  \\
    b_{2} +b_{4}x -b_{6}z \\
    b_{3} -b_{5}x +b_{6}y \\
  \end{array}
\right) \label{eq.e3p3.2}
\end{equation}
where $k_{1}, ..., k_{3}, b_{1}, ..., b_{6}$ are arbitrary constants and the generated KT $L_{(a;b)}= diag(k_{1}, k_{2}, k_{3})$.

We assume $k_{1}=k_{2}=k_{3}=k$ is an arbitrary constant. Then, the vector (\ref{eq.e3p3.2}) is the linear combination of the homothetic vector (HV) with the gradient and non-gradient KVs. The KT $L_{(a;b)}= k\delta_{ab}$ and the time-dependent QFI $I_{(3)}$ becomes
\begin{equation}
I= e^{\lambda t} \left( -k \dot{q}^{a}\dot{q}_{a} +\lambda L_{a}\dot{q}^{a} +L_{a}V^{,a} \right). \label{eq.e3p3.3}
\end{equation}
The condition (\ref{eq.e3p3.1}) is
\begin{equation}
\left( L_{b} V^{,b} +2kV \right)_{,a} +\lambda^{2} L_{a} =0. \label{eq.e3p3.4}
\end{equation}
From the integrability condition of (\ref{eq.e3p3.4}), we get:
\[
L_{a,b} -L_{b,a}= 0 \implies L_{a,b}= k\delta_{ab} \implies b_{4}=b_{5}=b_{6}=0.
\]
This implies that only the HV and the gradient KVs survive, that is, the vector (\ref{eq.e3p3.2}) becomes
\begin{equation}
L_{a}=
\left(
  \begin{array}{c}
    kx +b_{1} \\
    ky +b_{2} \\
    kz +b_{3} \\
  \end{array}
\right). \label{eq.e3p3.5}
\end{equation}

We consider the following special cases.

1) \underline{Case $k=0$, $b_{3}=0$ and $b_{1}\neq0$.}

The vector $L_{a}= \left( b_{1}, b_{2}, 0 \right)$. Then, equation (\ref{eq.e3p3.4}) gives the potential
\begin{equation}
V(x,y,z)= \frac{\lambda^{2}}{2} \left( c_{1}^{2} -1 \right)x^{2} +c_{2}x -c_{1}\lambda^{2}xy +F( y -c_{1}x, z) \label{eq.e3p3.6}
\end{equation}
where $c_{1}\equiv \frac{b_{2}}{b_{1}}$, $c_{2}$ are arbitrary constants and $F$ is an arbitrary smooth function of its arguments.

The associated time-dependent LFI is
\begin{equation}
I =e^{\lambda t} \left( \lambda \dot{x} +c_{1} \lambda \dot{y} -\lambda^{2}x -c_{1}\lambda^{2}y +c_{2} \right). \label{eq.e3p3.7}
\end{equation}
We note that $\{H, I\}= \frac{\partial I}{\partial t}= \lambda I$.

- For $c_{1}=0$, the potential (\ref{eq.e3p3.6}) becomes
\begin{equation}
V(x,y,z)= -\frac{\lambda^{2}}{2}x^{2} +c_{2}x +F(y,z) \label{eq.e3p3.8}
\end{equation}
and the associated LFI (\ref{eq.e3p3.7}) is
\begin{equation}
I= e^{\lambda t} \left( \lambda \dot{x} -\lambda^{2}x +c_{2} \right). \label{eq.e3p3.9}
\end{equation}
In the case that $F(y,z)= F_{1}(y) +F_{2}(z)$, the potential (\ref{eq.e3p3.8}) is separable; therefore, it is minimally superintegrable due to the additional independent LFI (\ref{eq.e3p3.9}).

2) \underline{Case $k=0$ and $b_{1}=b_{2}=b_{3}$.}

We have $L_{a}=(1,1,1)$.

The potential (after the transformation $x \leftrightarrow z$)
\begin{equation}
V(x,y,z)= \frac{\lambda^{2}}{2}x^{2} +kx -\lambda^{2}(y+z)x +F( x-z, y-z) \label{eq.e3p3.10}
\end{equation}
where $k$ is an arbitrary constant and $F$ is an arbitrary smooth function of its arguments.

The associated LFI is
\begin{equation}
I= e^{\lambda t} \left[ \lambda (\dot{x} +\dot{y} +\dot{z}) -\lambda^{2}(x+y+z) +k \right]. \label{eq.e3p3.11}
\end{equation}

3) \underline{Case $k\neq0$.}

We find the potential
\begin{equation}
V(x,y,z)= -\frac{\lambda^{2}}{8}r^{2} -\frac{\lambda^{2}}{4} \left( \frac{b_{1}}{k}x +\frac{b_{2}}{k}y +\frac{b_{3}}{k}z \right) +\frac{1}{\left( z +\frac{b_{3}}{k} \right)^{2}} F \left( \frac{y +\frac{b_{2}}{k}}{x +\frac{b_{1}}{k}}, \frac{z +\frac{b_{3}}{k}}{x +\frac{b_{1}}{k}} \right) \label{eq.e3p3.11.1}
\end{equation}
where $F$ is an arbitrary function of its arguments.

The associated QFI is
\begin{eqnarray}
I&=& e^{\lambda t} \left[ \left( \dot{x} -\frac{\lambda}{2}x \right)^{2} + \left( \dot{y} -\frac{\lambda}{2}y \right)^{2} + \left( \dot{z} -\frac{\lambda}{2}z \right)^{2} -\lambda \left( \frac{b_{1}}{k}\dot{x} +\frac{b_{2}}{k}\dot{y} +\frac{b_{3}}{k}\dot{z} \right) + \right. \notag \\
&& \left. +\frac{\lambda^{2}}{2} \left( \frac{b_{1}}{k}x +\frac{b_{2}}{k}y +\frac{b_{3}}{k}z +\frac{b_{1}^{2}}{2k^{2}} +\frac{b_{2}^{2}}{2k^{2}} +\frac{b_{3}^{2}}{2k^{2}} \right) +\frac{2F}{\left(z +\frac{b_{3}}{k}\right)^{2}} \right]. \label{eq.e3p3.11.2}
\end{eqnarray}

3.1. For $b_{1}=b_{2}=b_{3}=0$.

The potential
\begin{equation}
V(x,y,z)= -\frac{\lambda^{2}}{8}r^{2} +\frac{F \left( \frac{y}{x}, \frac{z}{x} \right)}{z^{2}} \label{eq.e3p3.12}
\end{equation}
and the associated QFI is
\begin{equation}
I= e^{\lambda t} \left[ \dot{x}^{2} +\dot{y}^{2} +\dot{z}^{2} -\lambda(x\dot{x} +y\dot{y} +z\dot{z}) +\frac{\lambda^{2}}{4}r^{2} +\frac{2F \left( \frac{y}{x}, \frac{z}{x} \right)}{z^{2}} \right]. \label{eq.e3p3.13}
\end{equation}

3.2. For $b_{1}=k$ and $b_{2}=b_{3}=0$.

The potential
\begin{equation}
V(x,y,z)= -\frac{\lambda^{2}}{8}r^{2} -\frac{\lambda^{2}}{4}x +\frac{F\left( \frac{y}{x+1}, \frac{z}{x+1} \right)}{z^{2}} \label{eq.e3p3.17}
\end{equation}
and the associated QFI is
\begin{equation}
I= e^{\lambda t} \left[ \left( \dot{x} -\frac{\lambda}{2}x \right)^{2} + \left( \dot{y} -\frac{\lambda}{2}y \right)^{2} + \left( \dot{z} -\frac{\lambda}{2}z \right)^{2} -\lambda\dot{x} +\frac{\lambda^{2}}{2}x +\frac{\lambda^{2}}{4} +\frac{2F}{z^{2}} \right]. \label{eq.e3p3.18}
\end{equation}

3.3. For $b_{1}=b_{2}=k$ and $b_{3}=0$.

The potential
\begin{equation}
V(x,y,z)= -\frac{\lambda^{2}}{8}r^{2} -\frac{\lambda^{2}}{4}(x+y) +\frac{1}{z^{2}} F\left( \frac{y+1}{x+1}, \frac{z}{x+1} \right) \label{eq.e3p3.19}
\end{equation}
and the associated QFI is
\begin{equation}
I= e^{\lambda t} \left[ \left( \dot{x} -\frac{\lambda}{2}x \right)^{2} + \left( \dot{y} -\frac{\lambda}{2}y \right)^{2} + \left( \dot{z} -\frac{\lambda}{2}z \right)^{2} -\lambda(\dot{x} +\dot{y}) +\frac{\lambda^{2}}{2}(x+y+1) +\frac{2F}{z^{2}} \right]. \label{eq.e3p3.20}
\end{equation}

3.4. For $b_{1}=b_{2}=b_{3}=k$.

The potential
\begin{equation}
V(x,y,z)= -\frac{\lambda^{2}}{8}r^{2} -\frac{\lambda^{2}}{4} \left( x+y+z \right) +\frac{1}{\left( z +1 \right)^{2}} F \left( \frac{y +1}{x +1}, \frac{z +1}{x +1} \right) \label{eq.e3p3.21}
\end{equation}
and the associated QFI is
\begin{eqnarray}
I&=& e^{\lambda t} \left[ \left( \dot{x} -\frac{\lambda}{2}x \right)^{2} + \left( \dot{y} -\frac{\lambda}{2}y \right)^{2} + \left( \dot{z} -\frac{\lambda}{2}z \right)^{2} -\lambda \left( \dot{x} +\dot{y} +\dot{z} \right) + \right. \notag \\
&& \left. +\frac{\lambda^{2}}{2} \left( x+y+z +\frac{3}{2} \right) +\frac{2F}{\left(z +1\right)^{2}} \right]. \label{eq.e3p3.21}
\end{eqnarray}

3.5. For $F \left( \frac{y +\frac{b_{2}}{k}}{x +\frac{b_{1}}{k}}, \frac{z +\frac{b_{3}}{k}}{x +\frac{b_{1}}{k}} \right) = F_{1} \left( \frac{y +\frac{b_{2}}{k}}{x +\frac{b_{1}}{k}} \frac{x +\frac{b_{1}}{k}}{z +\frac{b_{3}}{k}} \right) +\frac{c_{0}}{\left( x +\frac{b_{1}}{k} \right)^{2}}= F_{1} \left( \frac{y +\frac{b_{2}}{k}}{z +\frac{b_{3}}{k}} \right) +\frac{c_{0}}{\left( x +\frac{b_{1}}{k} \right)^{2}}$, where $c_{0}$ is an arbitrary constant.

The potential
\begin{equation}
V(x,y,z)= -\frac{\lambda^{2}}{8}r^{2} -\frac{\lambda^{2}}{4} \left( c_{1}x +c_{2}y +c_{3}z \right) +\frac{c_{0}}{\left( x +c_{1} \right)^{2}} +\frac{1}{\left( z +c_{3} \right)^{2}} F_{1}\left( \frac{y +c_{2}}{z +c_{3}} \right) \label{eq.e3p3.21.1}
\end{equation}
where $c_{i}= \frac{b_{i}}{k}$.

The associated QFI consists of the independent QFIs:
\begin{eqnarray}
I_{1}&=& e^{\lambda t} \left[ \left( \dot{x} -\frac{\lambda}{2}x \right)^{2} -\lambda c_{1} \left( \dot{x} -\frac{\lambda}{2} x -\frac{\lambda c_{1}}{4} \right) +\frac{2c_{0}}{\left(x +c_{1}\right)^{2}} \right] \label{eq.e3p3.21.2} \\
I_{2}&=& e^{\lambda t} \left[ \left( \dot{y} -\frac{\lambda}{2}y \right)^{2} + \left( \dot{z} -\frac{\lambda}{2}z \right)^{2} -\lambda \left( c_{2}\dot{y} +c_{3}\dot{z} \right) +\frac{\lambda^{2}}{2} \left( c_{2}y +c_{3}z +\frac{c_{2}^{2}}{2} +\frac{c_{3}^{2}}{2} \right) +\frac{2F_{1}}{\left(z +c_{3}\right)^{2}} \right]. \label{eq.e3p3.21.3}
\end{eqnarray}

4) \underline{Case $k_{1}=k_{2}=k_{3}=k$ and $L_{a}=V_{,a}$.}

We find the potential
\begin{equation}
V(x,y,z)= \frac{k}{2}r^{2} +b_{1}x +b_{2}y +b_{3}z. \label{eq.e3p3.14}
\end{equation}
Then, equation (\ref{eq.e3p3.4}) gives $k=-\frac{\lambda^{2}}{4}$ and the potential (\ref{eq.e3p3.14}) becomes
\begin{equation}
V(x,y,z)= -\frac{\lambda^{2}}{8}r^{2} +b_{1}x +b_{2}y +b_{3}z. \label{eq.e3p3.15}
\end{equation}

The associated QFI is
\begin{equation}
I= e^{\lambda t} \left[ \frac{\lambda^{2}}{4} \sum^{3}_{i=1} \left( \dot{q}^{i} -\frac{\lambda}{2}q^{i} \right)^{2} +\lambda (b_{i}\dot{q}^{i}) -\frac{\lambda^{2}}{2} b_{i}q^{i} +\sum^{3}_{i=1} b_{i}^{2} \right]. \label{eq.e3p3.16}
\end{equation}
This QFI consists of the independent QFIs:
\[
I_{1}= e^{\lambda t} \left[ \frac{\lambda^{2}}{4} \left( \dot{x} -\frac{\lambda}{2}x \right)^{2} +\lambda b_{1}\dot{x} -\frac{\lambda^{2}}{2} b_{1}x +b_{1}^{2} \right], \enskip I_{2}= e^{\lambda t} \left[ \frac{\lambda^{2}}{4} \left( \dot{y} -\frac{\lambda}{2}y \right)^{2} +\lambda b_{2}\dot{y} -\frac{\lambda^{2}}{2} b_{2}y +b_{2}^{2} \right],
\]
\[
I_{3}= e^{\lambda t} \left[ \frac{\lambda^{2}}{4} \left( \dot{z} -\frac{\lambda}{2}z \right)^{2} +\lambda b_{3}\dot{z} -\frac{\lambda^{2}}{2} b_{3}z +b_{3}^{2} \right].
\]
Therefore, the potential (\ref{eq.e3p3.15}) is maximally superintegrable (see Table \ref{Table.e23.3}).

5) \underline{Case $k_{1}k_{2}k_{3}\neq0$ and $b_{4}=b_{5}=b_{6}=0$.}

The potential
\begin{equation}
V(x,y,z)= -\frac{\lambda^{2}}{8}r^{2} -\frac{\lambda^{2}}{4} \left( \frac{b_{1}}{k_{1}}x +\frac{b_{2}}{k_{2}}y +\frac{b_{3}}{k_{3}}z \right) +\frac{c_{1}}{\left( x +\frac{b_{1}}{k_{1}} \right)^{2}} +\frac{c_{2}}{\left( y +\frac{b_{2}}{k_{2}} \right)^{2}} +\frac{c_{3}}{\left( z +\frac{b_{3}}{k_{3}} \right)^{2}} \label{eq.e3p3.22}
\end{equation}
where $c_{1}, c_{2}$, and $c_{3}$ are arbitrary constants.

The associated QFI gives the following three independent QFIs
\begin{equation}
I_{i}= e^{\lambda t} \left[ \left( \dot{q}^{i} -\frac{\lambda}{2}q^{i} \right)^{2} -\lambda \frac{b_{i}}{k_{i}}\left( \dot{q}^{i} -\frac{\lambda}{2}q^{i} -\frac{\lambda b_{i}}{4k_{i}} \right) +\frac{2c_{i}}{\left( q^{i} +\frac{b_{i}}{k_{i}} \right)^{2}} \right] \label{eq.e3p3.23}
\end{equation}
where $i=1,2,3$ and $q^{i}=(x,y,z)$.

Therefore, the separable potential (\ref{eq.e3p3.22}) is maximally superintegrable (see Table \ref{Table.e23.3}).

We note that, as expected, for $k_{1}=k_{2}=k_{3}=k$ the resulting potential (\ref{eq.e3p3.22}) belongs to the family of potentials (\ref{eq.e3p3.11.1}) if we set
\[
F \left( \frac{y +\frac{b_{2}}{k}}{x +\frac{b_{1}}{k}}, \frac{z +\frac{b_{3}}{k}}{x +\frac{b_{1}}{k}} \right) = c_{1} \left( \frac{z +\frac{b_{3}}{k}}{x +\frac{b_{1}}{k}} \right)^{2} +c_{2} \left( \frac{y +\frac{b_{2}}{k}}{x +\frac{b_{1}}{k}} \right)^{-2} \left( \frac{z +\frac{b_{3}}{k}}{x +\frac{b_{1}}{k}} \right)^{2} +c_{3}.
\]

6) \underline{Case $k_{1}b_{2}\neq0$, $k_{2}=k_{3}=0$ and $b_{4}=b_{5}=b_{6}=0$.}

The vector $L_{a}= \left( k_{1}x +b_{1}, b_{2}, b_{3} \right)$.

The potential
\begin{equation}
V(x,y,z)= -\frac{\lambda^{2}}{8} \left[ x^{2} +4(1 -c_{1}^{2}) y^{2} \right] -\frac{\lambda^{2}}{4} \left( c_{2}x +4c_{1}yz \right) +c_{3}y +\frac{c_{4}}{(x +c_{2})^{2}} + F(z -c_{1}y) \label{eq.e3p3.24}
\end{equation}
where $c_{1}= \frac{b_{3}}{b_{2}}$, $c_{2}= \frac{b_{1}}{k_{1}}$, $c_{3}$, $c_{4}$ are arbitrary constants and $F$ is an arbitrary smooth function of its arguments.

We find the independent FIs:
\begin{eqnarray}
I_{1}&=& e^{\lambda t} \left[ \left( \dot{x} -\frac{\lambda}{2}x \right)^{2} -\lambda c_{2}\left( \dot{x} -\frac{\lambda}{2}x -\frac{\lambda}{4}c_{2} \right) +\frac{2c_{4}}{\left( x +c_{2} \right)^{2}} \right] \label{eq.e3p3.25.1} \\
I_{2}&=& e^{\lambda t} \left[ \dot{y} +c_{1}\dot{z} -\lambda (y+c_{1}z) +\frac{c_{3}}{\lambda} \right]. \label{eq.e3p3.25.2}
\end{eqnarray}

We note that for $c_{1}=0$ we obtain the separable potential
\begin{equation}
V(x,y,z)= -\frac{\lambda^{2}}{8} \left( x^{2} +4y^{2} \right) -\frac{\lambda^{2}}{4} c_{2}x +c_{3}y +\frac{c_{4}}{(x +c_{2})^{2}} + F(z) \label{eq.e3p3.26}
\end{equation}
which is a new maximally superintegrable potential due to the additional time-dependent FIs (\ref{eq.e3p3.25.1}) and (\ref{eq.e3p3.25.2}). The potential (see Table \ref{Table.e3.3})
\begin{equation}
V(x,y,z)= -\frac{\lambda^{2}}{8} \left( R^{2} +4z^{2} \right) +\frac{c_{4}}{x^{2}} +\frac{c_{0}}{y^{2}} +c_{3}z. \label{eq.e3p3.27}
\end{equation}
is a subcase of (\ref{eq.e3p3.26}) for $y \leftrightarrow z$, $c_{2}=0$ and $F(z)= -\frac{\lambda^{2}}{8}z^{2} +\frac{c_{0}}{z^{2}}$.

\subsection{Parameters $a_{17}, a_{19}, a_{20}$: The components $L_{(a;b)}$ are constant and non-diagonal}

\label{sec.e3.pot.3.2}

In the following cases, the only non-vanishing parameters are the $a_{17}, a_{19}$, and $a_{20}$.

1) \underline{Case $a_{17}\neq0$, $a_{20}=0$ and $a_{19}$ is free.}

The vector $L_{a}= \left( 0, 2a_{17}x, 2a_{19}x \right)$ and the KT $L_{(a;b)}=
\left(
  \begin{array}{ccc}
    0 & a_{17} & a_{19} \\
    a_{17} & 0 & 0 \\
    a_{19} & 0 & 0 \\
  \end{array}
\right)$.

Then, equation (\ref{eq.e3p3.1}) gives the potential
\begin{equation}
V(x,y,z)= -\frac{\lambda^{2}}{2}x^{2} +F\left( z -cy \right) \label{eq.e3p3.28}
\end{equation}
where $c= \frac{a_{19}}{a_{17}}$ and $F$ is an arbitrary smooth function.

The associated QFI is
\begin{equation}
I= e^{\lambda t} (\dot{x} -\lambda x)(\dot{y} +c\dot{z}). \label{eq.e3p3.29}
\end{equation}

From Table \ref{Table.e23.2}, the potential (\ref{eq.e3p3.28}) admits the additional autonomous FIs:  $I_{1}= \frac{1}{2}\dot{x}^{2} -\frac{\lambda^{2}}{2}x^{2}$ and $I_{2}= \dot{y} +c\dot{z}$. Therefore, the QFI (\ref{eq.e3p3.29}) contains the independent LFI $I_{3}= e^{\lambda t} (\dot{x} -\lambda x)$.

We conclude that (\ref{eq.e3p3.28}) is a new minimally superintegrable potential.

2) \underline{Case $a_{17}=\frac{\alpha}{2}\neq0$, $a_{19}=0$ and $a_{20}=\frac{\beta}{2}$.}

The vector $L_{a}= (0, \alpha x, \beta y)$, where $\alpha$ and $\beta$ are arbitrary constants and the KT $L_{(a;b)}=
\left(
  \begin{array}{ccc}
    0 & \frac{\alpha}{2} & 0 \\
    \frac{\alpha}{2} & 0 & \frac{\beta}{2} \\
    0 & \frac{\beta}{2} & 0 \\
  \end{array}
\right)$.

The potential
\begin{equation}
V(x,y,z)= -\frac{\lambda^{2}}{2(1+c_{1}^{2})} \left( x^{2} +c_{1}^{2}y^{2} \right) -\frac{\lambda^{2}}{2(1+c_{1}^{2})} \left( z -2c_{1}x \right)^{2} +c_{2} \left( z -2c_{1}x \right) \label{eq.e3p3.30}
\end{equation}
where $c_{1}= \frac{\beta}{\alpha}$ and $c_{2}$ are arbitrary constants.

The associated QFI is
\begin{equation}
I= e^{\lambda t} \left[ (\dot{x} -\lambda x)\dot{y} +c_{1} (\dot{y} -\lambda y) \dot{z} -\frac{\lambda^{2} c_{1}}{1 +c_{1}^{2}} (c_{1}x -z) y -c_{1}c_{2}y \right]. \label{eq.e3p3.31}
\end{equation}

Moreover, the potential (\ref{eq.e3p3.30}) admits the additional autonomous QFI
$I_{1}= \frac{1}{2}\dot{y}^{2} -\frac{\lambda^{2}c_{1}^{2}}{2(1+c_{1}^{2})} y^{2}$ because the $y$-coordinate is separated from the coordinates $x$ and $z$.

\subsection{Parameters $a_{2}, a_{5}, a_{8}, a_{11}, a_{12}, a_{15}, a_{16}, a_{18}$: The components $L_{(a;b)}$ are linear on $x,y,z$}

\label{sec.e3.pot.3.3}

We consider the following cases:

1) \underline{$a_{15}$ is the only non-vanishing parameter.}

The vector $L_{a}= a_{15} (-y^{2}, xy, 0)$ and the KT $L_{(a;b)}= a_{15}
\left(
  \begin{array}{ccc}
    0 & -\frac{y}{2} & 0 \\
    -\frac{y}{2} & x & 0 \\
    0 & 0 & 0 \\
  \end{array}
\right)$.

The potential
\begin{equation}
V(x,y,z)= -\frac{\lambda^{2}}{2}R^{2} +\frac{c_{1}x}{y^{2}R} +\frac{c_{2}}{y^{2}} +F(z) \label{eq.e3p3.32}
\end{equation}
where $c_{1}, c_{2}$ are arbitrary constants and $F(z)$ is an arbitrary smooth function.

The associated QFI is
\begin{equation}
I= e^{\lambda t} \left[ M_{3}(\dot{y} -\lambda y) +\frac{2c_{2}x}{y^{2}} +\frac{c_{1} (y^{2} +2x^{2})}{y^{2}R} \right]. \label{eq.e3p3.33}
\end{equation}

We note that the potential (\ref{eq.e3p3.32}) is of the integrable form (see Table \ref{Table.e23.1}) $V= \frac{F_{1}\left( \frac{y}{x}\right) }{R^{2}} +F_{2}(R) +F_{3}(z)$ with
\begin{equation}
F_{1}\left( \frac{y}{x}\right)= \left( \frac{c_{1}}{\sqrt{1 +\frac{y^{2}}{x^{2}}}} +c_{2} \right) \left( 1 + \frac{x^{2}}{y^{2}} \right), \enskip F_{2}(R)= -\frac{\lambda^{2}}{2}R^{2}. \label{eq.e3p3.33.1}
\end{equation}
Therefore, it is a new minimally superintegrable potential due to the additional autonomous QFIs:
\[
I_{1}= \frac{1}{2}\dot{z}^{2} +F_{3}(z), \enskip I_{2}= \frac{1}{2}M_{3}^{2} +\frac{(c_{1}R +c_{2}x)x}{y^{2}}.
\]

Moreover, for $F(z)= -\frac{\lambda^{2}}{2}z^{2} +\frac{c_{3}}{z^{2}}$, where $c_{3}$ is an arbitrary constant, the resulting potential
\begin{equation}
V(x,y,z)= -\frac{\lambda^{2}}{2}r^{2} +\frac{c_{1}x}{y^{2}R} +\frac{c_{2}}{y^{2}} +\frac{c_{3}}{z^{2}} \label{eq.e3p3.34}
\end{equation}
is a subcase of the minimally superintegrable potential (\ref{eq.e3pot.24a}) with $F_{1}\left( \frac{y}{x}\right)$ as given in (\ref{eq.e3p3.33.1}). Hence, (\ref{eq.e3p3.34}) is a new maximally superintegrable potential due to the additional autonomous QFI (see Table \ref{Table.e3.2})
\begin{equation}
I_{3}= \frac{1}{2} \mathbf{M}^{2} +\frac{c_{1}xr^{2}}{y^{2}R} +c_{2}\frac{r^{2}}{y^{2}} +\frac{c_{3}R^{2}}{z^{2}}. \label{eq.e3p3.35}
\end{equation}

2) \underline{$a_{2}$ and $a_{12}$ are the only non-vanishing parameters.}

The vector $L_{a}= (a_{2}xz, a_{12}yz, -a_{2}x^{2} -a_{12}y^{2})$ and the KT $L_{(a;b)}=
\left(
  \begin{array}{ccc}
    a_{2}z & 0 & -\frac{a_{2}}{2}x \\
    0 & a_{12}z & -\frac{a_{12}}{2}y \\
    -\frac{a_{2}}{2}x & -\frac{a_{12}}{2}y & 0 \\
  \end{array}
\right)$.

The potential (see Table \ref{Table.e23.3})
\begin{equation}
V(x,y,z)= -\frac{\lambda^{2}}{2}r^{2} +\frac{k_{1}}{x^{2}} +\frac{k_{2}}{y^{2}} \label{eq.e3p3.36}
\end{equation}
where $k_{1}$ and $k_{2}$ are arbitrary constants.

The associated QFI consists of the independent QFIs:
\[
I_{1}= e^{\lambda t} \left[ M_{2}(\dot{x} -\lambda x) +\frac{2k_{1}z}{x^{2}} \right], \enskip I_{2}= e^{\lambda t} \left[ M_{1}(\dot{y} -\lambda y) -\frac{2k_{2}z}{y^{2}} \right].
\]
Therefore, the separable potential (\ref{eq.e3p3.36}) is maximally superintegrable.

3) \underline{Case $a_{2}=a_{12}$.}

The vector $L_{a}= a_{2}(xz, yz, -R^{2})$ and the KT $L_{(a;b)}= a_{2}
\left(
  \begin{array}{ccc}
    z & 0 & -\frac{x}{2} \\
    0 & z & -\frac{y}{2} \\
    -\frac{x}{2} & -\frac{y}{2} & 0 \\
  \end{array}
\right)$.

The potential
\begin{equation}
V(x,y,z)= -\frac{\lambda^{2}}{2}r^{2} +\frac{c_{1}z}{rR^{2}} +\frac{F\left( \frac{y}{x} \right)}{R^{2}} \label{eq.e3p3.37}
\end{equation}
where $c_{1}$ is an arbitrary constant and $F\left( \frac{y}{x} \right)$ is an arbitrary smooth function.

The associated QFI is
\begin{equation}
I_{1}= e^{\lambda t} \left[ M_{2} \left( \dot{x} -\lambda x\right) -M_{1} \left( \dot{y} -\lambda y \right) +\frac{c_{1}}{r} +\frac{2c_{1}z^{2}}{rR^{2}} +\frac{2zF\left(\frac{y}{x} \right)}{R^{2}} \right]. \label{eq.e3p3.38}
\end{equation}

We note that the potential (\ref{eq.e3p3.37}) belongs to the general family of potentials (\ref{eq.e3pot.21c}); hence, it admits the additional autonomous QFI (see Table \ref{Table.e3.1})
\begin{equation}
I_{2}= \frac{1}{2}M_{3}^{2} + F\left( \frac{y}{x} \right). \label{eq.e3p3.39}
\end{equation}

If $c_{1}=0$, the resulting potential
\begin{equation}
V(x,y,z)= -\frac{\lambda^{2}}{2}r^{2} +\frac{F\left( \frac{y}{x} \right)}{R^{2}} \label{eq.e3p3.40}
\end{equation}
is a new maximally superintegrable potential due to the additional autonomous QFIs (see Table \ref{Table.e3.2}):
\[
I_{3}= \frac{1}{2}\dot{z}^{2} -\frac{\lambda^{2}}{2}z^{2}, \enskip I_{4}= \frac{1}{2} \mathbf{M}^{2} +\frac{r^{2} F\left( \frac{y}{x} \right)}{R^{2}}.
\]
We note that the potential (\ref{eq.e3p3.40}) is of the form (\ref{eq.e3pot.24a}) for $k_{1}= -\frac{\lambda^{2}}{2}$ and $k_{2}=0$.

4) \underline{$a_{3}, a_{6}, a_{10}, a_{14}$ are non-vanishing and $a_{2}a_{13}\neq0$.}

The vector $L_{a}=
\left(
  \begin{array}{c}
    a_{2}xz +a_{3}x +a_{6} \\
    a_{13}y +a_{14} \\
    -a_{2}x^{2} +a_{10} \\
  \end{array}
\right)$ and the KT
$L_{(a;b)}=
\left(
  \begin{array}{ccc}
    a_{2}z +a_{3} & 0 & -\frac{a_{2}}{2}x  \\
    0 & a_{13} & 0 \\
    -\frac{a_{2}}{2}x & 0 & 0 \\
  \end{array}
\right)$.

The potential (see Table \ref{Table.e23.3})
\begin{equation}
V(x,y,z)= -\frac{\lambda^{2}}{8}(4x^{2} +4z^{2} +y^{2}) -\lambda^{2}c_{1}z -\frac{\lambda^{2}}{4}c_{2}y +\frac{k}{(y+c_{2})^{2}} \label{eq.e3p3.41}
\end{equation}
where $k, c_{1}= \frac{a_{3}}{a_{2}}$, and $c_{2}= \frac{a_{14}}{a_{13}}$ are arbitrary constants.

The associated QFI consists of the independent FIs:
\begin{eqnarray}
I_{1}&=& e^{\lambda t} \left( \dot{x} -\lambda x\right) \label{eq.e3p3.42.1} \\
I_{2}&=& e^{\lambda t} \left\{ \left[ \dot{y} -\frac{\lambda}{2}(y+c_{2}) \right]^{2} +\frac{2k}{(y+c_{2})^{2}} \right\} \label{eq.e3p3.42.2} \\
I_{3}&=& e^{\lambda t} \left[ \dot{z} -\lambda (z+c_{1}) \right] \label{eq.e3p3.42.3} \\
I_{4}&=& M_{2} +c_{1}\dot{x}. \label{eq.e3p3.42.4}
\end{eqnarray}
We note that $\{I_{2}, I_{p}\}=0$ where $p=1,2,3,4$, $\{I_{1},I_{3}\}=0$, $\{I_{1}, I_{4}\}= I_{3}$ and $\{I_{4}, I_{3}\}= I_{1}$.

The potential (\ref{eq.e3p3.41}) is integrable because the independent FIs $I_{1}, I_{2}, I_{3}$ are in involution or, directly, because it is separable. It is also maximally superintegrable due to the additional independent FIs $I_{4}$ and $H$, where $H$ is the Hamiltonian.

5) \underline{Case $a_{2}\neq0$ and $a_{3}, a_{13}$ are non-vanishing.}

The vector $L_{a}=
\left(
  \begin{array}{c}
    a_{2}xz +a_{3}x \\
    a_{13}y \\
    -a_{2}x^{2} \\
  \end{array}
\right)$ and the KT
$L_{(a;b)}=
\left(
  \begin{array}{ccc}
    a_{2}z +a_{3} & 0 & -\frac{a_{2}}{2}x \\
    0 & a_{13} & 0 \\
    -\frac{a_{2}}{2}x & 0 & 0 \\
  \end{array}
\right)$.

The potential
\begin{equation}
V(x,y,z)= -\frac{\lambda^{2}}{2} \left( x^{2} +z^{2} \right) -\frac{\lambda^{2}}{8}y^{2} +\frac{c_{1}}{x^{2}} +\frac{c_{2}}{y^{2}} -\lambda^{2}c_{3}z +\frac{k (z +c_{3})}{x^{2} \sqrt{(z+c_{3})^{2} +x^{2}}} \label{eq.e3p3.43}
\end{equation}
where $k, c_{1}, c_{2}$, and $c_{3}= \frac{a_{3}}{a_{2}}$ are arbitrary constants.

The associated QFI consists of the following independent QFIs:
\begin{eqnarray}
I_{1}&=& e^{\lambda t} \left[ \left( \dot{y} -\frac{\lambda}{2}y \right)^{2} +\frac{2c_{2}}{y^{2}} \right] \label{eq.e3p3.44.1} \\
I_{2}&=& e^{\lambda t} \left[ (M_{2} +c_{3}\dot{x}) (\dot{x} -\lambda x) +\frac{2c_{1}(z+c_{3})}{x^{2}} +k \frac{x^{2} +2 (z+c_{3})^{2}}{x^{2} \sqrt{x^{2} +(z+c_{3})^{2}}} \right]. \label{eq.e3p3.44.2}
\end{eqnarray}

It is well-known that the dynamical equations (and hence the associated FIs) of a regular Lagrangian system are preserved if: \newline
a. We add an arbitrary constant $c$ to the potential $V$ of the system. \newline
b. We apply a canonical transformation.

Then, the potential (\ref{eq.e3p3.43}) is a subcase of the minimally superintegrable potential (\ref{eq.e3p3.32}). Indeed, by adding the constant $c= -\frac{\lambda^{2}}{2}c_{3}^{2}$ to (\ref{eq.e3p3.43}), we obtain the equivalent potential
\begin{equation}
V(x,y,z)= -\frac{\lambda^{2}}{2} \left[ x^{2} +(z+c_{3})^{2} \right] +\frac{k (z +c_{3})}{x^{2} \sqrt{(z+c_{3})^{2} +x^{2}}} +\frac{c_{1}}{x^{2}} -\frac{\lambda^{2}}{8}y^{2} +\frac{c_{2}}{y^{2}}. \label{eq.e3p3.45}
\end{equation}
If we apply the canonical transformation $x \rightarrow y$, $y \rightarrow z$ and $z \rightarrow x -c_{3}$, the potential (\ref{eq.e3p3.45}) becomes
\begin{equation}
V(x,y,z)= -\frac{\lambda^{2}}{2} R^{2} +\frac{kx}{y^{2}R} +\frac{c_{1}}{y^{2}} -\frac{\lambda^{2}}{8}z^{2} +\frac{c_{2}}{z^{2}} \label{eq.e3p3.46}
\end{equation}
which is a subcase of (\ref{eq.e3p3.32}) for $F(z)= -\frac{\lambda^{2}}{8}z^{2} +\frac{c_{2}}{z^{2}}$.

The potential (\ref{eq.e3p3.46}) is a new maximally superintegrable potential due to the following independent QFIs:
\begin{eqnarray}
I_{1}&=& e^{\lambda t} \left[ \left( \dot{z} -\frac{\lambda}{2}z \right)^{2} +\frac{2c_{2}}{z^{2}} \right] \label{eq.e3p3.47.1} \\
I_{2}&=& e^{\lambda t} \left[ M_{3} (\dot{y} -\lambda y) +\frac{2c_{1}x}{y^{2}} +\frac{k(y^{2} +2x^{2})}{y^{2}R} \right]. \label{eq.e3p3.47.2} \\
I_{3}&=& \frac{1}{2}\dot{z}^{2} -\frac{\lambda^{2}}{8}z^{2} +\frac{c_{2}}{z^{2}} \label{eq.e3p3.47.3} \\
I_{4}&=& \frac{1}{2}M_{3}^{2} + \frac{(kR +c_{1}x)x}{y^{2}}. \label{eq.e3p3.47.4}
\end{eqnarray}
We recall that the potential (\ref{eq.e3p3.34}) is another maximally superintegrable potential which is also a subcase of (\ref{eq.e3p3.32}) but for a different choice of the function $F(z)$. If we rename $\lambda \to 2\lambda$, the QFI (\ref{eq.e3p3.47.1}) is admitted also by (\ref{eq.e3p3.34}) because the $z$-coordinate is separated from $x$ and $y$.
\bigskip

We collect the results of section \ref{sec.e3.pot.3} in Tables \ref{Table.e3.7} - \ref{Table.e3.9}.

\newpage

\begin{longtable}{|l|l|}
\hline
{\large Potential} & {\large LFIs and
QFIs} \\ \hline
\makecell[l]{$V= \frac{\lambda^{2}}{2} \left( c_{1}^{2} -1 \right)x^{2} +c_{2}x -$ \\ \qquad $-c_{1}\lambda^{2}xy +F( y -c_{1}x, z)$} &
$I =e^{\lambda t} \left( \lambda \dot{x} +c_{1} \lambda \dot{y} -\lambda^{2}x -c_{1}\lambda^{2}y +c_{2} \right)$ \\ \hline
\makecell[l]{$V= \frac{\lambda^{2}}{2}x^{2} +kx -\lambda^{2}(y+z)x+$ \\ \qquad $+F( x-z, y-z)$} & $I= e^{\lambda t} \left[ \lambda (\dot{x} +\dot{y} +\dot{z}) -\lambda^{2}(x+y+z) +k \right]$ \\ \hline
\makecell[l]{$V= -\frac{\lambda^{2}}{8}r^{2} -\frac{\lambda^{2}}{4} \left( c_{1}x +c_{2}y +c_{3}z \right)+$ \\ \qquad $+\frac{1}{\left( z +c_{3} \right)^{2}} F \left( \frac{y +c_{2}}{x +c_{1}}, \frac{z +c_{3}}{x +c_{1}} \right)$} & \makecell[l]{$I= e^{\lambda t} \left[ \left( \dot{x} -\frac{\lambda}{2}x \right)^{2} + \left( \dot{y} -\frac{\lambda}{2}y \right)^{2} + \left( \dot{z} -\frac{\lambda}{2}z \right)^{2} - \right.$ \\ \qquad $\left. -\lambda \left( c_{1}\dot{x} +c_{2}\dot{y} +c_{3}\dot{z} \right) + \right.$ \\ \qquad $\left. +\frac{\lambda^{2}}{2} \left( c_{1}x +c_{2}y +c_{3}z +\frac{c_{1}^{2}}{2} +\frac{c_{2}^{2}}{2} +\frac{c_{3}^{2}}{2} \right) +\frac{2F}{\left(z +c_{3}\right)^{2}} \right]$} \\ \hline
$V= -\frac{\lambda^{2}}{8}r^{2} +\frac{F \left( \frac{y}{x}, \frac{z}{x} \right)}{z^{2}}$ & $I= e^{\lambda t} \left[ \dot{x}^{2} +\dot{y}^{2} +\dot{z}^{2} -\lambda(x\dot{x} +y\dot{y} +z\dot{z}) +\frac{\lambda^{2}}{4}r^{2} +\frac{2F \left( \frac{y}{x}, \frac{z}{x} \right)}{z^{2}} \right]$ \\ \hline
$V= -\frac{\lambda^{2}}{8}r^{2} -\frac{\lambda^{2}}{4}x +\frac{F\left( \frac{y}{x+1}, \frac{z}{x+1} \right)}{z^{2}}$ & \makecell[l]{$I= e^{\lambda t} \left[ \left( \dot{x} -\frac{\lambda}{2}x \right)^{2} + \left( \dot{y} -\frac{\lambda}{2}y \right)^{2} + \left( \dot{z} -\frac{\lambda}{2}z \right)^{2} -\lambda\dot{x} + \right.$ \\ \qquad $\left. +\frac{\lambda^{2}}{2}x +\frac{\lambda^{2}}{4} +\frac{2F}{z^{2}} \right]$} \\ \hline
\makecell[l]{$V= -\frac{\lambda^{2}}{8}r^{2} -\frac{\lambda^{2}}{4}(x+y)+$ \\ \qquad $+\frac{1}{z^{2}} F\left( \frac{y+1}{x+1}, \frac{z}{x+1} \right)$} & \makecell[l]{$I= e^{\lambda t} \left[ \left( \dot{x} -\frac{\lambda}{2}x \right)^{2} + \left( \dot{y} -\frac{\lambda}{2}y \right)^{2} + \left( \dot{z} -\frac{\lambda}{2}z \right)^{2} - \right.$ \\ \qquad $\left. -\lambda(\dot{x} +\dot{y}) +\frac{\lambda^{2}}{2}(x+y+1) +\frac{2F}{z^{2}} \right]$} \\ \hline
\makecell[l]{$V= -\frac{\lambda^{2}}{8}r^{2} -\frac{\lambda^{2}}{4} \left( x+y+z \right) +$ \\ \qquad $+\frac{1}{\left( z +1 \right)^{2}} F \left( \frac{y +1}{x +1}, \frac{z +1}{x +1} \right)$} & \makecell[l]{$I= e^{\lambda t} \left[ \left( \dot{x} -\frac{\lambda}{2}x \right)^{2} + \left( \dot{y} -\frac{\lambda}{2}y \right)^{2} + \left( \dot{z} -\frac{\lambda}{2}z \right)^{2} -\lambda \left( \dot{x} +\dot{y} +\dot{z} \right) + \right.$ \\ \qquad $\left. +\frac{\lambda^{2}}{2} \left( x+y+z +\frac{3}{2} \right) +\frac{2F}{\left(z +1\right)^{2}} \right]$} \\ \hline
\makecell[l]{$V= -\frac{\lambda^{2}}{8}r^{2} -\frac{\lambda^{2}}{4} \left( c_{1}x +c_{2}y +c_{3}z \right) +$ \\ \qquad  $+\frac{c_{0}}{\left( x +c_{1} \right)^{2}} +\frac{1}{\left( z +c_{3} \right)^{2}} F_{1}\left( \frac{y +c_{2}}{z +c_{3}} \right)$} & \makecell[l]{$I_{1}= e^{\lambda t} \left[ \left( \dot{x} -\frac{\lambda}{2}x \right)^{2} -\lambda c_{1} \left( \dot{x} -\frac{\lambda}{2} x -\frac{\lambda c_{1}}{4} \right) +\frac{2c_{0}}{\left(x +c_{1}\right)^{2}} \right]$ \\ $I_{2}= e^{\lambda t} \left[ \left( \dot{y} -\frac{\lambda}{2}y \right)^{2} + \left( \dot{z} -\frac{\lambda}{2}z \right)^{2} -\lambda \left( c_{2}\dot{y} +c_{3}\dot{z} \right) + \right.$ \\ \qquad $\left. +\frac{\lambda^{2}}{2} \left( c_{2}y +c_{3}z +\frac{c_{2}^{2}}{2} +\frac{c_{3}^{2}}{2} \right) +\frac{2F_{1}}{\left(z +c_{3}\right)^{2}} \right]$} \\ \hline
\makecell[l]{$V= -\frac{\lambda^{2}}{8} \left[ x^{2} +4(1 -c_{1}^{2}) y^{2} \right] -$ \\ \qquad $-\frac{\lambda^{2}}{4} \left( c_{2}x +4c_{1}yz \right) +$ \\ \qquad $+c_{3}y +\frac{c_{4}}{(x +c_{2})^{2}} + F(z -c_{1}y)$} & \makecell[l]{$I_{1}= e^{\lambda t} \left[ \left( \dot{x} -\frac{\lambda}{2}x \right)^{2} -\lambda c_{2}\left( \dot{x} -\frac{\lambda}{2}x -\frac{\lambda}{4}c_{2} \right) +\frac{2c_{4}}{\left( x +c_{2} \right)^{2}} \right]$ \\
$I_{2}= e^{\lambda t} \left[ \dot{y} +c_{1}\dot{z} -\lambda (y+c_{1}z) +\frac{c_{3}}{\lambda} \right]$} \\ \hline
\makecell[l]{$V= -\frac{\lambda^{2}}{2(1+c_{1}^{2})} \left( x^{2} +c_{1}^{2}y^{2} \right)-$ \\ \qquad $-\frac{\lambda^{2}}{2(1+c_{1}^{2})} \left( z -2c_{1}x \right)^{2} +$ \\ \qquad $+c_{2} \left( z -2c_{1}x \right)$} & \makecell[l]{$I_{1}= \frac{1}{2}\dot{y}^{2} -\frac{\lambda^{2}c_{1}^{2}}{2(1+c_{1}^{2})} y^{2}$ \\
$I_{2}= e^{\lambda t} \left[ (\dot{x} -\lambda x)\dot{y} +c_{1} (\dot{y} -\lambda y) \dot{z} -\frac{\lambda^{2} c_{1}}{1 +c_{1}^{2}} (c_{1}x -z) y -c_{1}c_{2}y \right]$} \\ \hline
$V= -\frac{\lambda^{2}}{2}r^{2} +\frac{c_{1}z}{rR^{2}} +\frac{F\left( \frac{y}{x} \right)}{R^{2}}$ & \makecell[l]{$I_{1}= e^{\lambda t} \left[ M_{2} \left( \dot{x} -\lambda x\right) -M_{1} \left( \dot{y} -\lambda y \right) +\frac{c_{1}}{r} +\frac{2c_{1}z^{2}}{rR^{2}} +\frac{2zF\left(\frac{y}{x} \right)}{R^{2}} \right]$ \\ $I_{2}= \frac{1}{2}M_{3}^{2} + F\left( \frac{y}{x} \right)$} \\ \hline
\caption{\label{Table.e3.7} Possibly non-integrable potentials $V(x,y,z)$ in $E^{3}$ that admit time-dependent LFIs/QFIs of the form $I_{(3)}$.}
\end{longtable}

\newpage

\begin{longtable}{|l|c|l|}
\hline
\multicolumn{3}{|c|}{{\large{Minimally superintegrable potentials}}} \\ \hline
{\large Potential} & {\large Ref \cite{Evans 1990}} & {\large LFIs and
QFIs} \\ \hline
$V= -\frac{\lambda^{2}}{2}x^{2} +c_{2}x +F_{1}(y) +F_{2}(z)$ & New & \makecell[l]{$I_{1}= \frac{1}{2}\dot{x}^{2} -\frac{\lambda^{2}}{2}x^{2} +c_{2}x$ \\ $I_{2}= \frac{1}{2}\dot{y}^{2} +F_{1}(y)$ \\ $I_{3}= \frac{1}{2}\dot{z}^{2} +F_{2}(z)$ \\ $I_{4}= e^{\lambda t} \left( \lambda \dot{x} -\lambda^{2}x +c_{2} \right)$} \\ \hline
$V= -\frac{\lambda^{2}}{2}x^{2} +F\left( z -cy \right)$ & New & \makecell[l]{$I_{1}= \frac{1}{2}\dot{x}^{2} -\frac{\lambda^{2}}{2}x^{2}$ \\ $I_{2}= \dot{y} +c\dot{z}$ \\ $I_{3}= e^{\lambda t} (\dot{x} -\lambda x)$} \\ \hline
$V= -\frac{\lambda^{2}}{2}R^{2} +\frac{c_{1}x}{y^{2}R} +\frac{c_{2}}{y^{2}} +F(z)$ & New & \makecell[l]{$I_{1}= \frac{1}{2}\dot{z}^{2} +F(z)$ \\ $I_{2}= \frac{1}{2}M_{3}^{2} +\frac{(c_{1}R +c_{2}x)x}{y^{2}}$ \\ $I_{3}= e^{\lambda t} \left[ M_{3}(\dot{y} -\lambda y) +\frac{2c_{2}x}{y^{2}} +\frac{c_{1} (y^{2} +2x^{2})}{y^{2}R} \right]$} \\ \hline
\caption{\label{Table.e3.8} Minimally superintegrable potentials $V(x,y,z)$ in $E^{3}$ that admit time-dependent LFIs/QFIs of the form $I_{(3)}$.}
\end{longtable}

\newpage

\begin{longtable}{|l|c|l|}
\hline
\multicolumn{3}{|c|}{{\large{Maximally superintegrable potentials}}} \\ \hline
{\large Potential} & {\large Ref \cite{Evans 1990}} & {\large LFIs and
QFIs} \\ \hline
$V= -\frac{\lambda^{2}}{8}r^{2} +b_{1}x +b_{2}y +b_{3}z$ & New &
\makecell[l]{$I_{i}= \frac{\dot{q}_{i}^{2}}{2} -\frac{\lambda^{2}}{8}q_{i}^{2} +b_{i}q_{i}$ \\ $J_{i}= e^{\lambda t} \left[ \frac{\lambda^{2}}{4} \left( \dot{q}_{i} -\frac{\lambda}{2}q_{i} \right)^{2} +\lambda b_{i}\dot{q}_{i} -\frac{\lambda^{2}}{2} b_{i}q_{i} +b_{i}^{2} \right]$} \\ \hline
\makecell[l]{$V= -\frac{\lambda^{2}}{8}r^{2} -\frac{\lambda^{2}}{4} \left( b_{1}x +b_{2}y +b_{3}z \right) +$ \\ \qquad $+\frac{c_{1}}{\left( x +b_{1} \right)^{2}} +\frac{c_{2}}{\left( y +b_{2} \right)^{2}} +\frac{c_{3}}{\left( z +b_{3} \right)^{2}}$} & New & \makecell[l]{$I_{i}= \frac{\dot{q}_{i}^{2}}{2} -\frac{\lambda^{2}}{8}q_{i}^{2} -\frac{\lambda^{2}}{4} b_{i}q_{i} +\frac{c_{i}}{\left( x +b_{i} \right)^{2}}$ \\ $J_{i}= e^{\lambda t} \left[ \left( \dot{q}^{i} -\frac{\lambda}{2}q^{i} \right)^{2} -\lambda b_{i}\left( \dot{q}^{i} -\frac{\lambda}{2}q^{i} -\frac{\lambda b_{i}}{4} \right) +\frac{2c_{i}}{\left( q^{i} +b_{i} \right)^{2}} \right]$} \\ \hline
\makecell[l]{$V= -\frac{\lambda^{2}}{8} \left( x^{2} +4y^{2} \right) -\frac{\lambda^{2}}{4} c_{2}x +c_{3}y +$ \\ \qquad $+\frac{c_{4}}{(x +c_{2})^{2}} + F(z)$} & New & \makecell[l]{$I_{1}= \frac{\dot{x}^{2}}{2} -\frac{\lambda^{2}}{8}x^{2} -\frac{\lambda^{2}}{4} c_{2}x +\frac{c_{4}}{(x +c_{2})^{2}}$ \\ $I_{2}= \frac{\dot{y}^{2}}{2} -\frac{\lambda^{2}}{2} y^{2} +c_{3}y$ \\ $I_{3}= \frac{\dot{z}^{2}}{2} +F(z)$ \\ $I_{4}= e^{\lambda t} \left[ \left( \dot{x} -\frac{\lambda}{2}x \right)^{2} -\lambda c_{2}\left( \dot{x} -\frac{\lambda}{2}x -\frac{\lambda}{4}c_{2} \right) +\frac{2c_{4}}{\left( x +c_{2} \right)^{2}} \right]$ \\
$I_{5}= e^{\lambda t} \left( \dot{y} -\lambda y +\frac{c_{3}}{\lambda} \right)$} \\ \hline
$V= -\frac{\lambda^{2}}{2}r^{2} +\frac{c_{1}x}{y^{2}R} +\frac{c_{2}}{y^{2}} +\frac{c_{3}}{z^{2}}$ & New & \makecell[l]{$I_{1}= \frac{1}{2}\dot{z}^{2} -\frac{\lambda^{2}}{2}z^{2} +\frac{c_{3}}{z^{2}}$ \\ $I_{2}= \frac{1}{2}M_{3}^{2} +\frac{(c_{1}R +c_{2}x)x}{y^{2}}$ \\ $I_{3}= e^{\lambda t} \left[ M_{3}(\dot{y} -\lambda y) +\frac{2c_{2}x}{y^{2}} +\frac{c_{1} (y^{2} +2x^{2})}{y^{2}R} \right]$ \\ $I_{4}= \frac{1}{2} \mathbf{M}^{2} +\frac{c_{1}xr^{2}}{y^{2}R} +c_{2}\frac{r^{2}}{y^{2}} +\frac{c_{3}R^{2}}{z^{2}}$} \\ \hline
$V= -\frac{\lambda^{2}}{2}r^{2} +\frac{k_{1}}{x^{2}} +\frac{k_{2}}{y^{2}}$ & New & \makecell[l]{$I_{1}= \frac{\dot{x}^{2}}{2} -\frac{\lambda^{2}}{2}x^{2} +\frac{k_{1}}{x^{2}}$ \\ $I_{2}= \frac{\dot{y}^{2}}{2} -\frac{\lambda^{2}}{2}y^{2} +\frac{k_{2}}{y^{2}}$ \\ $I_{3}= \frac{\dot{z}^{2}}{2} -\frac{\lambda^{2}}{2}z^{2}$ \\
$I_{4}= e^{\lambda t} \left[ M_{2}(\dot{x} -\lambda x) +\frac{2k_{1}z}{x^{2}} \right]$ \\ $I_{5}= e^{\lambda t} \left[ M_{1}(\dot{y} -\lambda y) -\frac{2k_{2}z}{y^{2}} \right]$} \\ \hline
$V= -\frac{\lambda^{2}}{2}r^{2} +\frac{F\left( \frac{y}{x} \right)}{R^{2}}$ & New & \makecell[l]{$I_{1}= e^{\lambda t} \left[ M_{2} \left( \dot{x} -\lambda x\right) -M_{1} \left( \dot{y} -\lambda y \right) +\frac{2zF\left(\frac{y}{x} \right)}{R^{2}} \right]$ \\ $I_{2}= \frac{1}{2}M_{3}^{2} + F\left( \frac{y}{x} \right)$ \\ $I_{3}= \frac{1}{2}\dot{z}^{2} -\frac{\lambda^{2}}{2}z^{2}$ \\ $I_{4}= \frac{1}{2} \mathbf{M}^{2} +\frac{r^{2} F\left( \frac{y}{x} \right)}{R^{2}}$} \\ \hline
\makecell[l]{$V= -\frac{\lambda^{2}}{8}(4x^{2} +4z^{2} +y^{2})-$ \\ \qquad $-\lambda^{2}c_{1}z -\frac{\lambda^{2}}{4}c_{2}y +\frac{k}{(y+c_{2})^{2}}$} & New & \makecell[l]{$I_{1}= e^{\lambda t} \left( \dot{x} -\lambda x\right)$ \\ $I_{2}= e^{\lambda t} \left\{ \left[ \dot{y} -\frac{\lambda}{2}(y+c_{2}) \right]^{2} +\frac{2k}{(y+c_{2})^{2}} \right\}$ \\ $I_{3}= e^{\lambda t} \left[ \dot{z} -\lambda (z+c_{1}) \right]$ \\ $I_{4}= M_{2} +c_{1}\dot{x}$} \\ \hline
\makecell[l]{$V= -\frac{\lambda^{2}}{2} R^{2} +\frac{kx}{y^{2}R} +\frac{c_{1}}{y^{2}}-$ \\ \qquad $-\frac{\lambda^{2}}{8}z^{2} +\frac{c_{2}}{z^{2}}$} & New & \makecell[l]{$I_{1}= e^{\lambda t} \left[ \left( \dot{z} -\frac{\lambda}{2}z \right)^{2} +\frac{2c_{2}}{z^{2}} \right]$ \\ $I_{2}= e^{\lambda t} \left[ M_{3} (\dot{y} -\lambda y) +\frac{2c_{1}x}{y^{2}} +\frac{k(y^{2} +2x^{2})}{y^{2}R} \right]$ \\
$I_{3}= \frac{1}{2}\dot{z}^{2} -\frac{\lambda^{2}}{8}z^{2} +\frac{c_{2}}{z^{2}}$ \\ $I_{4}= \frac{1}{2}M_{3}^{2} + \frac{(kR +c_{1}x)x}{y^{2}}$} \\ \hline
\caption{\label{Table.e3.9} Maximally superintegrable potentials $V(x,y,z)$ in $E^{3}$ that admit time-dependent LFIs/QFIs of the form $I_{(3)}$.}
\end{longtable}

\section{Comparison with existing results}

\label{Comparison with other works}

As we have remarked in section \ref{sec.3dint}, the main review works in this topic are the works of Evans in \cite{Evans 1990} and Kalnins in \cite{Kalnins 2006Pr}. Therefore, it is imperative to discuss how the present review is related to these.

\subsection{Evans work \cite{Evans 1990}}

Evans in \cite{Evans 1990}, using the separability of the Hamilton-Jacobi equation in $E^{3}$, determined all minimally and maximally superintegrable potentials with autonomous QFIs of the form $I= K_{ab}(q)\dot{q}^{a}\dot{q}^{b} +G(q)$. The author did not consider (autonomous or time-dependent) LFIs and time-dependent QFIs. In particular, in Table I of \cite{Evans 1990} are given five maximally superintegrable potentials and in Table II of \cite{Evans 1990} eight minimally superintegrable potentials.

As it can be seen from Tables \ref{Table.e23.1} - \ref{Table.e3.9}, all the results of \cite{Evans 1990} have been recovered plus new ones. Therefore, the claim made in \cite{Evans 1990} that all second order superintegrable potentials in $E^{3}$ are determined is not valid.

Furthermore, it should be noted that there are misprints in some results of \cite{Evans 1990}. Indeed, we have: \newline
1) In eq. (3.43) of \cite{Evans 1990}, the leading term of the QFI $I_{4}$ must be $L_{2}P_{1} -P_{2}L_{1}$. \newline
2) In Table II of \cite{Evans 1990}, the leading part of the QFIs $I_{3}$ associated with the potentials (\ref{eq.e3pot.27a}) and (\ref{eq.e3pot.27b}) should be $L_{2}P_{1} -P_{2}L_{1}$. \newline
3) The QFI $I_{2}$ in eq. (3.57) of \cite{Evans 1990} should be replaced with the QFI (\ref{eq.vs3}). \newline
4) The QFI $I_{3}$ in eq. (3.57) of \cite{Evans 1990} should be replaced with the QFI (\ref{eq.vs4}).

\subsection{Kalnins et all work \cite{Kalnins 2006Pr}}

In \cite{Kalnins 2006Pr}, the authors discussed classical 3d superintegrable nondegenerate (i.e. four-parameter) potentials on a conformally flat real or complex space. They proved that the quadratic algebra always closes at order six (the `$5 \implies 6$ Theorem'), that is, the space of autonomous QFIs is 6d. Moreover, using the St\"{a}ckel transformation (an invertible conformal mapping between superintegrable structures on distinct spaces), they gave strong evidence (no proof) that all nondegenerate 3d superintegrable systems are St\"{a}ckel transforms of constant curvature systems (i.e. the complex Euclidean space or the the complex 3-sphere). This means that in order to obtain all nondegenerate conformally flat superintegrable systems, it is sufficient to classify those in the complex Euclidean space and on the complex 3-sphere. Finally, they found eight families of superintegrable systems that are separable in generic coordinates.

Comparing the results of \cite{Kalnins 2006Pr} with the results of the present work, we note the following: \newline
1) All seven maximally superintegrable Euclidean potentials given in eqs. (10) - (16) of \cite{Kalnins 2006Pr} are recovered (see Table \ref{Table.e3.3}). \newline
2) The potentials given in eqs. (10) and (13) of \cite{Kalnins 2006Pr} have been found earlier in Table I of \cite{Evans 1990}. The potential (13) is more general from the one found by Evans. \newline
3) It is proved in section \ref{sec.e3.pot.1} that the potentials (11), (12), (14), (16) of \cite{Kalnins 2006Pr} are subcases of the more general potential (\ref{eq.kal25}) for specific forms of the arbitrary smooth functions $F_{1}(w,z)$ and $F_{2}(w)$. This justifies the fact that these potentials admit a QFI of the form $I= \dot{w}^{2} +G(x,y,z)$ where $w=x +iy$. \newline
4) The potential (15) of \cite{Kalnins 2006Pr} is a subcase of (\ref{eq.kal30}) and hence admits a QFI of the form $I= \dot{\bar{w}}^{2} +G(x,y,z)$. \newline
5) The potentials (12), (16) of \cite{Kalnins 2006Pr} are of the integrable form (\ref{eq.kal35}); therefore, they admit a QFI of the form $I= \dot{z}\dot{w} +G(x,y,z)$. \newline
6) The potentials (11), (12) of \cite{Kalnins 2006Pr} are of the integrable form (\ref{eq.kal43}); therefore, they admit a QFI of the form $I= \left( M_{2} -iM_{1} \right)^{2} +G(x,y,z)$. \newline
7) The potential (15) of \cite{Kalnins 2006Pr} is a subcase of the new minimally superintegrable potential (\ref{eq.kal14}) for $F(z)= k_{1}z^{2} +\frac{k_{4}}{z^{2}}$. For this reason, it admits an additional QFI of the form $I= \frac{1}{4}\dot{w}^{2} +iM_{3}\dot{\bar{w}} +G(x,y,z)$. \newline
8) The two additional maximally superintegrable potentials given in eq. (17) of \cite{Kalnins 2006Pr} are just subcases of the last maximally superintegrable potential in Table \ref{Table.e23.3} for $k_{1}=k_{2}=0$ when $F(z)= -\frac{\lambda^{2}}{8}z^{2} +c_{3}z$ and $F(z)= -\frac{\lambda^{2}}{32}z^{2} +\frac{c_{3}}{z^{2}}$.

Therefore, with the systematic application of Theorem \ref{The first integrals of an autonomous holonomic dynamical system}, we have found all the results of \cite{Kalnins 2006Pr} plus new ones; especially time-dependent QFIs.

\section{Conclusions}

\label{Conclusions}

The aim of the present work was twofold: a. To assess the second order integrability of autonomous conservative dynamical systems of the form $q^{a}=-V^{,a}(q)$ where $a=1,2,3$ in a systematic, i.e. algorithmic, way; and b. To enrich, if possible,  the existing results of the main sources on this topic which are found in the  review papers \cite{Evans 1990} and \cite{Kalnins 2006Pr}. Therefore, the present work should be approached as an updated review of the integrable/superintegrable 3d Newtonian autonomous conservative dynamical systems that admit LFIs/QFIs.

We have considered two types of integrable and superintegrable 3d Newtonian potentials. Potentials of the form $\Phi(x,y)+ F(z)$ which are $2+1$ decomposable and hence their QFIs follow from the QFIs of the 2d potentials $\Phi(x,y)$; and non-decomposable potentials $V(x,y,z)$ in $E^{3}$ which cannot be treated in this way. These latter potentials we have searched using the algorithm of Theorem \ref{The first integrals of an autonomous holonomic dynamical system}.

After a detailed study of the three types of QFIs $I_{(1,\ell)}, I_{(2,\ell)}, I_{(3)}$ considered in Theorem \ref{The first integrals of an autonomous holonomic dynamical system}, we have recovered all known integrable/superintegrable potentials together with new ones. It has also been shown that many of the existing results are in fact special cases of more general ones for specific values of the free parameters/functions. For convenience, the results in each case have been collected in tables which contain the known results with the appropriate reference and the new ones found in the present work. These results can be used in many ways in the study of the dynamical systems and, especially, in the case of more complex systems. One such study will be given elsewhere.

\end{document}